\documentclass[a4paper,10pt]{article}

\usepackage{latexsym}
\usepackage{amsmath}
\usepackage{amsthm}
\usepackage{amssymb}
\usepackage{enumerate}
\usepackage{multicol}
\usepackage{graphicx}

\allowdisplaybreaks

\theoremstyle{plain}
\newtheorem{theorem}{Theorem}[section]
\newtheorem{proposition}[theorem]{Proposition}
\newtheorem{lemma}[theorem]{Lemma}
\newtheorem{corollary}[theorem]{Corollary}

\theoremstyle{definition}
\newtheorem{definition}[theorem]{Definition}

\newtheorem{remark}[theorem]{Remark}
\numberwithin{equation}{section}

\def\C{{\mathbb{C}}}
\def\P{{\mathbb{P}}}
\def\T{{\mathbb{T}}}
\def\R{{\mathbb{R}}}
\def\N{{\mathbb{N}}}
\def\Z{{\mathbb{Z}}}
\def\X{{\mathcal{X}}}
\def\cZ{{\mathcal{Z}}}
\def\<{{\langle}}
\def\>{{\rangle}}

\def\Tr{\mathop{\mathrm{Tr}}}

\def\Res{\mathop{\mathrm{Res}}}
\def\opPi{\mathop{\mathrm{\Pi}}}
\def\sgn{\mathop{\mathrm{sgn}}\nolimits}

\def\mO{{\mathcal{O}}}
\def\b0{{\mathbf{0}}}

\def\be{{\mathbf{e}}}
\def\bk{{\mathbf{k}}}
\def\bl{{\mathbf{l}}}

\def\bx{{\mathbf{x}}}
\def\by{{\mathbf{y}}}
\def\bz{{\mathbf{z}}}
\def\bu{{\mathbf{u}}}
\def\bv{{\mathbf{v}}}
\def\bw{{\mathbf{w}}}
\def\bp{{\mathbf{p}}}

\def\bs{{\mathbf{s}}}

\def\bC{{\mathbf{C}}}

\def\opsi{{\overline{\psi}}}

\def\oeta{{\overline{\eta}}}

\def\spin{{\{\uparrow,\downarrow\}}}
\def\ua{{\uparrow}}
\def\da{{\downarrow}}
\def\la{{\lambda}}
\def\o{{\omega}}
\def\eps{{\varepsilon}}
\def\G{{\Gamma}}
\def\s{{\sigma}}

\begin{document}

\title{\large\bf A RIGOROUS TREATMENT OF THE PERTURBATION \\ \smallskip THEORY FOR MANY-ELECTRON SYSTEMS}
\author{\small YOHEI KASHIMA\smallskip \\
\small Institut f\"ur Theoretische Physik, Universit\"at Heidelberg\\
\small Philosophenweg 19, 69120 Heidelberg, Germany\\ 
\small y.kashima@thphys.uni-heidelberg.de}
\date{}

\maketitle

\begin{abstract}
\noindent
Four point correlation functions for many electrons at finite temperature in periodic lattice of dimension $d\ (\ge 1)$ are analyzed by the perturbation theory with respect to the coupling constant. The correlation functions are characterized as a limit of finite dimensional Grassmann integrals. 
A lower bound on the radius of convergence and an upper bound on the perturbation series are obtained by evaluating the Taylor expansion of logarithm of the finite dimensional Grassmann Gaussian integrals. The perturbation series up to second order is numerically implemented along with the volume-independent upper bounds on the sum of the higher order terms in 2 dimensional case.
\end{abstract}

\tableofcontents

\section{Introduction}
The thermal average of an observable $\mO$ for many electrons in a solid is expressed as $\Tr e^{-\beta H}\mO/\Tr e^{-\beta H}$, where $H$ is a Hamiltonian representing the total energy of the system, $\beta$ is the inverse temperature and the trace operation $\Tr$ is taken over the Fermionic Fock space, the Hilbert space of all the possible states of electrons. If the movements of electrons are confined in finite lattice sites under periodic boundary condition, the Fermionic Fock space becomes finite dimensional. The thermal average $\Tr e^{-\beta H}\mO/\Tr e^{-\beta H}$ is defined as a quotient of finite sums over the orthonormal basis spanning the space.
Though the expectation value $\Tr e^{-\beta H}\mO/\Tr e^{-\beta H}$ has a clear mathematical meaning in this setting, to rigorously control its behavior for interacting electrons poses a challenge. The purpose of this paper is to analyze the thermal expectation value for 4 point functions modeling paired electrons' condensation by means of the perturbation theory.

In the earlier article \cite{KT} Koma and Tasaki rigorously proved upper bounds on 2 point and 4 point correlation functions for the Hubbard model and concluded the decay properties in 1 and 2 dimensional cases. In an abstract general context, on the other hand, Feldman, Kn\"orrer and Trubowitz gave a concise representation of the Schwinger functionals formulating the correlation functions via Grassmann integral and established upper bounds of the Schwinger functionals in \cite{FKTrep}. Let us also remark the intensive renormalization group study by the same authors in \cite{FKTf3}, which analyzes the Grassmann integral formulation corresponding to the temperature zero limit of the correlation function for the momentum distribution function. The work \cite{FKTf3} was presented as the 11th paper in the series of Feldman, Kn\"orrer and Trubowitz's 2-d Fermi liquid construction. A flow chart showing the hierarchical relation between these 11 papers is found in the digest \cite{FKTf1}.

In this paper we focus on the correlation function $\Tr e^{-\beta H}\mO/\Tr e^{-\beta H}$ for 4 point functions $\mO=\psi_{\bx_1\ua}^*\psi_{\bx_2\da}^*\psi_{\by_2\da}\psi_{\by_1\ua}+\psi_{\by_1\ua}^*\psi_{\by_2\da}^*\psi_{\bx_2\da}\psi_{\bx_1\ua}$ and the Hubbard model $H$ defined on a finite lattice. We expand the 4 point correlation function as a perturbation series with respect to the coupling constant and study the properties of the perturbation series. We especially aim at establishing upper bounds on the sum of higher order terms of the perturbation series so that one can numerically measure the error between the correlation function and the low order terms of the perturbation series. More precisely, our goal is set to 
\begin{enumerate}[(1)]
\item\label{goal_radius} find a constant $r>0$ such that for any $U\in\R$ with $|U|\le r$
$$\frac{\Tr (e^{-\beta H}\mO)}{\Tr e^{-\beta H}}=\sum_{n=0}^{\infty}a_nU^n,$$
where $U$ denotes the coupling constant and $a_n\in\R$ $(\forall n\in\N\cup\{0\})$, and to 
\item\label{goal_estimate} establish an inequality of the form that for any $U\in\R$ with $|U|\le r$ and $m\in\N\cup \{0\}$
$$\left|\frac{\Tr (e^{-\beta H}\mO)}{\Tr e^{-\beta H}}-\sum_{n=0}^{m}a_nU^n\right|\le R_{m+1}(|U|),$$
where $ R_{m+1}(|U|)=O(|U|^{m+1})$ as $|U|\searrow 0$.
\end{enumerate}
The inequality claimed in \eqref{goal_estimate} is proved in Theorem \ref{thm_estimate_correlation_function} as our main result and a volume-independent $r$ required in \eqref{goal_radius}-\eqref{goal_estimate} is obtained in Proposition \ref{pro_evaluation_decay_constant} for 2 dimensional case.

Our strategy is based on the discretization of the integrals over the interval of temperature appearing in the temperature-ordered perturbation series. By replacing the integrals by finite Riemann sums we obtain a fully discrete analog of the perturbation series in which all the variables run in finite sets. The discretized perturbation series is formulated in a finite dimensional Grassmann Gaussian integral, which is rigorously defined as a linear functional on the finite dimensional linear space of Grassmann algebras. See \cite{Sc} for another approach to the finite dimensional Grassmann integral formulation based on the Lie-Trotter type formula. We then rewrite the 4 point correlation function as the Taylor series expansion of logarithm of the Grassmann Gaussian integral. By evaluating the partial derivatives of logarithm of the Grassmann Gaussian integral, which were characterized as the tree expansion by Salmhofer and Wieczerkowski in \cite{SW}, and passing the parameter defining the Riemann sum to infinity, we obtain an upper bound on each term of the perturbation series of the original correlation function. For completeness of the paper and convenience for readers, the derivation of the temperature-ordered perturbation series is presented in Appendices.

As a key lemma we make use of the volume- and temperature-independent upper bound on the determinant of the covariance matrix recently established by Pedra and Salmhofer in \cite{PS}. Pedra-Salmhofer's determinant bound enables us to find a numerical upper bound on the Fermionic perturbation theory in a simple argument. As one aim, this paper intends to show a practical application of Pedra-Salmhofer's determinant bound.

Let us note that the lower bound on the radius of convergence of the perturbation series proved in Theorem \ref{thm_estimate_correlation_function} and Proposition \ref{pro_evaluation_decay_constant} below for 2 dimensional case is proportional to $\beta^{-3}$. By applying advanced multi-scale, renormalization techniques to the correlation functions of the 2 dimensional Hubbard model, Rivasseau (\cite{R}) and Afchain, Magnen and Rivasseau (\cite{AMR}) proved that a lower bound on the radius of convergence is proportional to $(\log\beta)^{-2}$, which is larger than our lower bound for large $\beta$, i.e, small temperature. In this article, however, we feature calculating the quantities in a simple manner so that readers can verify the construction of the theory by themselves, rather than improving the temperature-dependency of the convergence of the perturbation theory via large machinery. 

Our motivation to implement the perturbation theory for many electrons with rigorous error estimate numerically was grown amid active research of numerical analysis for high temperature superconductivity. The macroscopic behavior of electromagnetic fields around a type-II superconductor is governed by a system of nonlinear Maxwell equations called the macroscopic critical-state models. Prigozhin initiated the variational formulation of the Bean critical-state model for type-II superconductivity and reported numerical simulations by finite element method in \cite{P}. Following Prigozhin's preceding work \cite{P}, finite element approximations of various macroscopic models have been studied in rigorous levels up until today. See \cite{BP}, \cite{K} for the latest developments on this subject. In a smaller length scale, the density of superconducting charge carriers, the induced magnetic field and motions of the quantized vortices in a type-II superconductor under an applied magnetic field can be simulated by solving the mesoscopic Ginzburg-Landau models. Numerical approximation schemes for the Ginzburg-Landau models such as finite element method, finite difference method and finite volume method are summarized in the review article \cite{D}, which also explains extensions of the Ginzburg-Landau models to describe high temperature superconductivity characterized by $d$-wave pairing symmetry. We now turn our attention to microscopic models governing many electrons in a solid and try to approximate the 4 point correlation functions, which are believed to exhibit the off-diagonal long-range order as explained by Yang in \cite{Ya} if superconductivity is happening in the system. However, the concept of error estimate for the numerical computation of the correlation functions formulated in the Fermionic Fock space is not yet seen in a mathematical literature as we can see for the macroscopic critical-state models and the mesoscopic Ginzburg-Landau models today. Hence, in this paper we attempt to propose an error analysis for the numerical approximation of the correlation functions defined in microscopic quantum theory and implement our numerical scheme in practice.

The contents of this paper are outlined as follows. In Section \ref{sec_perturbation_theory} the model Hamiltonian and the correlation function of our interest are defined. The perturbation series of the correlation function is derived. In Section \ref{sec_grassmann_gaussian_integral_formulation} the temperature-ordered perturbation series of the partition function is discretized and the discretized partition function is formulated in a finite dimensional Grassmann Gaussian integral. In Section \ref{sec_upper_bound} each coefficient of the perturbation series of the correlation function is evaluated and upper bounds on the sum over higher order terms are obtained as our main result. In Section \ref{sec_numerical_results} the perturbation series up to 2nd order is numerically implemented together with the error estimates between the 2nd order perturbation and the correlation function in 2 dimensional case. In Appendix \ref{appendix_Fock_space} the standard properties of the Fermionic Fock space are reviewed. A self-contained proof for the temperature-ordered perturbation series expansion is presented in Appendix \ref{appendix_perturbation_series}. Finally, the temperature-discrete covariance matrix is diagonalized and its determinant is calculated in Appendix \ref{appendix_covariance_matrix}.  

\section{The perturbation theory}\label{sec_perturbation_theory}
In this section we define the Hamiltonian operator, formulate 4 point correlation function governed by the Hamiltonian under finite temperature and expand the correlation function as a power series of the coupling constant. To analyze the properties of the power series of the 4 point correlation function derived in this section is set to be the main purpose of this paper.

\subsection{The Hubbard model}
First of all we define the Hubbard model $H$ as the field Hamiltonian operator on the Fermionic Fock space along with various notations and parameters treated in this paper. 

The spacial lattice $\G$ is defined by $\G:=\Z^d/(L\Z)^d$, where $L(\in \N)$ is the length of one edge of the rectangular lattice and $d(\in\N)$ stands for the space dimension.  

On any set $S$ we define Kronecker's delta $\delta_{x,y}$ $(x,y\in S)$ by $\delta_{x,y}:=1$ if $x$ is identical to $y$ in $S$, $\delta_{x,y}:=0$ otherwise. For example, $\delta_{(0,0),(L,L)}=1$ for $(0,0),(L,L)\in \Z^2/(L\Z)^2$.

For any proposition $A$ the function $1_A$ is defined by
$$1_A:=\left\{\begin{array}{ll} 1 &\text{ if }A\text{ is true,}\\ 0 &\text{ otherwise. }\end{array}\right.
$$

Using the annihilation operator $\psi_{\bx\s}$ and the creation operator $\psi_{\bx\s}^*$, which is the adjoint operator of $\psi_{\bx\s}$, at site $\bx\in\G$ and spin $\s\in\spin$, the free part $H_0$ and the interacting part $V$ of the Hubbard model $H$ are defined as follows.

\begin{equation}\label{eq_H_0}
H_0:= \sum_{\bx,\by\in \G}\sum_{\s,\tau\in\spin}F(\bx\s,\by\tau)
\psi_{\bx\s}^*\psi_{\by\tau},\ V:=U\sum_{\bx\in\G}\psi_{\bx\ua}^*\psi_{\bx\da}^*\psi_{\bx\da}\psi_{\bx\ua},
\end{equation}
where  
\begin{equation}\label{eq_matrix_F}
\begin{split}
F&(\bx\s,\by\tau):=\delta_{\s,\tau}\Bigg(-t\sum_{j=1}^d(\delta_{\bx,\by-\be_j}+\delta_{\bx,\by+\be_j})\\
&-t'\cdot 1_{d\ge 2}\sum_{j,k=1 \atop j<k}^d(\delta_{\bx,\by-\be_j-\be_k}+\delta_{\bx,\by-\be_j+\be_k}+\delta_{\bx,\by+\be_j-\be_k}+\delta_{\bx,\by+\be_j+\be_k})-\mu \delta_{\bx,\by}\Bigg),
\end{split}
\end{equation}
the vectors $\be_j\in\G$ $(j\in\{1,\cdots,d\})$ are given by $\be_j(l)=\delta_{j,l}$ for all $j,l\in \{1,\cdots,d\}$. The parameters $t,t',\mu,U\in \R$ are called the nearest neighbor hopping amplitude, the next to nearest neighbor hopping amplitude, the chemical potential and the coupling constant, respectively. Note that the term representing the next to nearest neighbor hopping in $F(\bx\s,\by\tau)$ is effective only for $d\ge 2$.

The Hubbard model $H$ is defined by $H:=H_0+V$ and is a self-adjoint operator on the Fermionic Fock space $F_f(L^2(\G\times\spin;\C))$. We summarize the definitions and the basic properties of the Fermionic Fock space, the annihilation, creation operators in Appendix \ref{appendix_Fock_space}. Here we note the fact that $\dim F_f(L^2(\G\times\spin;\C))=2^{2L^d}<+\infty$, which means that any linear operator on $F_f(L^2(\G\times\spin;\C))$ can be considered as a matrix.

Let us prepare some more notations used in this paper. For any linear operator $A:F_f(L^2(\G\times\spin;\C))\to F_f(L^2(\G\times\spin;\C))$, $\Tr A$ is defined by
$$\Tr A:= \sum_{l=1}^{2^{2L^d}}\<\phi_l,A\phi_l\>_{F_f},$$
where $\<\cdot,\cdot\>_{F_f}$ is the inner product of $F_f(L^2(\G\times\spin;\C))$ (see Appendix \ref{appendix_Fock_space}) and $\{\phi_l\}_{l=1}^{2^{2L^d}}$ is any orthonormal system of $F_f(L^2(\G\times\spin;\C))$. The correlation function $\<A\>$ under the finite temperature $T$ is defined by
$$\<A\>:=\frac{\Tr(e^{-\beta H}A)}{\Tr e^{-\beta H}},$$
where $\beta:=1/(k_{B}T)>0$ with the Boltzmann constant $k_B>0$.

The momentum lattice $\G^*$ is defined by $\G^*:=(2\pi\Z/L)^d/(2\pi\Z)^d$.

For any vectors $\alpha,\gamma$ of algebra of length $n$, let $\<\alpha,\gamma\>$ denote $\sum_{l=1}^n\alpha(l)\gamma(l)$. Let $\<\cdot,\cdot\>_{\C^n}$ denote the inner product of $\C^n$ defined by $\<\bu,\bv\>_{\C^n}:=\sum_{l=1}^n\overline{\bu(l)}\bv(l)$ for any $\bu,\bv\in\C^n$.

For any finite set $S$, $\sharp S$ stands for the number of elements contained in $S$.
 
Let $S_n$ denote the set of all the permutations on $n$ elements for $n\in\N$.

\subsection{The correlation function}
Our goal is to analyze the $4$ point correlation function\\
$\<\psi_{\bx_1\ua}^*\psi_{\bx_2\da}^*\psi_{\by_2\da}\psi_{\by_1\ua}+\psi_{\by_1\ua}^*\psi_{\by_2\da}^*\psi_{\bx_2\da}\psi_{\bx_1\ua}\>$ by means of the perturbation method with respect to the coupling constant $U$. The correlation function  of our interest can be derived from the logarithm of the partition function. Let us substitute real parameters $\{\lambda_{\bx,\by,\bz,\bw}\}_{\bx,\by,\bz,\bw\in\G}(\subset \R)$ into our Hamiltonian $H$ and define the parameterized Hamiltonian $H_{\lambda}$ by
\begin{equation}\label{eq_V_lambda}
H_{\lambda}:=H_0+V_{\lambda},\ V_{\lambda}:=\sum_{\bx,\by,\bz,\bw\in\G}U_{\bx,\by,\bz,\bw}\psi_{\bx\ua}^*\psi_{\by\da}^*\psi_{\bw\da}\psi_{\bz\ua},
\end{equation}
where we set 
\begin{equation}\label{eq_relation_U_lambda}
U_{\bx,\by,\bz,\bw}:=U\delta_{\bx,\by}\delta_{\bz,\bw}\delta_{\bx,\bz}+\lambda_{\bx,\by,\bz,\bw}+\lambda_{\bz,\bw,\bx,\by},
\end{equation}
for all $\bx,\by,\bz,\bw\in \G$. Note that $H_{\lambda}$ still keeps the self-adjoint property and that $H_{\la}|_{\lambda_{\bx,\by,\bz,\bw}=0,\forall \bx,\by,\bz,\bw\in\G}=H$.

To simplify notations, let $\X$ represent a vector in $\G^4$ in our argument unless otherwise stated. From now we fix $4$ sites $\bx_1,\bx_2,\by_1,\by_2\in\G$ to define the correlation function $\<\psi_{\bx_1\ua}^*\psi_{\bx_2\da}^*\psi_{\by_2\da}\psi_{\by_1\ua}+\psi_{\by_1\ua}^*\psi_{\by_2\da}^*\psi_{\bx_2\da}\psi_{\bx_1\ua}\>$ and write $\tilde{\X}_1=(\bx_1,\bx_2,\by_1,\by_2)$ and $\tilde{\X}_2=(\by_1,\by_2,\bx_1,\bx_2)$.

\begin{lemma}\label{lem_derivation_correlation_function}
The following equality holds. 
\begin{equation}\label{eq_derivation_correlation_function_1}
\<\psi_{\bx_1\ua}^*\psi_{\bx_2\da}^*\psi_{\by_2\da}\psi_{\by_1\ua}+\psi_{\by_1\ua}^*\psi_{\by_2\da}^*\psi_{\bx_2\da}\psi_{\bx_1\ua}\>=-\frac{1}{\beta}\frac{\partial}{\partial\lambda_{\tilde{\X}_1}}\log\left(\frac{\Tr e^{-\beta H_{\lambda}}}{\Tr e^{-\beta H_{0}}}\right)\Big|_{\lambda_{\X}=0\atop\forall \X\in\G^4}.
\end{equation}
\end{lemma}
\begin{remark}
Since $H_{\lambda}$ is self-adjoint, its spectrum $\s(H_{\lambda})$ is a subset of $\R$. The spectral mapping theorem (see, e.g,  \cite[\mbox{Section VIII-7,Corollary 1}]{Y}) shows that $\{e^{-\beta x}\}_{x\in\s(H_{\lambda})}$ is the spectrum of $e^{-\beta H_{\lambda}}$. Thus, $\Tr e^{-\beta H_{\lambda}}>0$. For the same reason as above the inequality $\Tr e^{-\beta H_{0}}>0$ holds. Therefore, $\log(\Tr e^{-\beta H_{\lambda}}/\Tr e^{-\beta H_{0}})$ is well-defined.
\end{remark}

Let $\mathcal{L}(F_f(L^2(\G\times\spin;\C)))$ denote the space of linear operators on $F_f(L^2(\G\times\spin;\C))$. The proof of Lemma \ref{lem_derivation_correlation_function} is based on the following lemma.
\begin{lemma}\label{lem_derivative_exponential}
Let $(a,b)$ be an interval of $\R$. Assume that $A:(a,b)\to  \mathcal{L}(F_f(L^2(\G\times\spin;\C)))$ is an operator-valued $C^1$-class function. The following equality holds. For all $s\in (a,b)$
$$
\frac{d}{ds}e^{A(s)}=\int_0^1e^{(1-t)A(s)}\frac{d}{ds}A(s)e^{tA(s)}dt.
$$
\end{lemma}
\begin{proof}
Fix any $s\in(a,b)$ and take small $\eps>0$ such that $[s-\eps,s+\eps]\subset (a,b)$. For any $s'\in (s-\eps,s+\eps)$
\begin{equation*}
\begin{split}
e^{A(s)}-e^{A(s')}&=[-e^{(1-t)A(s)}e^{tA(s')}]_{t=0}^{t=1}=-\int_0^1\frac{d}{dt}(e^{(1-t)A(s)}e^{tA(s')})dt\\
&=\int_0^1e^{(1-t)A(s)}(A(s)-A(s'))e^{tA(s')}dt.
\end{split}
\end{equation*}
Moreover, we see that
\begin{equation*}
\begin{split}
\frac{d}{ds}&e^{A(s)}=\lim_{s'\to s\atop s'\in (s-\eps,s+\eps)}\frac{e^{A(s)}-e^{A(s')}}{s-s'}\\
&=\lim_{s'\to s\atop s'\in (s-\eps,s+\eps)}\int_0^1e^{(1-t)A(s)}\frac{A(s)-A(s')}{s-s'}e^{tA(s')}dt=\int_0^1e^{(1-t)A(s)}\frac{d}{ds}A(s)e^{tA(s)}dt,
\end{split}
\end{equation*}
where we have used the inequality
$$\left\|\frac{A(s)-A(s')}{s-s'}e^{tA(s')}\right\|\le\sup_{\theta\in
 [s-\eps,s+\eps]}\left\|\frac{d}{ds}A(\theta)\right\|e^{\sup_{\theta \in [s-\eps,s+\eps]}\|A(\theta)\|}$$ 
with the operator norm $\|\cdot\|$ and Lebesgue's dominated convergence theorem to exchange the order of the limit operation and the integral.
\end{proof}
\begin{proof}[Proof of Lemma \ref{lem_derivation_correlation_function}] 
Since the operator-valued function $\lambda_{\tilde{\X}_1}\mapsto H_{\lambda}$ is continuously differentiable on any interval containing $0$ inside, we can apply Lemma \ref{lem_derivative_exponential} to have 
\begin{equation*}
\begin{split}
-\frac{1}{\beta}&\frac{\partial}{\partial\lambda_{\tilde{\X}_1}}\log\left(\frac{\Tr e^{-\beta H_{\lambda}}}{\Tr e^{-\beta H_{0}}}\right)\Big|_{\lambda_{\X}=0\atop\forall \X\in\G^4}\\
&=-\frac{1}{\beta}\frac{\Tr\left(\displaystyle\int_0^1 e^{(1-t)(-\beta H)}
\frac{\partial}{\partial\lambda_{\tilde{\X}_1}}(-\beta H_{\lambda})\Big|_{\lambda_{\X}=0\atop\forall \X\in\G^4}e^{t(-\beta H)}dt\right)}{\Tr e^{-\beta H}}
\end{split}
\end{equation*}
\begin{equation*}
\begin{split}
&=\int_0^1dt \frac{\Tr\left(e^{(1-t)(-\beta H)}(\psi_{\bx_1\ua}^*\psi_{\bx_2\da}^*\psi_{\by_2\da}\psi_{\by_1\ua}+\psi_{\by_1\ua}^*\psi_{\by_2\da}^*\psi_{\bx_2\da}\psi_{\bx_1\ua})e^{t(-\beta H)}\right)}{\Tr e^{-\beta H}}\\
&=\<\psi_{\bx_1\ua}^*\psi_{\bx_2\da}^*\psi_{\by_2\da}\psi_{\by_1\ua}+\psi_{\by_1\ua}^*\psi_{\by_2\da}^*\psi_{\bx_2\da}\psi_{\bx_1\ua}\>,
\end{split}
\end{equation*}
where we have used the equality that $\Tr(AB)=\Tr(BA)$ for any operators $A,B$.
\end{proof}

\subsection{The perturbation series}
The partition function $\Tr e^{-\beta H_{\lambda}}/\Tr e^{-\beta H_{0}}$ can be expanded as a power series of the parameter $\{U_{\X}\}_{\X\in\G^4}$. We give the derivation of the temperature-ordered perturbation series in Appendix \ref{appendix_perturbation_series}. Here we only state the result. 
\begin{proposition}\label{pro_perturbation_series}
For any $U\in \R$ and $\{\lambda_{\X}\}_{\X\in\G^4}\subset \R$,
\begin{equation}\label{eq_perturbation_series}
\begin{split}
&\frac{\Tr e^{-\beta
H_{\lambda}}}{\Tr e^{-\beta H_0}}\\
&\
 =1+\sum_{n=1}^{\infty}\frac{1}{n!}\opPi_{j=1}^n\Bigg(-\sum_{\bx_{2j-1},\bx_{2j},\by_{2j-1},\by_{2j}\in\G}\sum_{\s_{2j-1},\s_{2j}\in \spin}\delta_{\s_{2j-1},\ua}\delta_{\s_{2j},\da}\int_0^{\beta}dx_{2j-1}\\
&\quad\quad \cdot U_{\bx_{2j-1},\bx_{2j}, \by_{2j-1},\by_{2j}} \Bigg)\det(C(\bx_j\s_jx_{j},\by_k\s_kx_{k}))_{1\le j,k\le 2n}\Big|_{x_{2j}=x_{2j-1}\atop\forall j\in\{1,2,\cdots,n\}},
\end{split}
\end{equation}
where the constraint $x_{2j}=x_{2j-1}$ requires the variable $x_{2j}$ to take the same value as $x_{2j-1}$ for all $j\in\{1,2,\cdots,n\}$ and each component of the covariance matrix \\
$(C(\bx_j\s_jx_{j},\by_k\s_kx_{k}))_{1\le j,k\le 2n}$ is defined by
\begin{equation}\label{eq_covariance_matrix}
C(\bx\s x,\by\tau y):=\frac{\delta_{\s,\tau}}{L^d}\sum_{\bk\in\G^*}e^{i\<\bk,\by-\bx\>}e^{-(y-x)E_{\bk}}\left(\frac{1_{y-x\le
0}}{1+e^{\beta E_{\bk}}}-\frac{1_{y-x>
0}}{1+e^{-\beta E_{\bk}}}\right)
\end{equation}
with the dispersion relation
\begin{equation}\label{eq_dispersion_relation}
E_{\bk}:=-2t\sum_{j=1}^d\cos(\<\bk,\be_j\>) - 4t'\cdot 1_{d\ge 2}\sum_{j,k = 1\atop j<k}^d\cos(\<\bk,\be_j\>)\cos(\<\bk,\be_k\>)-\mu.
\end{equation}
\end{proposition}

Lemma \ref{lem_derivation_correlation_function} indicates that we can construct the power series of \\
$\<\psi_{\bx_1\ua}^*\psi_{\bx_2\da}^*\psi_{\by_2\da}\psi_{\by_1\ua}+\psi_{\by_1\ua}^*\psi_{\by_2\da}^*\psi_{\bx_2\da}\psi_{\bx_1\ua}\>$ by substituting the series \eqref{eq_perturbation_series} into the Taylor series expansion of the function $\log(x)$ around $x=1$. Since the radius of convergence of the Taylor series of $\log(x)$ around $1$ is $1$, we need to know when the inequality $|\Tr e^{-\beta H_{\lambda}}/\Tr e^{-\beta H_{0}}-1|<1$ holds beforehand. An answer will be given to this question in Proposition \ref{pro_perturbation_series_P} below.

It will be more convenient for our analysis to generalize the problems so that the variables $U$, $\{\lambda_{\X}\}_{\X\in\G^4}$, $\{U_{\X}\}_{\X\in\G^4}$ are allowed to be complex. We will then recover the statements on our original problem by restricting the variables to be real. For $\{U_{\X}\}_{\X\in\G^4}\subset \C$ we define a function $P(\{U_{\X}\}_{\X\in\G^4})$ by the power series of the right hand side of \eqref{eq_perturbation_series}. Let us recall that the real function $\log(x)$ $(x>0)$ is extended to be the complex analytic function $\log(z)$ in the domain $\{z\in\C\ |\ |z-1|<1\}$ by the power series
$$\log(z)=\sum_{n=1}^{\infty}\frac{(-1)^{n-1}}{n}(z-1)^n.$$

In our argument to clarify when the inequality
\begin{equation}\label{eq_domain_P}
|P(\{U_{\X}\}_{\X\in\G^4})-1|<1
\end{equation}
holds as well as in the proofs of other lemmas in this paper, the following lemma on the determinant bound on the covariance matrix plays essential roles.
\begin{lemma}\cite[\mbox{Theorem 2.4}]{PS}\label{lem_determinant_bound}
For any $n\in\N$, $(\bx_j,\s_j,x_{j}),(\by_j,\tau_j,y_{j})\in \G\times \spin\times [0,\beta)$ $(\forall j\in\{1,\cdots,n\})$,
\begin{equation*}
\sup_{\bu_j,\bv_j\in\C^n\text{ with }\|\bu_j\|_{\C^n},\|\bv_j\|_{\C^n}\le 1\atop\forall j\in \{1,\cdots,n\}}|\det(\<\bu_j,\bv_k\>_{\C^n}C(\bx_j\s_j x_{j},\by_k\tau_k y_{k}))_{1\le j,k \le n}|\le 4^n,
\end{equation*}
where $\|\bu\|_{\C^n}:=\<\bu,\bu\>_{\C^n}^{1/2}$ for all $\bu\in\C^n$.
\end{lemma}
\begin{remark} The statement of \cite[\mbox{Theorem 2.4}]{PS} is on the determinant bound of the covariance matrices independent of the spin coordinate. It is, however, straightforward to derive the bound claimed in Lemma \ref{lem_determinant_bound} on our spin-dependent covariance matrix from \cite[\mbox{Theorem 2.4}]{PS}.
\end{remark}

We can expand $-1/\beta \partial/\partial\lambda_{\tilde{\X}_1}\log(P(\{U_{\X}\}_{\X\in\G^4}))|_{\lambda_{\X}=0,\forall\X\in\G^4}$ as a power series of $U$ as follows. 

\begin{proposition}\label{pro_perturbation_series_P}
Assume that $U\in\C$ satisfies $|U|<\log 2/(16\beta L^{4d})$.
Then there exists $\eps>0$ such that if $\{\lambda_{\X}\}_{\X\in\G^4}$ satisfies $|\lambda_{\X}|\le \eps$ for all $\X\in\G^4$, the inequality \eqref{eq_domain_P} holds. Moreover, we have
\begin{equation}\label{eq_perturbation_series_P_1}
-\frac{1}{\beta}\frac{\partial}{\partial\lambda_{\tilde{\X}_1}}\log(P(\{U_{\X}\}_{\X\in\G^4}))\Big|_{\lambda_{\X}=0\atop\forall\X\in\G^4}=\sum_{n=0}^{\infty}a_nU^n,
\end{equation}
where the coefficients $\{a_n\}_{n=0}^{\infty}$ are given by 
\begin{equation}\label{eq_perturbation_series_P_1_1}
a_n := -\frac{1}{\beta}\frac{\partial}{\partial \la_{\tilde{\X}_1}}\left(
\sum_{j=1}^{n+1}\frac{(-1)^{j-1}}{j}\sum_{m_1+\cdots+m_j=n+1\atop m_k\ge 1,\forall k\in\{1,\cdots,j\}}\opPi_{k=1}^jG_{m_k}\right)\Bigg|_{\la_{\X}=0\atop\forall \X\in\G^4},
\end{equation}
with $\{G_n\}_{n=1}^{\infty}$ defined by 
\begin{equation}\label{eq_perturbation_series_P_2}
\begin{split}
G_n:=& \frac{1}{n!}\opPi_{j=1}^n\Bigg(-\sum_{\bx_{2j-1},\bx_{2j},\by_{2j-1},\by_{2j}\in\G}\sum_{\s_{2j-1},\s_{2j}\in \spin}\int_0^{\beta}dx_{2j-1}\delta_{\s_{2j-1},\ua}\delta_{\s_{2j},\da}\\
&\cdot(\delta_{\bx_{2j-1},\bx_{2j}}\delta_{\by_{2j-1},\by_{2j}}\delta_{\bx_{2j-1}, \by_{2j-1}}+\la_{\bx_{2j-1},\bx_{2j},\by_{2j-1},\by_{2j}}+\la_{\by_{2j-1},\by_{2j},\bx_{2j-1},\bx_{2j}})\Bigg)\\
&\cdot\det(C(\bx_j\s_jx_{j},\by_k\s_kx_{k}))_{1\le j,k\le 2n}\Big|_{x_{2j}=x_{2j-1}\atop\forall j\in\{1,2,\cdots,n\}}.
\end{split}
\end{equation}
\end{proposition}
\begin{proof}
Let us fix $U\in\C$ with $|U|<\log 2/(16\beta L^{4d})$. Take any \\
$\eps\in (0,\log 2/(32\beta L^{4d})-|U|/2)$ and assume that $\{\lambda_{\X}\}_{\X\in\G^4}$ satisfies $|\lambda_{\X}|\le \eps$ for all $\X\in\G^4$. Then, we see that for all $\X\in\G^4$
\begin{equation}\label{eq_perturbation_series_P_3}
|U_{\X}|<\frac{\log 2}{16\beta L^{4d}}.
\end{equation}
By using the inequality \eqref{eq_perturbation_series_P_3} and Lemma \ref{lem_determinant_bound} we observe that
\begin{equation}\label{eq_perturbation_series_P_3'}
|P(\{U_{\X}\}_{\X\in\G^4})-1|<\sum_{n=1}^{\infty}\frac{1}{n!}\left(\beta L^{4d}\cdot\frac{\log 2}{16\beta L^{4d}}\cdot16\right)^n=e^{\log2}-1=1.
\end{equation}
The inequality \eqref{eq_perturbation_series_P_3'} allows us to consider $\log(P(\{U_{\X}\}_{\X\in\G^4}))$ as an analytic function of the multi-variable $\{U_{\X}\}_{\X\in\G^4}$ in the domain \eqref{eq_perturbation_series_P_3}. Moreover, we have
\begin{equation}\label{eq_perturbation_series_P_4}
\begin{split}
 -\frac{1}{\beta}&\frac{\partial}{\partial\lambda_{\tilde{\X}_1}}\log(P(\{U_{\X}\}_{\X\in\G^4}))\Big|_{\lambda_{\X}=0\atop\forall\X\in\G^4}\\
&=-\frac{1}{\beta}\sum_{m=0}^{\infty}(-1)^m\left(P(\{U_{\X}\}_{\X\in\G^4})\Big|_{\lambda_{\X}=0\atop\forall\X\in\G^4}-1\right)^m\frac{\partial}{\partial\lambda_{\tilde{\X}_1}} P(\{U_{\X}\}_{\X\in\G^4})\Big|_{\lambda_{\X}=0\atop\forall\X\in\G^4},
\end{split}
\end{equation}
where we used the equality that $d\log(z)/dz = \sum_{m=0}^{\infty}(-1)^m(z-1)^m$ ($\forall z\in\C$ with $|z-1|<1$). Furthermore, we can write
\begin{equation}\label{eq_Fn_Gn}
\begin{split}
&P(\{U_{\X}\}_{\X\in\G^4})\Big|_{\lambda_{\X}=0\atop\forall\X\in\G^4}-1=\sum_{n=1}^{\infty}G_n\Big|_{\la_{\X}=0\atop\forall \X\in\G^4}U^n,\\
&\frac{\partial}{\partial \la_{\tilde{\X}_1}}P(\{U_{\X}\}_{\X\in\G^4})\Big|_{\lambda_{\X}=0\atop\forall\X\in\G^4}=\sum_{n=1}^{\infty}\frac{\partial}{\partial \la_{\tilde{\X}_1}}G_n\Big|_{\la_{\X}=0\atop\forall \X\in\G^4}U^{n-1},
\end{split}
\end{equation}
where $G_n$ $(n\in\N)$ is defined in \eqref{eq_perturbation_series_P_2}. By substituting \eqref{eq_Fn_Gn} into \eqref{eq_perturbation_series_P_4} we obtain
\begin{equation}\label{eq_composition_Fn_Gn}
\begin{split}
-\frac{1}{\beta}&\frac{\partial}{\partial \la_{\tilde{\X}_1}}\log(P(\{U_{\X}\}_{\X\in\G^4}))\Big|_{\lambda_{\X}=0\atop\forall\X\in\G^4}\\
&=-\frac{1}{\beta}\sum_{m=0}^{\infty}\left(-\sum_{n=1}^{\infty}G_n\Big|_{\la_{\X}=0\atop\forall \X\in\G^4}U^n\right)^m\sum_{n=1}^{\infty}\frac{\partial}{\partial \la_{\tilde{\X}_1}}G_n\Big|_{\la_{\X}=0\atop\forall \X\in\G^4}U^{n-1}.
\end{split}
\end{equation}

Again by using Lemma \ref{lem_determinant_bound} we can show that for $U\in \C$ with \\
$|U|<\log 2/(16\beta L^{4d})$
\begin{equation}\label{eq_perturbation_series_P_5}
\sum_{n=1}^{\infty}\left|G_n\Big|_{\lambda_{\X}=0\atop\forall\X\in\G^4}\right||U|^n<1,\ \sum_{n=1}^{\infty}\left|\frac{\partial}{\partial \la_{\tilde{\X}_1}}G_n\Big|_{\lambda_{\X}=0\atop\forall\X\in\G^4}\right||U|^{n-1}<\infty.
\end{equation}
Since the radius of convergence of the power series $\sum_{m=0}^{\infty}z^m$ is $1$, the inequalities \eqref{eq_perturbation_series_P_5} provide a sufficient condition to reorder the right hand side of \eqref{eq_composition_Fn_Gn} (see, e.g, \cite[\mbox{Theorem 3.1, Theorem 3.4}]{La} for products and compositions of convergent power series) to deduce 
\begin{equation}\label{eq_perturbation_series_P_6}
\begin{split}
-\frac{1}{\beta}&\frac{\partial}{\partial\lambda_{\tilde{\X}_1}}\log(P(\{U_{\X}\}_{\X\in\G^4}))\Big|_{\lambda_{\X}=0\atop\forall\X\in\G^4}\\
&=-\frac{1}{\beta}\frac{\partial}{\partial \la_{\tilde{\X}_1}}G_1\Big|_{\lambda_{\X}=0\atop\forall\X\in\G^4} -\frac{1}{\beta}\sum_{n=1}^{\infty}\Bigg(\sum_{l=1}^{n}\frac{\partial}{\partial \la_{\tilde{\X}_1}}G_{l}\Big|_{\lambda_{\X}=0\atop\forall\X\in\G^4}
\sum_{j=1}^{n+1-l}\sum_{m_1+\cdots+m_j=n+1-l\atop m_k\ge 1,\forall k\in\{1,\cdots,j\}}\\
&\quad\cdot\opPi_{k=1}^j(-G_{m_k})\Big|_{\lambda_{\X}=0\atop\forall\X\in\G^4}+\frac{\partial}{\partial \la_{\tilde{\X}_1}}G_{n+1}\Big|_{\lambda_{\X}=0\atop\forall\X\in\G^4}\Bigg)U^n.
\end{split}
\end{equation}
Arranging \eqref{eq_perturbation_series_P_6} yields \eqref{eq_perturbation_series_P_1_1}.
\end{proof}

By restricting $U$ to be real in \eqref{eq_perturbation_series_P_1}, we obtain the power series expansion of the correlation function $\<\psi_{\bx_1\ua}^*\psi_{\bx_2\da}^*\psi_{\by_2\da}\psi_{\by_1\ua}+\psi_{\by_1\ua}^*\psi_{\by_2\da}^*\psi_{\bx_2\da}\psi_{\bx_1\ua}\>$. At this point, however, we only know that the series $\sum_{n=0}^{\infty}a_nU^n$ converges for $U\in\C$ with $|U|<\log 2/(16\beta L^{4d})$, which heavily depends on the volume factor $L^d$. With the aim of enlarging the radius of convergence and finding upper bounds of the power series $\sum_{n=0}^{\infty}a_nU^n$, we will construct our theory in the following sections.

\section{Grassmann Gaussian integral formulation}\label{sec_grassmann_gaussian_integral_formulation}
In this section we discretize the integrals over $[0,\beta]$ contained in the perturbation series $P(\{U_{\X}\}_{\X\in\G^4})$ so that the discretized perturbation series can be formulated in a Grassmann Gaussian integral involving only finite dimensional Grassmann algebras. Moreover, by showing that the discrete analog of $P$ uniformly converges to the original $P$, we characterize our partition function $\Tr e^{-\beta H}/\Tr e^{-\beta H_0}$ and the 4 point function $\<\psi_{\bx_1\ua}^*\psi_{\bx_2\da}^*\psi_{\by_2\da}\psi_{\by_1\ua}+\psi_{\by_1\ua}^*\psi_{\by_2\da}^*\psi_{\bx_2\da}\psi_{\bx_1\ua}\>$ as a limit of finite dimensional Grassmann integrals. The finite dimensional Grassmann Gaussian integral formulation will then enable us to apply the tree formula for the connected part of the exponential of Laplacian operator of the Grassmann left derivatives to express each term of the discretized perturbation series as a finite sum over trees in Section \ref{sec_upper_bound}.

\subsection{Discretization of the integral over $[0,\beta]$}
We define the fully discrete perturbation series by replacing the integral $\int_0^{\beta}dx$ in $P(\{U_{\X}\}_{\X\in\G^4})$ by the Riemann sum. Let us introduce finite sets $[0,\beta)_h$ and $[-\beta,\beta)_h$ parameterized by $h\in \N/\beta$ as follows. 
\begin{equation*}
\begin{split}
&[0,\beta)_h:= \left\{0,\frac{1}{h},\frac{2}{h},\cdots,\beta-\frac{1}{h}\right\},\\
&[-\beta,\beta)_h:=\left\{-\beta,-\beta +\frac{1}{h},-\beta +\frac{2}{h},\cdots,\beta-\frac{1}{h}\right\}.
\end{split}
\end{equation*}
Note that $\sharp [0,\beta)_h=\beta h$ and $\sharp [-\beta,\beta)_h=2\beta h$.
We define the function $P_h(\{U_{\X}\}_{\X\in\G^4})$ of the multi-variable $\{U_{\X}\}_{\X\in\G^4}(\subset\C)$ by 
\begin{equation}\label{eq_perturbation_series_P_h}
\begin{split}
&P_h(\{U_{\X}\}_{\X\in\G^4}) :=\\
&\quad1+\sum_{n=1}^{L^d\beta h}\frac{1}{n!}\opPi_{j=1}^n\Bigg(-\frac{1}{h}\sum_{\bx_{2j-1},\bx_{2j},\by_{2j-1},\by_{2j}\in\G}\sum_{\s_{2j-1},\s_{2j}\in \spin}\sum_{x_{2j-1},x_{2j}\in[0,\beta)_h}\\
&\quad\cdot\delta_{\s_{2j-1},\ua}\delta_{\s_{2j},\da}\delta_{x_{2j-1},x_{2j}}U_{\bx_{2j-1},\bx_{2j}, \by_{2j-1},\by_{2j}} \Bigg)\det(C(\bx_j\s_jx_{j},\by_k\s_kx_{k}))_{1\le j,k\le 2n}.
\end{split}
\end{equation}
Note that if $n>L^d\beta h$, $\det(C(\bx_j\s_jx_{j},\by_k\tau_ky_{k}))_{1\le j,k\le 2n}=0$ for any $(\bx_j,\s_j, x_{j}),$\\
$(\by_j,\tau_j,y_{j})\in \G\times\spin\times[0,\beta)_h$ $(j\in\{1,\cdots,2n\})$, since $\sharp \G\times \spin \times [0,\beta)_h=2L^d\beta h$. 

Let us summarize the properties of the function $P_h$ in the same manner as in Proposition \ref{pro_perturbation_series_P}
\begin{lemma}\label{lem_properties_P_h}
Assume that $U\in\C$ satisfies $|U|<\log2/(16\beta L^{4d})$. The following statements hold.
\begin{enumerate}[(i)]
\item\label{item_lem_properties_P_h_1} There exists $\eps >0$ such that for any $\{\lambda_{\X}\}_{\X\in\G^4}(\subset\C)$ with $|\lambda_{\X}|\le\eps$ $(\forall \X\in\G^4)$ and any $h\in \N/\beta$, the inequality $|P_h(\{U_{\X}\}_{\X\in\G^4})-1|<1$ holds.
\item\label{item_lem_properties_P_h_2}For any $h\in \N/\beta$
\begin{equation*}
-\frac{1}{\beta}\frac{\partial}{\partial\lambda_{\tilde{\X}_1}}\log(P_h(\{U_{\X}\}_{\X\in\G^4}))\Big|_{\lambda_{\X}=0\atop\forall\X\in\G^4}=\sum_{n=0}^{\infty}a_{h,n}U^n,
\end{equation*}
where the coefficients $\{a_{h,n}\}_{n=0}^{\infty}$ are given by 
\begin{equation}\label{eq_def_a_h_n}
a_{h,n} := -\frac{1}{\beta}\frac{\partial}{\partial \la_{\tilde{\X}_1}}\Bigg(
\sum_{j=1}^{n+1}\frac{(-1)^{j-1}}{j}\sum_{m_1+\cdots+m_j=n+1\atop m_k\ge 1,\forall k\in\{1,\cdots,j\}}\opPi_{k=1}^jG_{h,m_k}\Bigg)\Bigg|_{\la_{\X}=0\atop\forall \X\in\G^4},
\end{equation}
with $\{G_{h,n}\}_{n=1}^{\infty}$ defined by 
\begin{equation}\label{eq_def_F_h_n_G_h_n}
\begin{split}
&G_{h,n}:= \frac{1}{n!}\opPi_{j=1}^n\Bigg(-\frac{1}{h}\sum_{\bx_{2j-1},\bx_{2j},\by_{2j-1},\by_{2j}\in\G}\sum_{\s_{2j-1},\s_{2j}\in \spin}\sum_{x_{2j-1}, x_{2j}\in [0,\beta)_h}\\
&\cdot \delta_{\s_{2j-1},\ua}\delta_{\s_{2j},\da}\delta_{x_{2j-1},x_{2j}}\\
&\cdot(\delta_{\bx_{2j-1},\bx_{2j}}\delta_{\by_{2j-1},\by_{2j}}\delta_{\bx_{2j-1}, \by_{2j-1}}+\la_{\bx_{2j-1},\bx_{2j},\by_{2j-1},\by_{2j}}+\la_{\by_{2j-1},\by_{2j},\bx_{2j-1},\bx_{2j}})\Bigg)\\
&\cdot\det(C(\bx_j\s_jx_{j},\by_k\s_kx_{k}))_{1\le j,k\le 2n}.
\end{split}
\end{equation}
\item\label{item_lem_properties_P_h_3} For all $n\in \N\cup\{0\}$, $\lim_{h\to +\infty,h\in \N/\beta}a_{h,n}=a_n$, where $\{a_n\}_{n=0}^{\infty}$ is defined in \eqref{eq_perturbation_series_P_1_1}-\eqref{eq_perturbation_series_P_2}.
\end{enumerate}
\end{lemma}
\begin{proof}
The proofs for the claims \eqref{item_lem_properties_P_h_1} and \eqref{item_lem_properties_P_h_2} are parallel to that of Proposition \ref{pro_perturbation_series_P}, based on Lemma \ref{lem_determinant_bound}. By the definition \eqref{eq_covariance_matrix} $\det(C(\bx_j\s_jx_{j},\by_k\s_kx_{k}))_{1\le j,k\le 2n}$ is piece-wise smooth with respect to the variables $\{x_j\}_{j=1}^{2n}$, which implies that the Riemann sums over $[0,\beta)_h$ in $G_{h,n}$ all converge to the corresponding integrals in $G_n$ as $h\to +\infty$. Thus, the claim \eqref{item_lem_properties_P_h_3} is true.
\end{proof}

Lemma \ref{lem_properties_P_h} \eqref{item_lem_properties_P_h_3} tells us that establishing an $h$-dependent upper bound on $|a_{h,n}|$ and showing that the upper bound converges as $h\to +\infty$ lead to finding a bound on $|a_n|$. This goal will be achieved in Section \ref{sec_upper_bound}.

The main aim of this section is to formulate $P_h$ as a finite dimensional Grassmann Gaussian integral, which will be used in the characterization of the coefficients $\{a_{h,n}\}_{n=0}^{\infty}$ in Section \ref{sec_upper_bound}. Though it is not directly required in our search for the upper bound on $\sum_{n=0}^{\infty}a_nU^n$, to represent the original partition function $P$ and the 4 point function $\<\psi_{\bx_1\ua}^*\psi_{\bx_2\da}^*\psi_{\by_2\da}\psi_{\by_1\ua}+\psi_{\by_1\ua}^*\psi_{\by_2\da}^*\psi_{\bx_2\da}\psi_{\bx_1\ua}\>$ as a limit of the finite dimensional Grassmann integrals also interests us. The following uniform convergence property of $P_h$ provides a framework to this purpose. The following proposition will be referred in the proof of our main theorem Theorem \ref{thm_estimate_correlation_function} as well.

\begin{proposition}\label{pro_convergence_P_h}
For any $r>0$
\begin{equation}\label{eq_uniform_convergence_P_h}
\lim_{h\to +\infty\atop h\in \N/\beta}\sup_{U_{\X}\in\C\text{ with }|U_{\X}|\le r\atop\forall \X\in\G^4}|P_h(\{U_{\X}\}_{\X\in\G^4})-P(\{U_{\X}\}_{\X\in\G^4})|=0.
\end{equation}
\end{proposition}
\begin{remark}
For the same reason as for the convergence property Lemma \ref{lem_properties_P_h} \eqref{item_lem_properties_P_h_3}, each term of the series $P_h(\{U_{\X}\}_{\X\in\G^4})$ converges to the corresponding term of $P(\{U_{\X}\}_{\X\in\G^4})$ as $h\to +\infty$. By using this fact and Lebesgue's dominated convergence theorem for $l^1$-space, the convergence property \eqref{eq_uniform_convergence_P_h} can be shown. Below we present an elementary proof without employing the convergence theorem of the Lebesgue integration theory.
\end{remark}
\begin{proof}[Proof of Proposition \ref{pro_convergence_P_h}] 
By using Lemma \ref{lem_determinant_bound} and the inequality that $|U_{\X}|\le r\ (\forall \X\in\G^4)$, we have 
\begin{equation}\label{eq_convergence_P_h_1}
\begin{split}
&|P(\{U_{\X}\}_{\X\in\G^4})-P_h(\{U_{\X}\}_{\X\in\G^4})|\le \sum_{n=\beta h +
 1}^{\infty}\frac{2}{n!}(r\beta L^{4d})^n4^{2n}\\
&+ \sum_{n=2}^{\beta h}\frac{1}{n!}\opPi_{j=1}^n \left(\sum_{\bx_{2j-1},\bx_{2j},\by_{2j-1},\by_{2j} \in
 \G}\sum_{\s_{2j-1},\s_{2j}\in\spin}\delta_{\s_{2j-1},\ua}\delta_{\s_{2j},\da}
|U_{\bx_{2j-1},\bx_{2j},\by_{2j-1},\by_{2j}}|\right)\\
&\qquad\cdot\Bigg|\opPi_{j=1}^n\left(\int_0^{\beta}ds_{2j-1}\right)\det(C(\bx_j\s_js_j,\by_k\s_k s_k))_{1\le j,k\le 2n}\Big|_{s_{2j}=s_{2j-1}\atop\forall j\in\{1,\cdots,n\}}\\
&\qquad\quad-\opPi_{j=1}^n\left(\frac{1}{h}\sum_{x_{2j-1}\in [0,\beta)_h}\right)\det(C(\bx_j\s_jx_{j},\by_k\s_k x_{k}))_{1\le j,k\le 2n}\Big|_{x_{2j}=x_{2j-1}\atop\forall j\in\{1,\cdots,n\}}\Bigg|\\
&\le \sum_{n=\beta h +
 1}^{\infty}\frac{2}{n!}(r\beta L^{4d})^n4^{2n}\\
&\quad + \sum_{n= 2}^{\beta h}\frac{1}{n!}(r L^{4d})^n\sup_{\bx_j,\by_j\in\G,\s_j\in\spin\atop\forall j\in\{1,\cdots,2n\}}
\opPi_{j=1}^n\left(\sum_{l_{2j-1}=0}^{\beta h
 -1}\int_{l_{2j-1}/h}^{(l_{2j-1}+1)/h}ds_{2j-1}\right)\\
&\qquad\cdot\Big|\det(C(\bx_j\s_js_j,\by_k\s_ks_k))_{1\le j,k\le 2n}\Big|_{s_{2j}=s_{2j-1}\atop\forall j\in\{1,\cdots,n\}} \\
&\qquad\quad-\det(C(\bx_j\s_jl_j/h,\by_k\s_k
 l_{k}/h))_{1\le j,k\le 2n}\Big|_{l_{2j}=l_{2j-1}\atop\forall j\in\{1,\cdots,n\}}\Big|.
\end{split}
\end{equation}
We especially need to show that the second term of the right hand side of the
 inequality \eqref{eq_convergence_P_h_1} converges to 0 as $h\to +\infty$. 

Let us fix $n\in \{2,3,\cdots,\beta h\}$ and $\bx_j,\by_j\in \G$, $\s_j\in \spin$ $(\forall j\in \{1,\cdots,2n\})$. There exists a function 
$$
g:(-\beta,\beta)^{n(n-1)/2}\to \R,\ g\in
C^{\infty}(((-\beta,\beta)\backslash\{0\})^{n(n-1)/2})
$$
such that for all $s_{2j-1}\in [0,\beta)$ $(\forall j\in\{1,\cdots,n\})$
\begin{equation*}
\begin{split}
&g(s_1-s_3,s_1-s_5,\cdots,
 s_1-s_{2n-1},s_3-s_5,\cdots,s_3-s_{2n-1},\cdots,s_{2n-3}-s_{2n-1})\\
&\quad=\det(C(\bx_j\s_j s_j,\by_k\s_k s_k))_{1\le j,k\le
 2n}|_{s_{2j}=s_{2j-1}\atop\forall j\in\{1,\cdots,n\}}.
\end{split}
\end{equation*}
Note that by using the property that $E_{\bk}=E_{-\bk}$ for all $\bk\in\G^*$, we can show $C(\bx\s x,\by\tau y)\in\R$ for all $(\bx,\s,x),(\by,\tau,y)\in\G\times\spin\times [0,\beta)_h$. Thus, the function $g$ is chosen to be real-valued. Then we see that
\begin{equation}\label{eq_convergence_P_h_2}
\begin{split}
&\opPi_{j=1}^n\left(\sum_{l_{2j-1}=0}^{\beta h
 -1}\int_{l_{2j-1}/h}^{(l_{2j-1}+1)/h}ds_{2j-1}\right)\\
&\qquad\cdot\Big|\det(C(\bx_j\s_js_j,\by_k\s_k s_k))_{1\le j,k\le 2n}|_{s_{2j}=s_{2j-1}\atop\forall j\in\{1,\cdots,n\}}\\
&\qquad\quad-\det(C(\bx_j\s_jl_{j}/h,\by_k\s_k l_{k}/h))_{1\le j,k\le 2n}|_{l_{2j}=l_{2j-1}\atop\forall j\in\{1,\cdots,n\}}\Big|\\
 &=\opPi_{j=1}^n\left(\sum_{l_{2j-1}=0}^{\beta h
 -1}\int_{l_{2j-1}/h}^{(l_{2j-1}+1)/h}ds_{2j-1}\right)(\chi_{\bl} +
 (1-\chi_{\bl})\chi_{\bl,\bs})\\
&\qquad \cdot|g(s_1-s_3,\cdots,s_{2n-3}-s_{2n-1}) -
 g(l_1/h-l_3/h,\cdots,l_{2n-3}/h-l_{2n-1}/h)|,
 \end{split}
\end{equation}
where the functions $\chi_{\bl}$, $\chi_{\bl,\bs}$ are defined by
$$
\chi_{\bl}:=\left\{\begin{array}{ll}1 & \text{ if there exist
	 }j,k\in\{1,\cdots,n\}\text{ such that }j\neq k\text{ and }
	  l_{2j-1}=l_{2k-1},\\
0 & \text{ otherwise,}\end{array}
\right.
$$
and 
$$
\chi_{\bl,\bs}:=\left\{\begin{array}{ll}1 & \text{ if }s_{2j-1}-s_{2k-1}\neq
	     l_{2j-1}/h-l_{2k-1}/h\\ & \text{ for all }j,k\in
	      \{1,\cdots,n\}\text{ with }j\neq k, \\
0 & \text{ otherwise.}\end{array}\right.
$$

Let us fix $\bl=(l_1,l_3,\cdots,l_{2n-1})$ and
 $\bs=(s_1,s_3,\cdots,s_{2n-1})$ with \\
$l_{2j-1}\in \{0,1,\cdots,\beta h-1\}$, $s_{2j-1}\in
 (l_{2j-1}/h,(l_{2j-1}+1)/h)$ for all $j\in\{1,\cdots,n\}$ satisfying
 $\chi_{\bl}=0$ and $\chi_{\bl,\bs}=1$. In this case $l_{2j-1}\neq l_{2k-1}$ and $s_{2j-1}-s_{2k-1}\neq l_{2j-1}/h-l_{2k-1}/h$ for
 all $j,k\in\{1,\cdots,n\}$ with $j\neq k$. 
Note that if $l_{2j-1}<l_{2k-1}$, $l_{2j-1}/h-l_{2k-1}/h, s_{2j-1}-s_{2k-1}\in (-\beta,0)$. If $l_{2j-1}>l_{2k-1}$, $l_{2j-1}/h-l_{2k-1}/h, s_{2j-1}-s_{2k-1}\in (0
,\beta)$. 
Let us set the interval $I(b,c)$ for $b,c\in\R$ with $b\neq c$ by 
\begin{equation*}
I(b,c) := 
[b,c] \text{ if } b < c, [c,b]  \text{ if } b > c.
\end{equation*}
Then we see that $I(s_{2j-1}-s_{2k-1},l_{2j-1}/h-l_{2k-1}/h)\subset (-\beta,\beta)\backslash\{0\}$ 
for all $j,k\in\{1,\cdots,n\}$ with $j\neq k$.
Since  $g\in
C^{\infty}(((-\beta,\beta)\backslash\{0\})^{n(n-1)/2})$, the mean value theorem ensures that for any $j,k\in \{1,\cdots,n\}$ with $j<k$ there exists
 $\theta_{2j-1,2k-1}\in I(s_{2j-1}-s_{2k-1}, l_{2j-1}/h-l_{2k-1}/h)$
 such that
\begin{equation*}
\begin{split}
&g(s_1-s_3,\cdots,s_{2n-3}-s_{2n-1}) -
 g(l_1/h-l_3/h,\cdots,l_{2n-3}/h-l_{2n-1}/h)\\
&=\<\nabla g(\theta_{1,3},\cdots,\theta_{2n-3,2n-1}),\\
&\qquad(s_1-s_3-(l_1/h-l_3/h),\cdots,s_{2n-3}-s_{2n-1}-(l_{2n-3}/h-l_{2n-1}/h))^t\>,
\end{split}
\end{equation*}
which leads to 
\begin{equation}\label{eq_convergence_P_h_3}
\begin{split}
&|g(s_1-s_3,\cdots,s_{2n-3}-s_{2n-1}) -
 g(l_1/h-l_3/h,\cdots,l_{2n-3}/h-l_{2n-1}/h)|\\
&\ \le\frac{1}{h}
\left(\frac{n(n-1)}{2}\right)^{1/2}\sup_{\bs\in
 ((-\beta,\beta)\backslash\{0\})^{n(n-1)/2}}|\nabla g(\bs)|.
\end{split}
\end{equation}

Moreover, by using Lemma \ref{lem_determinant_bound} we
 see that for $j<k$,
\begin{equation}\label{eq_convergence_P_h_4}
\begin{split}
&\left|\frac{\partial}{\partial(s_{2j-1}-s_{2k-1})}g(s_1-s_3,\cdots,s_{2n-3}-s_{2n-1})\right|\le\sum_{p_1=0}^1\sum_{p_2=0}^1\\
&\cdot\Bigg(\left|\frac{\partial}{\partial(s_{2j-1}-s_{2k-1})}C(\bx_{2j-1+p_1}\s_{2j-1+p_1}s_{2j-1},\by_{2k-1+p_2}\s_{2k-1+p_2}s_{2k-1})\right|\\
&\quad+\left|\frac{\partial}{\partial(s_{2j-1}-s_{2k-1})}C(\bx_{2k-1+p_1}\s_{2k-1+p_1}s_{2k-1},\by_{2j-1+p_2}\s_{2j-1+p_2}s_{2j-1})\right|\Bigg)4^{2n-1}\\
&\le 8\cdot 4^{2n-1}\sup_{\bx\in \G,x\in
 (-\beta,\beta)\backslash\{0\}}\left|\frac{\partial}{\partial
 x}C(\bx\ua x,\b0\ua 0)\right|.
\end{split}
\end{equation}
By \eqref{eq_convergence_P_h_3}-\eqref{eq_convergence_P_h_4} we have
\begin{equation}\label{eq_convergence_P_h_5}
\begin{split}
&\opPi_{j=1}^n\left(\sum_{l_{2j-1}=0}^{\beta h
 -1}\int_{l_{2j-1}/h}^{(l_{2j-1}+1)/h}ds_{2j-1}\right)(1-\chi_{\bl})\chi_{\bl,\bs}\\
&\quad \cdot|g(s_1-s_3,\cdots,s_{2n-3}-s_{2n-1}) -
 g(l_1/h-l_3/h,\cdots,l_{2n-3}/h-l_{2n-1}/h)|\\
&\le \frac{n(n-1)\beta^n 4^{2n}}{h}\sup_{\bx\in \G,x\in
 (-\beta,\beta)\backslash\{0\}}\left|\frac{\partial}{\partial
 x}C(\bx\ua x,\b0\ua 0)\right|.
\end{split}
\end{equation}

On the other hand, note that
\begin{equation}\label{eq_convergence_P_h_6}
\begin{split}
\sharp\{\bl/h\in[0,\beta)_h^n\ |\ \chi_{\bl}=1 \}&= (\beta h)^n
-\sharp\{\bl/h\in [0,\beta)_h^n\ |\ \chi_{\bl}=0\}\\
&=(\beta h)^n -\left(\begin{array}{c}\beta h \\ n \end{array}
\right)n!\le n^2 (\beta h)^{n-1},
\end{split}
\end{equation}
where we used the inequality
$$N^n-\left(\begin{array}{c}N \\ n\end{array}
\right)n!\le n^2N^{n-1},$$
which holds for all $N\in \N$ and $n\in \{0,1,\cdots,N\}$. By using Lemma \ref{lem_determinant_bound} and \eqref{eq_convergence_P_h_6} we obtain
\begin{equation}\label{eq_convergence_P_h_7}
\begin{split}
&\opPi_{j=1}^n\left(\sum_{l_{2j-1}=0}^{\beta h
 -1}\int_{l_{2j-1}/h}^{(l_{2j-1}+1)/h}ds_{2j-1}\right)\chi_{\bl}\\
&\quad\cdot |g(s_1-s_3,\cdots,s_{2n-3}-s_{2n-1}) -
 g(l_1/h-l_3/h,\cdots,l_{2n-3}/h-l_{2n-1}/h)|\\
&\le 2 h^{-n}n^2(\beta
 h)^{n-1}4^{2n}=\frac{2}{h}n^2\beta^{n-1}4^{2n}.
\end{split}
\end{equation}
Combining \eqref{eq_convergence_P_h_2}, \eqref{eq_convergence_P_h_5}, \eqref{eq_convergence_P_h_7} with \eqref{eq_convergence_P_h_1} shows
\begin{equation*}
\begin{split}
&\sup_{U_{\X}\in\C\text{ with }|U_{\X}|\le r\atop\forall \X\in\G^4}|P(\{U_{\X}\}_{\X\in\G^4})-P_h(\{U_{\X}\}_{\X\in\G^4})|\\
&\le \sum_{n=\beta
 h+1}^{\infty}\frac{2}{n!}(r\beta L^{4d})^n4^{2n} + \frac{1}{h}\sum_{n=2}^{\beta h}\frac{1}{n!}(r L^{4d})^n\\
&\quad\cdot\left(n(n-1)\beta^n 4^{2n}\sup_{\bx\in \G,x\in (-\beta,\beta)\backslash\{0\}}\left|\frac{\partial}{\partial x}C(\bx\ua x,\b0\ua 0)\right|+2n^2\beta^{n-1}4^{2n}\right)\\
&\to 0,
\end{split}
\end{equation*}
as $h\to +\infty$, $h\in \N/\beta$.
\end{proof}

\begin{corollary}\label{cor_correlation_P_h}
For all $U\in\R$
\begin{equation}\label{eq_correlation_P_h}
\<\psi_{\bx_1\ua}^*\psi_{\bx_2\da}^*\psi_{\by_2\da}\psi_{\by_1\ua}+\psi_{\by_1\ua}^*\psi_{\by_2\da}^*\psi_{\bx_2\da}\psi_{\bx_1\ua}\>=-\frac{1}{\beta}\lim_{h\to+\infty\atop h\in\N/\beta}\frac{\frac{\partial}{\partial \la_{\tilde{\X}_1}}P_h(\{U_{\X}\}_{\X\in\G^4})}{P_h(\{U_{\X}\}_{\X\in\G^4})}\Bigg|_{\la_{\X}=0\atop\X\in\G^4}.
\end{equation}
\end{corollary}
\begin{proof}
The relation \eqref{eq_relation_U_lambda} and Cauchy's integral formula ensure that for any \\
$\{\tilde{U}_{\X}\}_{\X\in\G^4} \subset \C$ and $r>0$
\begin{equation}\label{eq_application_cauchy_formula}
\begin{split}
&\frac{\partial}{\partial \la_{\tilde{\X}_1}}(P_h-P)(\{\tilde{U}_{\X}\}_{\X\in\G^4})=\left(\frac{\partial}{\partial U_{\tilde{\X}_1}}+\frac{\partial}{\partial U_{\tilde{\X}_2}}\right)(P_h-P)(\{\tilde{U}_{\X}\}_{\X\in\G^4})\\
&=\frac{1}{2\pi i}\Bigg(\oint_{|U_{\tilde{\X}_1}-\tilde{U}_{\tilde{\X}_1}|=r}dU_{\tilde{\X}_1}\frac{(P_h-P)(\{U_{\X}\}_{\X\in\G^4})}{(U_{\tilde{\X}_1}-\tilde{U}_{\tilde{\X}_1})^2}\\
&\qquad\qquad +\oint_{|U_{\tilde{\X}_2}-\tilde{U}_{\tilde{\X}_2}|=r}dU_{\tilde{\X}_2}\frac{(P_h-P)(\{U_{\X}\}_{\X\in\G^4})}{(U_{\tilde{\X}_2}-\tilde{U}_{\tilde{\X}_2})^2}\Bigg)\Bigg|_{U_{\X}=\tilde{U}_{\X}\atop \forall \X\in\G^4}.
\end{split}
\end{equation}
By applying Proposition \ref{pro_convergence_P_h} to \eqref{eq_application_cauchy_formula} we can show that for any $\tilde{r}>0$ and any $\{\tilde{U}_{\X}\}_{\X\in\G^4}$ with $|\tilde{U}_{\X}|\le \tilde{r}$ ($\forall \X\in\G^4$)
\begin{equation}\label{eq_uniform_convergence_derivative_P_h}
\begin{split}
&\left|\frac{\partial}{\partial \la_{\tilde{\X}_1}}(P_h-P)(\{\tilde{U}_{\X}\}_{\X\in\G^4})\right|\\
&\le \frac{2}{r}\sup_{\{U_{\X}\}_{\X\in\G^4}\subset \C\atop |U_{\X}|\le r+\tilde{r},\forall \X\in\G^4}|P_h(\{U_{\X}\}_{\X\in\G^4}) -P(\{U_{\X}\}_{\X\in\G^4})|\to 0
\end{split}
\end{equation}
as $h\to +\infty$, $h\in \N/\beta$. Combining \eqref{eq_uniform_convergence_derivative_P_h} with Lemma \ref{lem_derivation_correlation_function} yields \eqref{eq_correlation_P_h}.
\end{proof}

\subsection{The Grassmann Gaussian integral}\label{subsec_grassmann_integral}
To deal with the discretized partition function $P_h$ rather than $P$ is advantageous since the variables run in the finite set $\G\times\spin\times[0,\beta)_h$ in every term of the power series $P_h$. Accordingly, we can formulate $P_h$ as a Grassmann Gaussian integral on finite Grassmann algebras. Elementary calculus on finite Grassmann algebras has been summarized in the books \cite{FKT}, \cite{S}. For a convenience of calculation, especially in order to refer to Proposition \ref{pro_determinant_covariance} shown in Appendix \ref{appendix_covariance_matrix}, we assume that $h\in2\N/\beta$ from now.

Let us number elements of the set $\G\times\spin\times[0,\beta)_h$ so that we can write $\G\times\spin\times[0,\beta)_h=\{(\bx_j,\s_j,x_{j})\ |\ j\in\{1,\cdots,N\}\}$ with $N:=2L^d\beta h$. We then introduce a set of Grassmann algebras denoted by $\{\psi_{\bx_j\s_jx_{j}},\opsi_{\bx_j\s_jx_{j}}\ |\ j\in\{1,\cdots,N\}\}$. Remind us that the Grassmann algebra $\{\psi_{\bx_j\s_jx_{j}},\opsi_{\bx_j\s_jx_{j}}\ |\ j\in\{1,\cdots,N\}\}$ satisfies the anti-commutation relations
\begin{equation*}
\begin{split}
&\psi_{\bx_j\s_jx_{j}}\psi_{\bx_k\s_kx_{k}}=-\psi_{\bx_k\s_kx_{k}}\psi_{\bx_j\s_jx_{j}},\ \psi_{\bx_j\s_jx_{j}}\opsi_{\bx_k\s_kx_{k}}=-\opsi_{\bx_k\s_kx_{k}}\psi_{\bx_j\s_jx_{j}},\\
&\opsi_{\bx_j\s_jx_{j}}\opsi_{\bx_k\s_kx_{k}}=-\opsi_{\bx_k\s_kx_{k}}\opsi_{\bx_j\s_jx_{j}}
\end{split}
\end{equation*}
for all $j,k\in\{1,\cdots,N\}$.  

Let $\C[\psi_{\bx_j\s_jx_{j}},\opsi_{\bx_j\s_jx_{j}}|j\in\{1,\cdots,N\}]$ denote the complex linear space spanned by all the monomials consisting of $\{\psi_{\bx_j\s_jx_{j}},\opsi_{\bx_j\s_jx_{j}}|j\in\{1,\cdots,N\}\}$. As a linear functional on $\C[\psi_{\bx_j\s_jx_{j}},\opsi_{\bx_j\s_jx_{j}}|j\in\{1,\cdots,N\}]$, the Grassmann integral \\
$\int\cdot d\opsi_{\bx_N\s_Nx_{N}}\cdots d\opsi_{\bx_1\s_1x_{1}}d\psi_{\bx_N\s_Nx_{N}}\cdots d\psi_{\bx_1\s_1x_{1}}$ is defined as follows.
\begin{equation*}
\begin{split}
&\int\psi_{\bx_1\s_1x_{1}}\cdots \psi_{\bx_N\s_Nx_{N}}\opsi_{\bx_1\s_1x_{1}}\cdots \opsi_{\bx_N\s_Nx_{N}}\\
&\qquad\qquad\qquad \cdot d\opsi_{\bx_N\s_Nx_{N}}\cdots d\opsi_{\bx_1\s_1x_{1}}d\psi_{\bx_N\s_Nx_{N}}\cdots d\psi_{\bx_1\s_1x_{1}}:=1,\\
&\int \psi_{\bx_{j_1}\s_{j_1}x_{j_1}}\cdots
\psi_{\bx_{j_n}\s_{j_n}x_{j_n}}\opsi_{\bx_{k_1}\s_{k_1}x_{k_1}}\cdots
\opsi_{\bx_{k_m}\s_{k_m}x_{k_m}}\\
&\qquad\qquad\qquad \cdot d\opsi_{\bx_N\s_Nx_{N}}\cdots
d\opsi_{\bx_1\s_1x_{1}}d\psi_{\bx_N\s_Nx_{N}}\cdots
d\psi_{\bx_1\s_1x_{1}}:=0
\end{split}
\end{equation*}
if $n\neq N$ or $m\neq N$, and linearly extended onto the whole space.

Let us simply write the vectors of the Grassmann algebras \\
$(\psi_{\bx_1\s_1x_{1}},\cdots,\psi_{\bx_N\s_Nx_{N}})$, $(\opsi_{\bx_1\s_1x_{1}},\cdots,\opsi_{\bx_N\s_Nx_{N}})$ as  
$$
\psi_X=(\psi_{\bx_1\s_1x_{1}},\cdots,\psi_{\bx_N\s_Nx_{N}}),\ \opsi_X=(\opsi_{\bx_1\s_1x_{1}},\cdots,\opsi_{\bx_N\s_Nx_{N}}).
$$

In order to indicate the dependency on the parameter $h$, we write the covariance matrix as
$$C_h:=(C(\bx_j\s_jx_{j},\bx_k\s_kx_{k}))_{1\le j,k\le N}$$
and define a $2N\times 2N$ skew symmetric matrix $\bC_h$ by
$$
\bC_h:=\left(\begin{array}{cc}0 & C_h \\ -C_h^t & 0 \end{array}\right).
$$
The diagonalization of $C_h$ is presented in Appendix \ref{appendix_covariance_matrix}. Here we note the fact that $\det C_h\neq0$ proved in Proposition \ref{pro_determinant_covariance} to see that $\bC_h$ is invertible.

For any $f(\psi_X,\opsi_X)\in \C[\psi_{\bx_j\s_jx_{j}},\opsi_{\bx_j\s_jx_{j}}|j\in\{1,\cdots,N\}]$, $e^{f(\psi_X,\opsi_X)}$ is defined by 
$$
e^{f(\psi_X,\opsi_X)}:=e^{f(0,0)}\left(\sum_{n=0}^{2N}\frac{1}{n!}(f(\psi_X,\opsi_X)-f(0,0))^n\right),
$$
where $f(0,0)$ denotes the constant part of $f(\psi_X,\opsi_X)$.

Let us also write in short
$$d\opsi_X = d\opsi_{\bx_N\s_Nx_{N}}\cdots d\opsi_{\bx_1\s_1x_{1}},\ d\psi_X= d\psi_{\bx_N\s_Nx_{N}}\cdots d\psi_{\bx_1\s_1x_{1}}.
$$
\begin{definition}\label{defn_grassmann_gaussian_integral}
As a linear functional on $\C[\psi_{\bx_j\s_jx_{j}},\opsi_{\bx_j\s_jx_{j}}|j\in\{1,\cdots,N\}]$, the Grassmann Gaussian integral $\int\cdot d\mu_{C_h}(\psi_X,\opsi_X)$ is defined by
\begin{equation}\label{eq_grassmann_gaussian_integral}
\int f(\psi_X,\opsi_X)d\mu_{C_h}(\psi_X,\opsi_X):=\frac{\int
 f(\psi_X,\opsi_X)e^{-\frac{1}{2}\<(\psi_X,\opsi_X)^t,\bC_h^{-1}(\psi_X,\opsi_X)^t\>}d\opsi_X d\psi_X}{\int
 e^{-\frac{1}{2}\<(\psi_X,\opsi_X)^t,\bC_h^{-1}(\psi_X,\opsi_X)^t\>}d\opsi_X d\psi_X},
\end{equation}
for all $f(\psi_X,\opsi_X)\in \C[\psi_{\bx_j\s_jx_{j}},\opsi_{\bx_j\s_jx_{j}}|j\in\{1,\cdots,N\}]$.
\end{definition}
\begin{remark}
The denominator of \eqref{eq_grassmann_gaussian_integral} is non-zero. In fact, a direct calculation and the assumption $h\in 2\N/\beta$ show
\begin{equation*}
\int
 e^{-\frac{1}{2}\<(\psi_X,\opsi_X)^t,\bC_h^{-1}(\psi_X,\opsi_X)^t\>}d\opsi_X d\psi_X=(\det C_h)^{-1}(-1)^{N(N-1)/2}= (\det C_h)^{-1},
\end{equation*}
which takes a positive value independent of $h$ by Proposition \ref{pro_determinant_covariance}.
\end{remark}
 
The Grassmann Gaussian integral representation of $P_h$ is as follows.
\begin{proposition}\label{pro_grassmann_integral_formulation_P_h}
Assume that $\{U_{\X}\}_{\X\in\G^4}(\subset\C)$ satisfies the equality that for all $\bx,\by,\bz,\bw\in\G$
\begin{equation}\label{eq_constraint_U}
U_{\bx,\by,\bz,\bw}=U_{\bz,\bw,\bx,\by}.
\end{equation}
The following equality holds.
\begin{equation*}
P_h(\{U_{\X}\}_{\X\in\G^4})=\int
e^{\sum_{\bx,\by,\bz,\bw\in \G}U_{\bx,\by,\bz,\bw}V_{h,\bx,\by,\bz,\bw}(\psi_X,\opsi_X)}d\mu_{C_h}(\psi_X,\opsi_X),
\end{equation*}
where 
$$V_{h,\bx,\by,\bz,\bw}(\psi_X,\opsi_X):=-\frac{1}{h}\sum_{x\in [0,\beta)_h}\opsi_{\bx\ua x}\opsi_{\by\da x}\psi_{\bw\da x}\psi_{\bz\ua x}.$$
\end{proposition}

\begin{proof}
By substituting the equalities $\int1d\mu_{C_h}(\psi_X,\opsi_X)=1$ and 
\begin{equation*}
\begin{split}
\int &\psi_{\bx_{j_n}\s_{j_n}x_{j_n}}\cdots
\psi_{\bx_{j_1}\s_{j_1}x_{j_1}}\opsi_{\bx_{k_1}\s_{k_1}x_{k_1}}\cdots
\opsi_{\bx_{k_n}\s_{k_n}x_{k_n}}
d\mu_{C_h}(\psi_X,\opsi_X)\\
&\qquad=\det(C_h(\bx_{j_p}\s_{j_p}x_{j_p},\bx_{k_q}\s_{k_q}x_{k_q}))_{1\le
 p,q\le n}
\end{split}
\end{equation*}
for any $j_1,\cdots,j_n,k_1,\cdots,k_n\in\{1,2,\cdots,N\}$ (see \cite[\mbox{Problem I.13}]{FKT}) into \eqref{eq_perturbation_series_P_h}, we have 
\begin{equation}\label{eq_grassmann_integral_formulation_P_h_1}
\begin{split}
&P_h(\{U_{\X}\}_{\X\in\G^4}) =1+\sum_{n=1}^{L^d\beta
 h}\frac{1}{n!}\opPi_{j=1}^n\Bigg(-\frac{1}{h}\sum_{\bx_{2j-1},\bx_{2j},\by_{2j-1},\by_{2j}\in\G}\sum_{\s_{2j-1},\s_{2j}\in \spin}\\
&\qquad \cdot\sum_{x_{2j-1},x_{2j}\in [0,\beta)_h}\delta_{\s_{2j-1},\ua}\delta_{\s_{2j},\da}\delta_{x_{2j-1},x_{2j}}U_{\bx_{2j-1},\bx_{2j},\by_{2j-1},\by_{2j}} \Bigg)\\
&\qquad\cdot\int \psi_{\bx_{2n}\s_{2n}x_{2n}}\cdots
\psi_{\bx_{1}\s_{1}x_{1}}\opsi_{\by_{1}\s_{1}x_{1}}\cdots
\opsi_{\by_{2n}\s_{2n}x_{2n}}
d\mu_{C_h}(\psi_X,\opsi_X)\\
&=\int\Bigg(1+\sum_{n=1}^{L^d\beta
 h}\frac{1}{n!}\opPi_{j=1}^n\Bigg(-\frac{1}{h}\sum_{\bx_{2j-1},\bx_{2j},\by_{2j-1},\by_{2j}\in\G}\sum_{\s_{2j-1},\s_{2j}\in \spin}\sum_{x_{2j-1},x_{2j}\in [0,\beta)_h}\\
&\qquad\cdot\delta_{\s_{2j-1},\ua}\delta_{\s_{2j},\da}\delta_{x_{2j-1},x_{2j}}U_{\bx_{2j-1},\bx_{2j},\by_{2j-1},\by_{2j}}\\
&\qquad\cdot\psi_{\bx_{2j}\s_{2j}x_{2j}}\psi_{\bx_{2j-1}\s_{2j-1}x_{2j-1}}\opsi_{\by_{2j-1}\s_{2j-1}x_{2j-1}}\opsi_{\by_{2j}\s_{2j}x_{2j}}\Bigg)\Bigg)
d\mu_{C_h}(\psi_X,\opsi_X)\\ 
&=\int
 e^{-\frac{1}{h}\sum_{\bx,\by,\bz,\bw\in\G}\sum_{x\in [0,\beta)_h}U_{\bx,\by,\bz,\bw}\opsi_{\bx\ua x}\opsi_{\by\da x}\psi_{\bw\da x}\psi_{\bz\ua x}}d\mu_{C_h}(\psi_X,\opsi_X). 
\end{split}
\end{equation}
To obtain the last equality of \eqref{eq_grassmann_integral_formulation_P_h_1} we used the equality \eqref{eq_constraint_U}.
\end{proof}
As a corollary, our original partition functions and the correlation function are represented as a limit of the finite dimensional Grassmann integrals. 
\begin{corollary}
For any $U\in \R$ and $\{\lambda_{\X}\}_{\X\in\G^4}\subset\R$, the following equalities hold.
\begin{align}
&\frac{\Tr e^{-\beta H_{\lambda}}}{\Tr e^{-\beta H_0}}=\lim_{h\to +\infty\atop h\in 2\N/\beta}
\int e^{\sum_{\bx,\by,\bz,\bw\in \G}U_{\bx,\by,\bz,\bw}V_{h,\bx,\by,\bz,\bw}(\psi_X,\opsi_X)}d\mu_{C_h}(\psi_X,\opsi_X),\label{eq_grassmann_integral_formulation_Tr_H_lambda}\\
&\frac{\Tr e^{-\beta H}}{\Tr e^{-\beta H_0}}=\lim_{h\to +\infty\atop h\in 2\N/\beta}
\int e^{-\frac{U}{h}\sum_{\bx\in \G}\sum_{x\in[0,\beta)_h}\opsi_{\bx\ua x}\opsi_{\bx\da x}\psi_{\bx\da x}\psi_{\bx\ua x}}d\mu_{C_h}(\psi_X,\opsi_X),\label{eq_grassmann_integral_formulation_Tr_H}
\end{align}
\begin{equation}\label{eq_grassmann_integral_formulation_correlation_function}
\begin{split}
&\<\psi_{\bx_1\ua}^*\psi_{\bx_2\da}^*\psi_{\by_2\da}\psi_{\by_1\ua}+\psi_{\by_1\ua}^*\psi_{\by_2\da}^*\psi_{\bx_2\da}\psi_{\bx_1\ua}\>\\
&\quad=\lim_{h\to +\infty\atop h\in 2\N/\beta}\frac{1}{\beta h}\sum_{x\in [0,\beta)_h}\int(\opsi_{\bx_1\ua x}\opsi_{\bx_2\da x}\psi_{\by_2\da x}\psi_{\by_1\ua x}+\opsi_{\by_1\ua x}\opsi_{\by_2\da x}\psi_{\bx_2\da x}\psi_{\bx_1\ua x})\\
&\qquad\qquad\cdot e^{-\frac{U}{h}\sum_{\bx\in \G}\sum_{x\in[0,\beta)_h}\opsi_{\bx\ua x}\opsi_{\bx\da x}\psi_{\bx\da x}\psi_{\bx\ua x}}d\mu_{C_h}(\psi_X,\opsi_X)\\
&\qquad\qquad\Big/\int e^{-\frac{U}{h}\sum_{\bx\in \G}\sum_{x\in[0,\beta)_h}\opsi_{\bx\ua x}\opsi_{\bx\da x}\psi_{\bx\da x}\psi_{\bx\ua x}}d\mu_{C_h}(\psi_X,\opsi_X).
\end{split}
\end{equation}
\end{corollary}
\begin{proof}
Since the relation \eqref{eq_relation_U_lambda} implies the condition \eqref{eq_constraint_U}, we can apply Proposition \ref{pro_convergence_P_h} and Proposition \ref{pro_grassmann_integral_formulation_P_h} to deduce \eqref{eq_grassmann_integral_formulation_Tr_H_lambda}. The equality \eqref{eq_grassmann_integral_formulation_Tr_H} is \eqref{eq_grassmann_integral_formulation_Tr_H_lambda} for $\lambda_{\X}=0$ $(\forall \X\in \G^4)$. Note the fact that
\begin{equation}\label{eq_grassmann_exponential_derivative}
\begin{split}
&\frac{\partial}{\partial \la_{\tilde{\X}_1}}e^{\sum_{\X\in \G^4}U_{\X}V_{h,\X}(\psi_X,\opsi_X)}\Big|_{\la_{\X}=0\atop \forall \X\in\G^4}\\
\quad&=-\frac{1}{h}\sum_{x\in [0,\beta)_h} (\opsi_{\bx_1\ua x}\opsi_{\bx_2\da x}\psi_{\by_2\da x}\psi_{\by_1\ua x}+\opsi_{\by_1\ua x}\opsi_{\by_2\da x}\psi_{\bx_2\da x}\psi_{\bx_1\ua x})\\
&\qquad\quad \cdot e^{-\frac{U}{h}\sum_{\bx\in \G}\sum_{x\in[0,\beta)_h}\opsi_{\bx\ua x}\opsi_{\bx\da x}\psi_{\bx\da x}\psi_{\bx\ua x}},
\end{split}
\end{equation}
where the differential operator $\partial /\partial \lambda_{\tilde{\X}_1}$ is defined to act on every coefficient of Grassmann monomials in the expansion $e^{\sum_{\X\in \G^4}U_{\X}V_{h,\X}(\psi_X,\opsi_X)}$ (see \cite[\mbox{Problem I.3}]{FKT}). Moreover, by expanding $e^{\sum_{\X\in \G^4}U_{\X}V_{h,\X}(\psi_X,\opsi_X)}$ one can verify the equality
\begin{equation}\label{eq_grassmann_integral_exchange}
\begin{split}
&\frac{\partial}{\partial \lambda_{\tilde{\X}_1}}\int e^{\sum_{\X\in \G^4}U_{\X}V_{h,\X}(\psi_X,\opsi_X)}d\mu_{C_h}(\psi_X,\opsi_X)\\
&\quad=\int\frac{\partial}{\partial \lambda_{\tilde{\X}_1}} e^{\sum_{\X\in \G^4}U_{\X}V_{h,\X}(\psi_X,\opsi_X)}d\mu_{C_h}(\psi_X,\opsi_X).
\end{split}
\end{equation}
The equality \eqref{eq_grassmann_integral_formulation_correlation_function} follows from Corollary \ref{cor_correlation_P_h}, Proposition \ref{pro_grassmann_integral_formulation_P_h} and \eqref{eq_grassmann_exponential_derivative}-\eqref{eq_grassmann_integral_exchange}.
\end{proof}

\section{Upper bound on the perturbation series}\label{sec_upper_bound}
In this section we calculate upper bounds on our perturbation series $\sum_{n=0}^{\infty}a_nU^n$ by evaluating the tree formula for the connected part of the exponential of Laplacian operator. In order to employ the Grassmann Gaussian integral formulation of $P_h$ developed in Section \ref{subsec_grassmann_integral}, we assume that $h\in2\N/\beta$ throughout this section.

\subsection{The connected part of the exponential of Laplacian operator}
Our approach to find an upper bound on $|a_{n}|$ of our perturbation series$\sum_{n=0}^{\infty}a_nU^n$ is based on the characterization of the connected part of the exponential of the Laplacian operator of Grassmann left derivatives reported in \cite{SW}. Let us construct our argument step by step to reveal the structure of the problem.

The Grassmann integral $\int e^{\sum_{\X\in \G^4}U_{\X}V_{h,\X}(\psi_X,\opsi_X)}d\mu_{C_h}(\psi_X,\opsi_X)$ can be viewed as an analytic function of the multi-variable $\{U_{\X}\}_{\X\in\G^4}$. Since
$$\int e^{\sum_{\X\in \G^4}U_{\X}V_{h,\X}(\psi_X,\opsi_X)}d\mu_{C_h}(\psi_X,\opsi_X)\Bigg|_{U_{\X}=0\atop\forall\X\in\G^4}=1,
$$
if $|U_{\X}|$ is sufficiently small for all $\X\in\G^4$, the inequality
\begin{equation*}
\Big|\int e^{\sum_{\X\in \G^4}U_{\X}V_{h,\X}(\psi_X,\opsi_X)}d\mu_{C_h}(\psi_X,\opsi_X)-1\Big|<1
\end{equation*}
holds. Thus, we can define a function $W_h(\{U_{\X}\}_{\X\in\G^4})$ by
$$W_h(\{U_{\X}\}_{\X\in\G^4}):=\log\left(\int e^{\sum_{\X\in \G^4}U_{\X}V_{h,\X}(\psi_X,\opsi_X)}d\mu_{C_h}(\psi_X,\opsi_X)\right),
$$
which is analytic in a neighborhood of $0$ in $\C^{L^{4d}}$.
 
\begin{lemma}\label{lem_a_h_n_W_h}
For all $h\in2\N/\beta$ and $n\in\N\cup\{0\}$, the following equality holds.
\begin{equation}\label{eq_a_h_n_W_h_1}
\begin{split}
a_{h,n}=-\frac{1}{\beta n!}\sum_{\bz_1,\cdots,\bz_n\in\G}\Bigg(&\frac{\partial^{n+1} W_h}{\partial U_{\bx_1,\bx_2,\by_1,\by_2}\partial U_{\bz_1,\bz_1,\bz_1,\bz_1}\cdots\partial U_{\bz_n,\bz_n,\bz_n,\bz_n}}\Bigg|_{U_{\X}=0\atop\forall\X\in\G^4} \\
+&\frac{\partial^{n+1} W_h}{\partial U_{\by_1,\by_2,\bx_1,\bx_2}\partial U_{\bz_1,\bz_1,\bz_1,\bz_1}\cdots\partial U_{\bz_n,\bz_n,\bz_n,\bz_n}}\Bigg|_{U_{\X}=0\atop\forall\X\in\G^4}\Bigg),
\end{split}
\end{equation}
where $a_{h,n}$ was defined in \eqref{eq_def_a_h_n}-\eqref{eq_def_F_h_n_G_h_n}.
\end{lemma}

\begin{proof}
The Taylor expansion of $W_h(\{U_{\X}\}_{\X\in\G^4})$ around $0$ is given by 
\begin{equation}\label{eq_a_h_n_W_h_2}
W_h(\{U_{\X}\}_{\X\in\G^4})=W_h|_{U_{\X}=0\atop\forall \X\in\G^4}+\sum_{n=1}^{\infty}\frac{1}{n!}\sum_{\X_1,\cdots,\X_n\in\G^4}\frac{\partial^nW_h}{\partial U_{\X_1}\cdots \partial U_{\X_n}}\Bigg|_{U_{\X}=0\atop\forall \X\in\G^4}U_{\X_1}\cdots U_{\X_n}.
\end{equation}

Fix any $U\in\C$ with $|U|<\log 2/(16\beta L^{4d})$. Let $\eps>0$ be the constant claimed in Lemma \ref{lem_properties_P_h} \eqref{item_lem_properties_P_h_1}. By using a parameter $\{\lambda_{\X}\}_{\X\in\G^4}(\subset \C)$ with $|\lambda_{\X}|\le\eps$ $(\forall \X\in\G^4)$ we define the variable $\{U_{\X}\}_{\X\in\G^4}$ by the equality \eqref{eq_relation_U_lambda}. Then the inequality $|P_h(\{U_{\X}\}_{\X\in\G^4})-1|<1$ holds by Lemma \ref{lem_properties_P_h} \eqref{item_lem_properties_P_h_1} and the condition \eqref{eq_constraint_U} is satisfied. Thus, Proposition \ref{pro_grassmann_integral_formulation_P_h} ensures that $W_h(\{U_{\X}\}_{\X\in\G^4})=\log(P_h(\{U_{\X}\}_{\X\in\G^4}))$ for this $\{U_{\X}\}_{\X\in\G^4}$. Moreover, by the equalities \eqref{eq_relation_U_lambda} and \eqref{eq_a_h_n_W_h_2} we have that
\begin{equation*}
\begin{split}
&-\frac{1}{\beta}\frac{\partial}{\partial \lambda_{\tilde{\X}_1}}\log(P_h)\Big|_{\lambda_{\X}=0\atop\forall\X\in\G^4}=-\frac{1}{\beta}\frac{\partial}{\partial \lambda_{\tilde{\X_1}}}W_h\Big|_{\lambda_{\X}=0\atop\forall\X\in\G^4}=-\frac{1}{\beta}\left(\frac{\partial}{\partial U_{\tilde{\X}_1}}+\frac{\partial}{\partial U_{\tilde{\X}_2}}\right)W_h\Big|_{\lambda_{\X}=0\atop\forall\X\in\G^4}\\
&=-\frac{1}{\beta}\sum_{n=1}^{\infty}\frac{U^{n-1}}{(n-1)!}\sum_{\bz_1,\cdots,\bz_{n-1}\in\G}\Bigg(\frac{\partial^nW_h}{\partial U_{\tilde{\X}_1}\partial U_{\bz_1,\bz_1,\bz_1,\bz_1}\cdots\partial U_{\bz_{n-1},\bz_{n-1},\bz_{n-1},\bz_{n-1}}}\Big|_{U_{\X}=0\atop\forall\X\in\G^4}\\
&\qquad\qquad +\frac{\partial^nW_h}{\partial U_{\tilde{\X}_2}\partial U_{\bz_1,\bz_1,\bz_1,\bz_1}\cdots\partial U_{\bz_{n-1},\bz_{n-1},\bz_{n-1},\bz_{n-1}}}\Big|_{U_{\X}=0\atop\forall\X\in\G^4}\Bigg).
\end{split}
\end{equation*}
 By uniqueness of Taylor series and Lemma \ref{lem_properties_P_h} \eqref{item_lem_properties_P_h_2} we obtain \eqref{eq_a_h_n_W_h_1}.
\end{proof}

A message from Lemma \ref{lem_a_h_n_W_h} is that upper bounds on $|a_{h,n}|$ can be obtained by characterizing the partial derivatives of $W_h$ at $U_{\X}=0$ $(\forall \X \in \G^4)$, which is the way we follow from now. Since $|a_{h,0}|$ can be evaluated directly from \eqref{eq_def_a_h_n} by using Lemma \ref{lem_determinant_bound}, let us study the equality \eqref{eq_a_h_n_W_h_1} for $n\ge 1$. Fix any $n\ge 1$ and use the simplified notations defined as follows.
\begin{equation}\label{eq_notation_Gamma_4}
\cZ_j:=(\bz_j,\bz_j,\bz_j,\bz_j)\in\G^4\text{ for }\bz_j\in\G\ (\forall j\in\{1,\cdots,n\}),\ 
\cZ_0:=\tilde{\X}_1\in\G^4.
\end{equation}
Set $\N_{n+1}:=\{0,1,\cdots,n\}$. By noting that
\begin{equation*}
\begin{split}
&\opPi_{j\in Q}\frac{\partial}{\partial U_{\cZ_j}}\int e^{\sum_{\X\in \G^4}U_{\X}V_{h,\X}(\psi_X,\opsi_X)}d\mu_{C_h}(\psi_X,\opsi_X)\Bigg|_{U_{\X}=0\atop\forall\X\in\G^4}\\
&\qquad= \opPi_{j\in Q}\frac{\partial}{\partial U_{\cZ_j}}\int\opPi_{j\in\N_{n+1}}(1+U_{\cZ_j}V_{h,\cZ_j}(\psi_X,\opsi_X))d\mu_{C_h}(\psi_X,\opsi_X)\Bigg|_{U_{\cZ_j}=0\atop\forall j\in \N_{n+1}}
\end{split}
\end{equation*}
for any $Q\subset \N_{n+1}$, we see that
\begin{equation}\label{eq_derivative_W_h}
\begin{split}
&\opPi_{j\in \N_{n+1}}\frac{\partial }{\partial U_{\cZ_j}}W_h\Bigg|_{U_{\X}=0\atop\forall \X\in \G^4}\\
&\quad= \opPi_{j\in \N_{n+1}}\frac{\partial }{\partial U_{\cZ_j}}
\log\left(\int\opPi_{j\in\N_{n+1}}(1+U_{\cZ_j}V_{h,\cZ_j}(\psi_X,\opsi_X))d\mu_{C_h}(\psi_X,\opsi_X)\right)\Bigg|_{U_{\cZ_j}=0\atop\forall j\in\N_{n+1}}\\
&\quad= \opPi_{j\in \N_{n+1}}\frac{\partial }{\partial U_{\cZ_j}}
\\
&\quad\quad\cdot\log\left(1+\sum_{Q\subset \N_{n+1}\atop Q\neq \emptyset}\int\opPi_{j\in Q}V_{h,\cZ_j}(\psi_X,\opsi_X)d\mu_{C_h}(\psi_X,\opsi_X)\opPi_{j\in Q}U_{\cZ_j}\right)\Bigg|_{U_{\cZ_j}=0\atop \forall j\in\N_{n+1}}.
\end{split}
\end{equation}
The Grassmann Gaussian integral contained in the right hand side of \eqref{eq_derivative_W_h} can be rewritten as follows.
\begin{lemma}\label{lem_exponential_laplacian}
Introduce Grassmann algebras 
$\{\psi_{\bx_j\s_jx_{j}}^q,\opsi_{\bx_j\s_jx_{j}}^q|j\in\{1,\cdots,N\}\}$ indexed by $q\in \N_{n+1}$ and write 
$$\psi_X^q =(\psi_{\bx_1\s_1x_{1}}^q,\cdots,\psi_{\bx_N\s_Nx_{N}}^q),\  \opsi_X^q =(\opsi_{\bx_1\s_1x_{1}}^q,\cdots,\opsi_{\bx_N\s_Nx_{N}}^q)$$ 
for all $q\in\N_{n+1}$. Let $\partial/\partial \psi_X^q$, $\partial/\partial \opsi_X^q$ be the vectors of left derivatives associated with $\psi_X^q$, $\opsi_X^q$, respectively. Then, the following equality holds. For all $Q\subset \N_{n+1}$ with $Q\neq \emptyset$ 
\begin{equation}\label{eq_exponential_laplacian_1}
\int\opPi_{q\in Q}V_{h,\cZ_q}(\psi_X,\opsi_X)d\mu_{C_h}(\psi_X,\opsi_X)=e^{\Delta}\opPi_{q\in Q}V_{h,\cZ_q}(\psi_X^q,\opsi_X^q)\Bigg|_{\psi_X^q=\opsi_X^q=0\atop\forall q\in Q},
\end{equation}
where the Laplacian operator $\Delta$ and its exponential $e^{\Delta}$ are defined by
$$\Delta := -\sum_{p,q\in\N_{n+1}}\<\left(\frac{\partial}{\partial \psi_X^p}\right)^t,C_h\left(\frac{\partial}{\partial \opsi_X^q}\right)^t\>,\quad e^{\Delta}:=\sum_{l=0}^{2N(n+1)}\frac{1}{l!}\Delta^l.$$
\end{lemma}      
\begin{remark}
When we introduce another set of Grassmann algebras, let us think that the complex linear space spanned by monomials of all the Grassmann algebras introduced up to this point is defined on the assumption of multiplication satisfying the anti-commutation relations between these algebras. The notion of the Grassmann integral $\int\cdot d\opsi_X d\psi_X$ is naturally extended to be a linear map from the enlarged linear space of all the algebras to the subspace without $\psi_X$, $\opsi_X$.

For a monomial $\phi_{j_1}\cdots\phi_{j_n}$ of Grassmann algebras $\{\phi_l\}_{l=1}^m$, the left derivative $(\partial/\partial\phi_l)\phi_{j_1}\cdots\phi_{j_n}$ $(l\in\{1,\cdots,m\})$ is defined by
\begin{equation*}
\frac{\partial}{\partial \phi_l}\phi_{j_1}\cdots\phi_{j_n}:=\left\{\begin{array}{ll}(-1)^{k-1}\phi_{j_1}\cdots\phi_{j_{k-1}}\phi_{j_{k+1}}\cdots\phi_{j_n} &\text{if there uniquely exists }\\
 &  k\in \{1,\cdots,n\}\text{ s.t. }l=j_k,\\
0 & \text{otherwise.}\end{array}\right.
\end{equation*}
Then, the left derivative $\partial/\partial\phi_l$ is extended to be a linear map on the linear space of monomials of the algebras $\{\phi_l\}_{l=1}^m$.

The concepts of Grassmann integrals and left derivatives are generally defined as operators on Grassmann algebra with coefficients in a superalgebra (see \cite[\mbox{Chapter I}]{FKT}).
\end{remark}

\begin{proof}[Proof of Lemma \ref{lem_exponential_laplacian}]
We define another Grassmann algebra $\{\eta_X^q,\oeta_X^q\}$ indexed by the sets $\G\times\spin\times[0,\beta)_h$ and $\N_{n+1}$ and the associated left derivative \\
$\{\partial/\partial \eta_X^q, \partial/\partial \oeta_X^q\}$ in the same way as $\{\psi_X^q,\opsi_X^q\}$ and $\{\partial/\partial\psi_X^q,\partial/\partial\opsi_X^q\}$. Then, we see that for any subset $Q\subset \N_{n+1}$ with $Q\neq\emptyset$
\begin{equation}\label{eq_grassmann_deformation}
\begin{split}
&\int\opPi_{q\in Q}V_{h,\cZ_q}(\psi_X,\opsi_X)d\mu_{C_h}(\psi_X,\opsi_X)\\
&\quad=\int\opPi_{q\in Q}V_{h,\cZ_q}\left(\frac{\partial}{\partial \eta_X^q},\frac{\partial}{\partial\oeta_X^q}\right)e^{\<(\eta_X^q,\oeta_X^q)^t,(\psi_X,\opsi_X)^t\>}d\mu_{C_h}(\psi_X,\opsi_X)\Bigg|_{\eta_X^q=\oeta_X^q=0\atop\forall q\in Q}\\
&\quad=\opPi_{q\in Q}V_{h,\cZ_q}\left(\frac{\partial}{\partial \eta_X^q},\frac{\partial}{\partial\oeta_X^q}\right)\int e^{\<(\eta_X^q,\oeta_X^q)^t,(\psi_X,\opsi_X)^t\>}d\mu_{C_h}(\psi_X,\opsi_X)\Bigg|_{\eta_X^q=\oeta_X^q=0\atop\forall q\in Q}\\
&\quad=\opPi_{q\in Q}V_{h,\cZ_q}\left(\frac{\partial}{\partial \eta_X^q},\frac{\partial}{\partial\oeta_X^q}\right)e^{-\sum_{p,q\in Q}\<(\eta_X^p)^t,C_h(\oeta_X^q)^t\>}\Bigg|_{\eta_X^q=\oeta_X^q=0\atop\forall q\in Q}\\
&\quad=e^{\Delta}\opPi_{q\in Q}V_{h,\cZ_q}(\psi_X^q,\opsi_X^q)\Big|_{\psi_X^q=\opsi_X^q=0\atop\forall q\in Q},
\end{split}
\end{equation}
where we have used the equality that 
$$\int e^{\sum_{q\in Q}\<(\eta_X^q,\oeta_X^q)^t,(\psi_X,\opsi_X)^t\>}d\mu_{C_h}(\psi_X,\opsi_X)=e^{-\sum_{p,q\in Q}\<(\eta_X^p)^t,C_h(\oeta_X^q)^t\>}
$$
(see \cite[\mbox{Problem I.13}]{FKT}). To verify the equalities \eqref{eq_grassmann_deformation} in more detail, see the books \cite{FKT}, \cite{S} for the properties of left derivatives.
\end{proof}

By combining \eqref{eq_derivative_W_h} with \eqref{eq_exponential_laplacian_1} we obtain 
\begin{equation}\label{eq_W_h_exponential_laplacian}
\begin{split}
&\opPi_{j\in\N_{n+1}}\frac{\partial}{\partial U_{\cZ_j}}W_h\Bigg|_{U_{\X}=0\atop\forall \X\in\G^4}=\\
&\opPi_{j\in\N_{n+1}} \frac{\partial}{\partial U_{\cZ_j}}\log\left(1+\sum_{Q\subset \N_{n+1}\atop Q\neq\emptyset}e^{\Delta}\opPi_{q\in Q}V_{h,\cZ_q}(\psi_X^q,\opsi_X^q)\Big|_{\psi_X^q=\opsi_X^q=0\atop\forall q\in Q}\opPi_{p\in Q}U_{\cZ_p}\right)\Bigg|_{U_{\cZ_q}=0\atop\forall q\in \N_{n+1}}.
\end{split}
\end{equation}
In order to characterize the right hand side of \eqref{eq_W_h_exponential_laplacian}, let us review the general theory developed in \cite{SW} by translating in our setting. Consider a map $\alpha$ from the power set $\P(\N_{n+1})$ of $\N_{n+1}$ to $\C$ defined by $\alpha(\emptyset):=1$ and for $Q\in \P(\N_{n+1})\backslash\{\emptyset\}$
$$\alpha(Q):=e^{\Delta}\opPi_{q\in Q}V_{h,\cZ_q}(\psi_X^q,\opsi_X^q)\Big|_{\psi_X^q=\opsi_X^q=0\atop\forall q\in Q}.$$
By \cite[\mbox{Lemma 1}]{SW} there uniquely exists a map $\alpha_c:\P(\N_{n+1})\to\C$ such that for all $Q\in \P(\N_{n+1})\backslash \{\emptyset\}$
$$
\alpha(Q)=\sum_{Q_0\subset Q\atop\text{min}Q\in Q_0}\alpha_c(Q_0)\alpha(Q\backslash Q_0),
$$
where min$Q$ stands for the smallest number contained in $Q$. In \cite[\mbox{Lemma 2}]{SW} it was proved that the right hand side of \eqref{eq_W_h_exponential_laplacian} is equal to $\alpha_c(\N_{n+1})$, which is called the connected part of the operator $e^{\Delta}$. The formula for $\alpha_c(\N_{n+1})$ was given in \cite[\mbox{Theorem 3}]{SW}. We summarize the result below.
\begin{lemma}\cite[\mbox{Theorem 3}]{SW}\label{lem_tree_formula}
The following equality holds.
\begin{equation}\label{eq_tree_formula_1}
\begin{split}
&\opPi_{j\in\N_{n+1}}\frac{\partial}{\partial U_{\cZ_j}}W_h\Bigg|_{U_{\X}=0\atop\forall \X\in\G^4}\\
&\quad=\sum_{T\in\T(\N_{n+1})}\opPi_{\{q,q'\}\in T}(\Delta_{q,q'}+\Delta_{q',q})\int_{[0,1]^n}d\bs\sum_{\pi\in S_{n+1}(T)}\phi(T,\pi,\bs)e^{\Delta(M(T,\pi,\bs))}\\
&\quad\qquad\cdot\opPi_{q\in\N_{n+1}}V_{h,\cZ_q}(\psi_X^q,\opsi_X^q)\Bigg|_{\psi_X^q=\opsi_X^q=0\atop\forall q\in\N_{n+1}},
\end{split}
\end{equation}
where $\T(\N_{n+1})$ is the set of all the trees (connected graphs without loop) on $\N_{n+1}$,
$$\Delta_{q,q'}:=-\<\left(\frac{\partial}{\partial \psi_X^q}\right)^t,C_h\left(\frac{\partial}{\partial \opsi_X^{q'}}\right)^t\>,$$
$\bs := (s_1,\cdots,s_n)$, $S_{n+1}(T)$ is a subset of $S_{n+1}$ depending on $T$, $\phi(T,\pi,\bs)$ is a real-valued non-negative function of $\bs$ depending on $T$ and $\pi$ with the property that
\begin{equation}\label{eq_tree_formula_2}
\int_{[0,1]^{n}}d\bs \sum_{\pi\in S_{n+1}(T)}\phi(T,\pi,\bs)=1,
\end{equation}
$M(T,\pi,\bs)$ is an $(n+1)\times (n+1)$ real symmetric non-negative matrix depending on $T$, $\pi$, $\bs$ satisfying $M(T,\pi,\bs)_{q,q}=1$ for all $q\in \N_{n+1}$ and the operator $\Delta(M(T,\pi,\bs))$ is defined by 
$$\Delta(M(T,\pi,\bs)):=\sum_{p,q\in\N_{n+1}}M(T,\pi,\bs)_{p,q}\Delta_{p,q}.$$
\end{lemma}
In order to bound $|a_{h,n}|$, the tree formula \eqref{eq_tree_formula_1} will be evaluated in the rest of this section.

\subsection{Evaluation of upper bounds}
Here we evaluate the tree expansion given in Lemma \ref{lem_tree_formula}. Let us first prepare some necessary tools. The following lemma essentially uses the determinant bound Lemma \ref{lem_determinant_bound}.
\begin{lemma}\label{lem_determinant_bound_application}
For any $l\in \N$ and any $p_m,q_m\in \N_{n+1}$, $(\bx_{j_m},\s_{j_m},x_{j_m})$, \\
$(\bx_{k_m},\s_{k_m},x_{k_m})\in\G\times \spin\times [0,\beta)_h$ $(\forall m\in\{1,\cdots,l\})$
$$\left|e^{\Delta(M(T,\pi,\bs))}\psi_{\bx_{j_1}\s_{j_1}x_{j_1}}^{p_1}\cdots\psi_{\bx_{j_l}\s_{j_l}x_{j_l}}^{p_l}\opsi_{\bx_{k_1}\s_{k_1}x_{k_1}}^{q_1}\cdots\opsi_{\bx_{k_l}\s_{k_l}x_{k_l}}^{q_l}\Big|_{\psi_X^q=\opsi_X^q=0\atop\forall q\in\N_{n+1}}\right|\le 4^l.$$
\end{lemma}
\begin{proof}
Since $M(T,\pi,\bs)$ is a non-negative real symmetric matrix, there are constants $\gamma_q\ge 0$ $(q\in \N_{n+1})$ and projection matrices $P_q$ $(q\in\N_{n+1})$ satisfying that $P_pP_q=0$ for all $p,q\in \N_{n+1}$ with $p\neq q$ such that $M(T,\pi,\bs)=\sum_{q=0}^n\gamma_qP_q$. Define an $(n+1)\times (n+1)$ real matrix $\tilde{M}$ by $\tilde{M}:=\sum_{q=0}^n\sqrt{\gamma_q}P_q$. Then we see that
\begin{equation}\label{eq_determinant_bound_application_1}
M(T,\pi,\bs)=\tilde{M}^t\tilde{M}.
\end{equation}
By writing $\tilde{M}=(\bv_0,\cdots,\bv_n)$ with vectors $\bv_q\in\R^{n+1}$ $( q\in\N_{n+1})$, the equality \eqref{eq_determinant_bound_application_1} implies that $M(T,\pi,\bs)_{p,q}=\<\bv_p,\bv_q\>$ for all $p,q\in \N_{n+1}$. The property that $M(T,\pi,\bs)_{q,q}=1$ $(\forall q\in \N_{n+1})$ ensures that for all $q\in \N_{n+1}$
\begin{equation}\label{eq_determinant_bound_application_2}
|\bv_q|=1.
\end{equation}
Then we observe that
\begin{equation}\label{eq_determinant_bound_application_3}
\begin{split}
&e^{\Delta(M(T,\pi,\bs))}\psi_{\bx_{j_1}\s_{j_1}x_{j_1}}^{p_1}\cdots\psi_{\bx_{j_l}\s_{j_l}x_{j_l}}^{p_l}\opsi_{\bx_{k_1}\s_{k_1}x_{k_1}}^{q_1}\cdots\opsi_{\bx_{k_l}\s_{k_l}x_{k_l}}^{q_l}\Big|_{\psi_X^q=\opsi_X^q=0\atop\forall q\in\N_{n+1}}\\
&=(-1)^{l(l-1)/2}\frac{1}{l!}\left(\sum_{p,q\in\N_{n+1}}\<\bv_p,\bv_q\>\Delta_{p,q}\right)^l\opPi_{m=1}^l\psi_{\bx_{j_m}\s_{j_m}x_{j_m}}^{p_m}\opsi_{\bx_{k_m}\s_{k_m}x_{k_m}}^{q_m}\Big|_{\psi_X^q=\opsi_X^q=0\atop\forall q\in\N_{n+1}}\\
&=(-1)^{l(l-1)/2}\det(\<\bv_{p_s},\bv_{q_t}\>C_h(\bx_{j_s}\s_{j_s}x_{j_s}, \bx_{k_t}\s_{k_t}x_{k_t}))_{1\le s,t\le l}.
\end{split}
\end{equation}
By noting \eqref{eq_determinant_bound_application_2} we can apply Lemma \ref{lem_determinant_bound} to \eqref{eq_determinant_bound_application_3} to deduce the desired inequality.
\end{proof}
 
One point we need to carefully deal with in the evaluation of the right hand side of \eqref{eq_tree_formula_1} is the combinatorial factor, which comes in the expansion of \\
$\opPi_{\{q,q'\}\in T}(\Delta_{q,q'}+\Delta_{q',q})\opPi_{q\in\N_{n+1}}V_{h,\cZ_q}(\psi_X^q,\opsi_X^q)$. In order to count the combinatorial factor explicitely, we need to prepare some notions concerning trees. 

Take any $T\in \T(\N_{n+1})$ and for any $q\in \N_{n+1}$ let $d_q$ $(\in\N)$ denote the incidence number of the vertex $q$ (the number of lines connected to the vertex $q$). From now let us always think that any tree in $\T(\N_{n+1})$ starts from the vertex $0$. For any $q\in\N_{n+1}$ let $L_q(T)$ $(\subset T)$ be the set of lines from the vertex $q$ to the vertices of the later generation. We see that
$$\sharp L_0(T)=d_0,\ \sharp L_q(T)=d_q-1,\ \forall q\in \N_{n+1}\backslash \{0\}.$$

We define the combinatorial factor $N(T)$ we want to calculate as follows.
\begin{definition}\label{def_combinatorial_factor}
For any $T\in \T(\N_{n+1})$ the combinatorial factor  $N(T)(\in\N)$ is defined as the total number of monomials appearing in the expansion of 
\begin{equation}\label{eq_monomial_expansion}
\opPi_{\{q,q'\}\in T}(\Delta_{q,q'}+\Delta_{q',q})\opPi_{q\in\N_{n+1}}\opsi_{\bz_1^q\ua x_{1}^q}\opsi_{\bz_2^q\da x_{2}^q}\psi_{\bz_3^q\ua x_{3}^q}\psi_{\bz_4^q\da x_{4}^q}.
\end{equation}
\end{definition}
Note that $N(T)$ is independent of how to choose $\bz_{j}^q\in\G$, $x_{j}^q\in[0,\beta)_h$ $(\forall j\in\{1,2,3,4\},\forall q\in\N_{n+1})$.

The combinatorial factor $N(T)$ is counted as follows.     
\begin{lemma}\label{lem_combinatorial_factor}
For $T\in\T(\N_{n+1})$ let $d_q$ $(q\in\N_{n+1})$ denote the incidence number of the vertex $q$ in $T$. If there is $q\in\N_{n+1}$ such that $d_q>4$, $N(T)=0$. Otherwise,
$$N(T)=4\opPi_{q\in\N_{n+1}}\left(\begin{array}{c} 3 \\ d_q - 1 \end{array}\right)(d_q-1)!.$$
\end{lemma}
\begin{proof}
Set $W:=\opPi_{q\in\N_{n+1}}\opsi_{\bz_1^q\ua x_{1}^q}\opsi_{\bz_2^q\da x_{2}^q}\psi_{\bz_3^q\ua x_{3}^q}\psi_{\bz_4^q\da x_{4}^q}$. If there is $p\in\N_{n+1}$ such that $d_p>4$, every term in the expansion of the product $\opPi_{\{q,q'\}\in T}(\Delta_{q,q'}+\Delta_{q',q})$ contains more than $4$ derivatives with respect to the Grassmann algebras indexed by $p$. Since the number of the Grassmann algebras with index $p$ in $W$ is $4$, \eqref{eq_monomial_expansion} must vanish.

Let us consider the case that $d_q\in\{1,2,3,4\}$ for all $q\in\N_{n+1}$. The operator $\Delta_{q,q'}$ can be decomposed as $\Delta_{q,q'}=\sum_{\s\in\spin}\Delta_{q,q'}^{\s}$, where 
\begin{equation}\label{eq_decomposed_laplacian}
\Delta_{q,q'}^{\s}:=-\sum_{\bx,\by\in\G}\sum_{x,y\in[0,\beta)_h}C_h(\bx\s x,\by\s y)\frac{\partial}{\partial\psi_{\bx\s x}^q}\frac{\partial}{\partial\opsi_{\by\s y}^{q'}}
\end{equation}
 for $\s\in\spin$. Note that for any $\s\in\spin$ and $p,p',p''\in\N_{n+1}$
\begin{equation}\label{eq_combinatorial_factor_1}
\Delta_{p,p'}^{\s}\Delta_{p,p''}^{\s}W=\Delta_{p',p}^{\s}\Delta_{p'',p}^{\s}W=0.\end{equation}

By changing the numbering of vertices if necessary we may assume the following condition on $T$ without losing generality.
\begin{enumerate}[($\clubsuit$)]
\item The distance between the vertex $p$ and the initial vertex $0$ is less than equal to that between the vertex $q$ and the vertex $0$ for all $p,q\in\N_{n+1}\backslash \{0\}$ with $p\le q$.
\end{enumerate}

Note that
\begin{equation}\label{eq_combinatorial_factor_2}
 \opPi_{\{q,q'\}\in T}(\Delta_{q,q'}+\Delta_{q',q})W=\opPi_{q\in\N_{n+1}\atop L_q(T)\neq\emptyset}\opPi_{\{q,p\}\in L_q(T)}(\Delta_{q,p}^{\ua}+\Delta_{q,p}^{\da}+\Delta_{p,q}^{\ua}+\Delta_{p,q}^{\da})W.
\end{equation}
Let us count $N(T)$ recursively with respect to $q\in\N_{n+1}$ as follows. The expansion of the product $\opPi_{\{0,p\}\in L_0(T)}(\Delta_{0,p}^{\ua}+\Delta_{0,p}^{\da}+\Delta_{p,0}^{\ua}+\Delta_{p,0}^{\da})$ is a sum of $4^{\sharp L_0(T)}$ terms, each of which is a product of $\sharp L_0(T)$ Laplacians. By the property \eqref{eq_combinatorial_factor_1} any term containing the products $\Delta_{0,q}^{\s}\Delta_{0,q'}^{\s}$ or $\Delta_{q,0}^{\s}\Delta_{q',0}^{\s}$ for some $\s\in\spin$, $\{0,q\},\{0,q'\}\in L_0(T)$ does not contribute to the number of remaining monomials in \eqref{eq_combinatorial_factor_2}, thus, can be eliminated. Therefore, we only need to count 
$$\left(\begin{array}{c} 4 \\ \sharp L_0(T) \end{array}\right) \sharp L_0(T)!$$
terms in the expansion of $\opPi_{\{0,p\}\in L_0(T)}(\Delta_{0,p}^{\ua}+\Delta_{0,p}^{\da}+\Delta_{p,0}^{\ua}+\Delta_{p,0}^{\da})$. 

Take any $q\in \N_{n+1}\backslash\{0\}$ with $L_q(T)\neq\emptyset$. By the condition ($\clubsuit$) there uniquely exists $q'\in\N_{n+1}$ with $q'<q$ such that $\{q',q\}\in L_{q'}(T)$. Thus, every term in the expansion of the product
$$
\opPi_{j\in\N_{n+1},L_j(T)\neq\emptyset\atop j<q}\opPi_{\{j,p\}\in L_j(T)}(\Delta_{j,p}^{\ua}+\Delta_{j,p}^{\da}+\Delta_{p,j}^{\ua}+\Delta_{p,j}^{\da})
$$
contains one of the Laplacians of the form $\Delta_{q,q'}^{\s},\Delta_{q',q}^{\s}$ $(\s\in\spin)$. Therefore, by the property \eqref{eq_combinatorial_factor_1} only 
$$\left(\begin{array}{c} 3 \\ \sharp L_q(T) \end{array}\right) \sharp L_q(T)!
$$
terms in the expansion of the product $\opPi_{\{q,p\}\in L_q(T)}(\Delta_{q,p}^{\ua}+\Delta_{q,p}^{\da}+\Delta_{p,q}^{\ua}+\Delta_{p,q}^{\da})$ need to be counted.

By repeating this argument recursively for all $q\in \N_{n+1}\backslash\{0\}$ we can calculate $N(T)$ as follows.
\begin{equation*}
\begin{split}
&N(T)=\left(\begin{array}{c} 4 \\ \sharp L_0(T) \end{array}\right)\sharp L_0(T)! \opPi_{q\in \N_{n+1}\backslash\{0\}\atop L_q(T)\neq\emptyset}\left(\begin{array}{c} 3 \\ \sharp L_q(T) \end{array}\right)\sharp L_q(T)!\\
&\quad= \left(\begin{array}{c} 4 \\ d_0 \end{array}\right)d_0! \opPi_{q\in \N_{n+1}\backslash\{0\}\atop d_q\ge 2}\left(\begin{array}{c} 3 \\ d_q-1 \end{array}\right)(d_q-1)!=4\opPi_{q\in\N_{n+1}}\left(\begin{array}{c} 3 \\ d_q-1 \end{array}\right)(d_q-1)!,
\end{split}
\end{equation*}
where we used the fact that 
$$\left(\begin{array}{c} 3 \\ 0 \end{array}\right)0!=1.
$$
\end{proof}

Let $D_h$ denote the $L^1$-norm of the covariance matrix $C_h$, i.e,
$$D_h:=\frac{1}{h}\sum_{\bx\in\G}\sum_{x\in[-\beta,\beta)_h}|C_h(\bx\ua x,\b0\ua 0)|.$$
Now we can find an upper bound on $|a_{h,n}|$.
\begin{lemma}\label{lem_bound_a_h_n}
For any $h\in2\N/\beta$ and for all $n\in\N\cup\{0\}$ the following equality holds.
\begin{equation}\label{eq_bound_a_h_n_1}
|a_{h,n}|\le \frac{128}{3n+4}\left(\begin{array}{c}3n+4 \\ n \end{array}\right)(4D_h)^n.
\end{equation}
\end{lemma}
\begin{proof}
By using Lemma \ref{lem_determinant_bound} and \eqref{eq_def_a_h_n}-\eqref{eq_def_F_h_n_G_h_n} we can directly evaluate $|a_{h,0}|$ to obtain $|a_{h,0}|\le 32 $, which is \eqref{eq_bound_a_h_n_1} for $n=0$. Let us show \eqref{eq_bound_a_h_n_1} for $n\ge 1$. By \eqref{eq_a_h_n_W_h_1} and \eqref{eq_tree_formula_1}, we need to evaluate
\begin{equation}\label{eq_bound_a_h_n_2}
\begin{split}
\frac{1}{\beta n!}&\Bigg|\sum_{T\in\T(\N_{n+1})}\opPi_{\{q,q'\}\in T}(\Delta_{q,q'}+\Delta_{q',q})\int_{[0,1]^n}d\bs\sum_{\pi\in S_{n+1}(T)}\phi(T,\pi,\bs)e^{\Delta(M(T,\pi,\bs))}\\
&\quad\cdot V_{h,\cZ_0}(\psi_X^0,\opsi_X^0)\opPi_{q=1}^n\sum_{\bz_q\in\G}V_{h,\cZ_q}(\psi_X^q,\opsi_X^q)\Bigg|_{\psi_X^q=\opsi_X^q=0\atop\forall q\in\N_{n+1}}\Bigg|.
\end{split}
\end{equation}
By using the property \eqref{eq_tree_formula_2} we have
\begin{equation}\label{eq_bound_a_h_n_3}
\begin{split}
\eqref{eq_bound_a_h_n_2}\le \frac{1}{\beta n!}\sum_{T\in\T(\N_{n+1})}&\sup_{\bs\in[0,1]^n\atop\pi\in S_{n+1}(T)}\Bigg|e^{\Delta(M(T,\pi,\bs))}\opPi_{\{q,q'\}\in T}(\Delta_{q,q'}+\Delta_{q',q})\\
&\cdot V_{h,\cZ_0}(\psi_X^0,\opsi_X^0)\opPi_{q=1}^n\sum_{\bz_q\in\G}V_{h,\cZ_q}(\psi_X^q,\opsi_X^q)\Bigg|_{\psi_X^q=\opsi_X^q=0\atop\forall q\in\N_{n+1}}\Bigg|.
\end{split}
\end{equation}
Take any $T\in\T(\N_{n+1})$. If $T$ contains a vertex whose incidence number is larger than $4$, by Lemma \ref{lem_combinatorial_factor} the tree $T$ does not contribute to the sum in \eqref{eq_bound_a_h_n_3}. Thus, we may assume that the incidence numbers $d_0,d_1,\cdots,d_n$ of $T$ are less than equal to $4$. Moreover, as in the proof of Lemma \ref{lem_combinatorial_factor} without losing generality we may assume the condition ($\clubsuit$) on $T$.

Let $q_1,q_2,\cdots,q_l\in\N_{n+1}\backslash\{0\}$ be such that $q_1<q_2<\cdots<q_l$ and \\
$\{q_1,q_2,\cdots,q_l\}=\{q\in\N_{n+1}\backslash\{0\}\ |\ L_q(T)\neq\emptyset\}$. Every term of the expansion of the product $\opPi_{\{q,q'\}\in T}(\Delta_{q,q'}+\Delta_{q',q})$ has the form 
\begin{equation}\label{eq_bound_a_h_n_4}
\opPi_{\{0,p_0\}\in L_0(T)}\Delta_{\{0,p_0\}}^{\s}\opPi_{j=1}^l\opPi_{\{q_j,p_j\}\in L_{q_j}(T)}\Delta_{\{q_j,p_j\}}^{\s},
\end{equation}
where by using the notation \eqref{eq_decomposed_laplacian}, $\Delta_{\{q,p\}}^{\s}\in\{\Delta_{q,p}^{\tau},\Delta_{p,q}^{\tau}\ |\ \tau\in\spin\}$ for all $\{q,p\}\in L_q(T)$ with $q\in \N_{n+1}$ satisfying $L_q(T)\neq \emptyset$. For any $\{0,p\}\in L_0(T)$ we set 
\begin{equation*}
\begin{split}
&C_{\{0,p\}}^{\s}(x_{0},\bz_px_{p}):=\\
&\quad |C_h(\bz_p\ua x_{p},\bx_1\ua x_{0})|\text{ if }\Delta_{\{0,p\}}^{\s}=\Delta_{p,0}^{\ua},\ |C_h(\bz_p\da x_{p},\bx_2\da x_{0})|\text{ if }\Delta_{\{0,p\}}^{\s}=\Delta_{p,0}^{\da},\\
&\quad |C_h(\by_1\ua x_{0},\bz_p\ua x_{p})|\text{ if }\Delta_{\{0,p\}}^{\s}=\Delta_{0,p}^{\ua},\ |C_h(\by_2\da x_{0},\bz_p\da x_{p})|\text{ if }\Delta_{\{0,p\}}^{\s}=\Delta_{0,p}^{\da}
\end{split}
\end{equation*}
for all $x_0,x_p\in [0,\beta)_h$ and $\bz_p\in\G$. 
For any $j\in \{1,2,\cdots,l\}$ and any $\{q_j,p\}\in L_{q_j}(T)$ we define
\begin{equation*}
\begin{split}
C_{\{q_j,p\}}^{\s}(\bz_{q_j}x_{q_j},\bz_px_{p}):=&|C_h(\bz_{q_j}\tau x_{q_j},\bz_p\tau x_{p})|\text{ if }\Delta_{\{q_j,p\}}^{\s}=\Delta_{q_j,p}^{\tau}\text{ for some }\tau \in\spin,\\
&|C_h(\bz_p\tau x_{p}, \bz_{q_j}\tau x_{q_j})|\text{ if }\Delta_{\{q_j,p\}}^{\s}=\Delta_{p,q_j}^{\tau}\text{ for some }\tau \in\spin
\end{split}
\end{equation*}
for all $x_p,x_{q_j}\in [0,\beta)_h$ and $\bz_p, \bz_{q_j}\in\G$. 
By noting that \eqref{eq_bound_a_h_n_4} is the product of $n$ Laplacians and using Lemma \ref{lem_determinant_bound_application} and the condition ($\clubsuit$), we observe that
\begin{equation*} 
\begin{split}
\Bigg|&e^{\Delta(M(T,\pi,\bs))}\opPi_{\{0,p_0\}\in L_0(T)}\Delta_{\{0,p_0\}}^{\s}\opPi_{j=1}^l\opPi_{\{q_j,p_j\}\in L_{q_j}(T)}\Delta_{\{q_j,p_j\}}^{\s}\\
&\qquad\qquad\qquad\qquad\cdot V_{h,\cZ_0}(\psi_X^0,\opsi_X^0)\opPi_{q=1}^n\sum_{\bz_q\in\G}V_{h,\cZ_q}(\psi_X^q,\opsi_X^q)\Bigg|_{\psi_X^q=\opsi_X^q=0\atop\forall q\in\N_{n+1}}\Bigg|\\
&\le \frac{4^{n+2}}{h}\sum_{x_{0}\in [0,\beta)_h}\opPi_{q=1}^n\left(\frac{1}{h}\sum_{x_{q}\in [0,\beta)_h}\sum_{\bz_q\in\G}\right)\opPi_{\{0,p_0\}\in L_0(T)}C_{\{0,p_0\}}^{\s}(x_{0},\bz_{p_0}x_{p_0})\\
&\qquad\qquad\cdot\opPi_{j=1}^l\opPi_{\{q_j,p_j\}\in L_{q_j}(T)}C_{\{q_j,p_j\}}^{\s}(\bz_{q_j}x_{q_j},\bz_{p_j}x_{p_j})\\
&=\frac{4^{n+2}}{h}\sum_{x_{0}\in [0,\beta)_h}\opPi_{\{0,p_0\}\in L_0(T)}\left(\frac{1}{h}\sum_{x_{p_0}\in [0,\beta)_h}\sum_{\bz_{p_0}\in\G}C_{\{0,p_0\}}^{\s}(x_{0},\bz_{p_0}x_{p_0})\right)\\
&\quad\cdot\opPi_{\{q_1,p_1\}\in L_{q_1}(T)}\left(\frac{1}{h}\sum_{x_{p_1}\in [0,\beta)_h}\sum_{\bz_{p_1}\in\G}C_{\{q_1,p_1\}}^{\s}(\bz_{q_1}x_{q_1},\bz_{p_1}x_{p_1})\right)\\
&\quad\cdot\opPi_{\{q_2,p_2\}\in L_{q_2}(T)}\left(\frac{1}{h}\sum_{x_{p_2}\in [0,\beta)_h}\sum_{\bz_{p_2}\in\G}C_{\{q_2,p_2\}}^{\s}(\bz_{q_2}x_{q_2},\bz_{p_2}x_{p_2})\right)\\
&\quad\cdots \opPi_{\{q_l,p_l\}\in L_{q_l}(T)}\left(\frac{1}{h}\sum_{x_{p_l}\in [0,\beta)_h}\sum_{\bz_{p_l}\in\G}C_{\{q_l,p_l\}}^{\s}(\bz_{q_l}x_{q_l},\bz_{p_l}x_{p_l})\right)
\end{split}
\end{equation*}
\begin{equation}\label{eq_bound_a_h_n_5} 
\begin{split}
&\le 4^{n+2}\beta \opPi_{\{0,p_0\}\in L_0(T)}\left(\sup_{x_0\in [0,\beta)_h}\frac{1}{h}\sum_{x_{p_0}\in [0,\beta)_h}\sum_{\bz_{p_0}\in\G}C_{\{0,p_0\}}^{\s}(x_{0},\bz_{p_0}x_{p_0})\right)\\
&\quad\cdot\opPi_{\{q_1,p_1\}\in L_{q_1}(T)}\left(\sup_{x_{q_1}\in [0,\beta)_h,\bz_{q_1}\in\G}\frac{1}{h}\sum_{x_{p_1}\in [0,\beta)_h}\sum_{\bz_{p_1}\in\G}C_{\{q_1,p_1\}}^{\s}(\bz_{q_1}x_{q_1},\bz_{p_1}x_{p_1})\right)\\
&\quad\cdot\opPi_{\{q_2,p_2\}\in L_{q_2}(T)}\left(\sup_{x_{q_2}\in [0,\beta)_h,\bz_{q_2}\in\G}\frac{1}{h}\sum_{x_{p_2}\in [0,\beta)_h}\sum_{\bz_{p_2}\in\G}C_{\{q_2,p_2\}}^{\s}(\bz_{q_2}x_{q_2},\bz_{p_2}x_{p_2})\right)\\
&\quad\cdots \opPi_{\{q_l,p_l\}\in L_{q_l}(T)}\left(\sup_{x_{q_l}\in [0,\beta)_h,\bz_{q_l}\in\G}\frac{1}{h}\sum_{x_{p_l}\in [0,\beta)_h}\sum_{\bz_{p_l}\in\G}C_{\{q_l,p_l\}}^{\s}(\bz_{q_l}x_{q_l},\bz_{p_l}x_{p_l})\right)\\
&\le 4^{n+2}\beta D_h^n.
\end{split}
\end{equation}
To obtain the last inequality in \eqref{eq_bound_a_h_n_5} we have used the fact that for all $\{0,p_0\}\in L_0(T)$ and all $\{q_j,p_j\}\in L_{q_j}(T)$ $(\forall j\in \{1,2,\cdots,l\})$
\begin{equation*}
\begin{split}
&\sup_{x_{0}\in [0,\beta)_h}\frac{1}{h}\sum_{x_{p_0}\in [0,\beta)_h}\sum_{\bz_{p_0}\in\G}C^{\s}_{\{0,p_0\}}(x_{0},\bz_{p_0}x_{p_0})\le D_h,\\
&\sup_{x_{q_j}\in [0,\beta)_h,\bz_{q_j}\in\G}\frac{1}{h}\sum_{x_{p_j}\in [0,\beta)_h}\sum_{\bz_{p_j}\in\G}C^{\s}_{\{q_j,p_j\}}(\bz_{q_j}x_{q_j},\bz_{p_j}x_{p_j})\le D_h.
\end{split}
\end{equation*}

By combining \eqref{eq_bound_a_h_n_5} with \eqref{eq_bound_a_h_n_3} we have
\begin{equation}\label{eq_combine}
\eqref{eq_bound_a_h_n_2}\le \frac{16}{n!}(4D_h)^n\sum_{T\in \T(\N_{n+1})}N(T),
\end{equation}
where $N(T)$ is defined in Definition \ref{def_combinatorial_factor}. 

By repeating the same argument as above we can show the inequality \eqref{eq_combine} for the case that $\cZ_0=\tilde{\X}_2$. Thus, by recalling \eqref{eq_a_h_n_W_h_1} we arrive at
\begin{equation}\label{eq_bound_a_h_n_6}
|a_{h,n}|\le \frac{32}{n!}(4D_h)^n\sum_{T\in \T(\N_{n+1})}N(T).
\end{equation}

To complete the proof we calculate the sum $\sum_{T\in \T(\N_{n+1})}N(T)$ in \eqref{eq_bound_a_h_n_6}. As characterized in Lemma \ref{lem_combinatorial_factor}, the number $N(T)$ only depends on the incidence numbers of $T$. By using Lemma \ref{lem_combinatorial_factor} and Cayley's theorem on the number of trees with fixed incidence numbers (see, e.g, \cite[\mbox{Corollary 2.2.4}]{W}) we have
\begin{equation*}
\begin{split}
&\sum_{T\in\T(\N_{n+1})}N(T)\\
&=\sum_{d_q\in\{1,2,3,4\}\atop \forall q\in\N_{n+1}}1_{\sum_{q\in\N_{n+1}}d_q=2n}\cdot\frac{(n-1)!}{\opPi_{q\in\N_{n+1}}(d_q-1)!}\cdot 4\opPi_{q\in\N_{n+1}}\left(\begin{array}{c} 3 \\ d_q-1 \end{array}\right)(d_q-1)!\\
&=4(n-1)!\sum_{d_q\in\{1,2,3,4\}\atop\forall q\in\N_{n+1}}1_{\sum_{q\in\N_{n+1}}d_q=2n}\cdot\opPi_{q\in\N_{n+1}}\left(\begin{array}{c} 3 \\ d_q-1 \end{array}\right).
\end{split}
\end{equation*}
Moreover, by using Cauchy's integral formula and the residue theorem we see that for a positive $r>0$
\begin{equation}\label{eq_bound_a_h_n_7}
\begin{split}
&\sum_{T\in\T(\N_{n+1})}N(T)=\frac{4(n-1)!}{(2n)!}\left(\frac{d}{dz}\right)^{2n}\left(\sum_{d=1}^4\left(\begin{array}{c} 3 \\ d -1 \end{array}\right)z^d\right)^{n+1}\Bigg|_{z=0}\\
&\quad=\frac{4(n-1)!}{2\pi i}\oint_{|z|=r}dz\frac{z^{n+1}(1+z)^{3(n+1)}}{z^{2n+1}}=4(n-1)!\Res_{z=0}\left(\frac{(1+z)^{3(n+1)}}{z^n}\right)\\
&\quad=4(n-1)!\left(\begin{array}{c}3n+3 \\ n-1\end{array}\right)=\frac{4n!}{3n+4}\left(\begin{array}{c}3n+4 \\ n\end{array}\right).
\end{split}
\end{equation}
Combining \eqref{eq_bound_a_h_n_7} with \eqref{eq_bound_a_h_n_6} yields the result.
\end{proof}

The inequality \eqref{eq_bound_a_h_n_1} motivates us to know the properties of the power series $f(x)$ defined by 
\begin{equation}\label{eq_dominant_power_series}
f(x):=\sum_{n=0}^{\infty}\frac{4}{3n+4}\left(\begin{array}{c}3n+4 \\ n\end{array}\right)x^n.
\end{equation}
As the last lemma before our main theorem, the properties of $f(x)$ are summarized.
\begin{lemma}\label{lem_properties_dominant_power_series}
The radius of convergence of the power series $f(x)$ defined in \eqref{eq_dominant_power_series} is $4/27$ and $f(4/27)=81/16$. Moreover, for any $x\in (0,4/27]$ the following equality holds.
\begin{equation}\label{eq_exact_form_f}
f(x)=\frac{16}{9x^2}\cos^4\left(\displaystyle\frac{\tan^{-1}\left(\sqrt{\frac{4}{27x}-1}\right)+\pi}{3}\right),
\end{equation}
where the function $\tan^{-1}(\cdot)$ is defined as a bijective map from $\R$ to $(-\pi/2,\pi/2)$ satisfying $\tan^{-1}(\tan(\theta))=\theta$ for all $\theta \in (-\pi/2,\pi/2)$.
\end{lemma}
\begin{proof}
As a topic in generating functions the power series \eqref{eq_dominant_power_series} is commonly studied (see, e.g, \cite[\mbox{pp. 200--201}]{GKP}). However, we give a proof for the statements for completeness. Let us analyze the cubic equation
\begin{equation}\label{eq_properties_dominant_power_series_1}
X=1+xX^3
\end{equation}
for $x\in (0,4/27)$. We see that for any $x\in (0,4/27)$ and $z\in\C$ satisfying $|z-1|=1/2$, the inequality $|xz^3|<|z-1|$ holds. Thus, the Lagrange inversion theorem (see, e.g, \cite[\mbox{Theorem 2.3.1}]{KP}) implies that in the domain $\{z\in\C\ |\ |z-1|<1/2\}$ there is exactly one root $X=v(x)$ of \eqref{eq_properties_dominant_power_series_1} and 
\begin{equation}\label{eq_properties_dominant_power_series_2}
v(x)^4=1+\sum_{n=1}^{\infty}\frac{x^n}{n!}\left(\frac{d}{dz}\right)^{n-1}(4z^3z^{3n})\Big|_{z=1}=f(x).
\end{equation} 

On the other hand, by algebraically solving \eqref{eq_properties_dominant_power_series_1} and specifying a root contained in the domain $\{z\in\C\ |\ |z-1|<1/2\}$ we can determine the explicite form of $v(x)$ as follows.
\begin{equation}\label{eq_properties_dominant_power_series_3}
v(x)= \displaystyle\frac{2}{\sqrt{3x}} \cos\left(\displaystyle\frac{\tan^{-1}\left(\sqrt{\frac{4}{27x}-1}\right)+\pi}{3}\right),
\end{equation}
where $\tan^{-1}(\cdot)$ is defined as stated in Lemma \ref{lem_properties_dominant_power_series}. 

The equalities \eqref{eq_properties_dominant_power_series_2}-\eqref{eq_properties_dominant_power_series_3} give \eqref{eq_exact_form_f} for $x\in(0,4/27)$. The ratio test shows that the radius of convergence of $f$ is $4/27$. Moreover, by continuity we have $\lim_{x \nearrow 4/27}f(x)=v(4/27)^4=81/16$, which completes the proof.
\end{proof}

Define a constant $D>0$ by
\begin{equation}\label{eq_definition_decay_constant}
D:=\lim_{h\to+\infty,h\in\N/\beta}D_h=\int_{-\beta}^{\beta}dx\sum_{\bx\in\G}|C(\bx\ua x,\b0\ua 0)|.
\end{equation}
Our main result is stated as follows.
\begin{theorem}\label{thm_estimate_correlation_function}
For any $\bx_1,\bx_2,\by_1,\by_2\in\G$ and $m\in\N\cup \{0\}$ and any $U\in\R$ with $|U|\le 1/(27D)$, the following equality and inequalities hold.
\begin{align}
&\<\psi_{\bx_1\ua}^*\psi_{\bx_2\da}^*\psi_{\by_2\da}\psi_{\by_1\ua}+\psi_{\by_1\ua}^*\psi_{\by_2\da}^*\psi_{\bx_2\da}\psi_{\bx_1\ua}\>=\sum_{n=0}^{\infty}a_nU^n,\label{eq_estimate_correlation_function_1}\\
&|\<\psi_{\bx_1\ua}^*\psi_{\bx_2\da}^*\psi_{\by_2\da}\psi_{\by_1\ua}+\psi_{\by_1\ua}^*\psi_{\by_2\da}^*\psi_{\bx_2\da}\psi_{\bx_1\ua}\>|\le R(|U|),\label{eq_estimate_correlation_function_2}
\end{align}
\begin{equation}\label{eq_estimate_correlation_function_3}
\begin{split}
&\Big|\<\psi_{\bx_1\ua}^*\psi_{\bx_2\da}^*\psi_{\by_2\da}\psi_{\by_1\ua}+\psi_{\by_1\ua}^*\psi_{\by_2\da}^*\psi_{\bx_2\da}\psi_{\bx_1\ua}\>-\sum_{n=0}^ma_nU^n\Big|\\
&\qquad\le R(|U|)-\sum_{n=0}^m\frac{128}{3n+4}\left(\begin{array}{c} 3n+4 \\ n \end{array}\right)(4D|U|)^n,
\end{split}
\end{equation}
where $\{a_n\}_{n=0}^{\infty}$ is given in \eqref{eq_perturbation_series_P_1_1}-\eqref{eq_perturbation_series_P_2} and 
\begin{equation}\label{eq_estimate_correlation_function_4}
R(|U|):=\left\{\begin{array}{cl} 32 & \text{ if } U=0,\\ \displaystyle\frac{32}{9D^2|U|^2} \cos^4\left(\displaystyle\frac{\tan^{-1}\left(\sqrt{\frac{1}{27D|U|}-1}\right)+\pi}{3}\right) &\text{ if }0<|U|\le \frac{1}{27D},\end{array}\right.
\end{equation}
with the function $\tan^{-1}(\cdot):\R\to (-\pi/2,\pi/2)$ satisfying $\tan^{-1}(\tan\theta)=\theta$ for all $\theta\in (-\pi/2,\pi/2)$.
\end{theorem}
\begin{proof}
Since by Lemma \ref{lem_properties_P_h} \eqref{item_lem_properties_P_h_3} and Lemma \ref{lem_bound_a_h_n}
\begin{equation}\label{eq_main_theorem_proof_3}
|a_n|=\lim_{h\to +\infty\atop h\in 2\N/\beta}|a_{h,n}|\le \frac{128}{3n+4}\left(\begin{array}{c}3n+4 \\ n\end{array}\right)(4D)^n,
\end{equation}
Lemma \ref{lem_properties_dominant_power_series} implies that for all $U\in [-1/(27D),1/(27D)]$ 
\begin{equation}\label{eq_main_theorem_proof_4}
\sum_{n=0}^{\infty}|a_n||U|^n\le R(|U|),
\end{equation}
where $R(|U|)$ is defined in \eqref{eq_estimate_correlation_function_4}. The inequalities \eqref{eq_estimate_correlation_function_2}-\eqref{eq_estimate_correlation_function_3} follow from \eqref{eq_estimate_correlation_function_1} and \eqref{eq_main_theorem_proof_3}-\eqref{eq_main_theorem_proof_4}.

We show that the equality \eqref{eq_estimate_correlation_function_1} holds for $U\in\R$ with $|U|\le 1/(27|D|)$. Let us fix any $\eps\in (0,1/(27D))$. Since $P|_{\la_{\X}=0,\forall \X\in\G^4}=\Tr e^{-\beta H}/\Tr e^{-\beta H_0}>0$ for all $U\in\R$, Proposition \ref{pro_convergence_P_h} implies that there exists $N_0\in\N$ such that $|P_h|_{\la_{\X}=0,\forall \X\in\G^4}|>0$ for all $h\in 2\N/\beta$ with $h\ge N_0/\beta$ and all $U\in\R$ with $|U|\le 1/(27D)-\eps$. Moreover, since $P_h|_{\la_{\X}=0,\forall \X\in\G^4}$ is a polynomial of $U$ we can take a simply connected domain $O_h$ $(\subset \C)$ containing the interval $[-1/(27D)+\eps,1/(27D)-\eps]$ inside such that $|P_h|_{\la_{\X}=0,\forall \X\in\G^4}|>0$ for all $U\in O_h$. Thus, we see that $\partial P_h/\partial \la_{\tilde{\X}_1}/P_h|_{\la_{\X}=0,\forall \X\in\G^4}$ defines an analytic function of $U$ in the domain $O_h$. By Lemma \ref{lem_bound_a_h_n} and Lemma \ref{lem_properties_dominant_power_series} the series $\sum_{n=0}^{\infty}a_{h,n}U^n$ converges for all $U\in\C$ with $|U|\le 1/(27D_h)$. By choosing $N_0$ sufficiently large we may assume that $1/(27D)-\eps \le 1/(27D_h)$ for all $h\in 2\N/\beta$ with $h\ge N_0/\beta$. Therefore, Lemma \ref{lem_properties_P_h} \eqref{item_lem_properties_P_h_2} and the identity theorem for analytic functions ensure that
\begin{equation}\label{eq_main_theorem_proof_1}
-\frac{1}{\beta}\frac{\frac{\partial}{\partial \la_{\tilde{\X}_1}}P_h(\{U_{\X}\}_{\X\in \G^4})}{P_h(\{U_{\X}\}_{\X\in \G^4})}\Bigg|_{\la_{\X}=0\atop\X\in\G^4}
=\sum_{n=0}^{\infty}a_{h,n}U^n
\end{equation}
for all $U\in [-1/(27D)+\eps,1/(27D)-\eps]$.

Note that Lemma \ref{lem_bound_a_h_n} implies 
\begin{equation}\label{eq_main_theorem_proof_2}
|a_{h,n}U^n|\le \frac{128}{3n+4}\left(\begin{array}{c}3n+4 \\ n \end{array}\right)\left(\frac{4}{27}\right)^n
\end{equation}
for all $U\in [-1/(27D)+\eps,1/(27D)-\eps]$ and the right hand side of \eqref{eq_main_theorem_proof_2} is summable over $\N\cup \{0\}$. Thus, by Lemma \ref{lem_properties_P_h} \eqref{item_lem_properties_P_h_3}, Corollary \ref{cor_correlation_P_h} and Lebesgue's dominated convergence theorem for $l^1(\N\cup \{0\})$ we can pass $h\to +\infty$ in \eqref{eq_main_theorem_proof_1} to deduce the equality \eqref{eq_estimate_correlation_function_1} for all $U\in [-1/(27D)+\eps,1/(27D)-\eps]$. Then, by sending $\eps \searrow 0$ and continuity we obtain \eqref{eq_estimate_correlation_function_1} for all $U\in [-1/(27D),1/(27D)]$.
\end{proof}

\begin{remark}
In Proposition \ref{pro_evaluation_decay_constant} we give a volume-independent upper bound on the decay constant $D$ in 2 dimensional case. One can straightforwardly extend the calculation of Proposition \ref{pro_evaluation_decay_constant} to derive volume-independent upper bounds on $D$ in any dimension. By replacing $D$ by these upper bounds, Theorem \ref{thm_estimate_correlation_function} provides volume-independent upper bounds on the perturbation series $\sum_{n=0}^{\infty}a_nU^n$.
\end{remark}

\section{Numerical results in 2D}\label{sec_numerical_results}
In this section we compute the perturbation series of the correlation function \\
$\<\psi_{\bx_1\ua}^*\psi_{\bx_2\da}^*\psi_{\by_2\da}\psi_{\by_1\ua}+\psi_{\by_1\ua}^*\psi_{\by_2\da}^*\psi_{\bx_2\da}\psi_{\bx_1\ua}\>$ up to 2nd order term in 2 dimensional case. We also implement the upper bound obtained in Theorem \ref{thm_estimate_correlation_function} and report the error between the correlation function and the 2nd order perturbation. Throughout this section it is assumed that $d=2$. 
\subsection{The decay constant for $d=2$}
In order to estimate the radius of convergence of the perturbation series $\sum_{n=0}^{\infty}a_nU^n$ and compute the upper bound on the sum of the higher order terms numerically, first we need to evaluate the decay constant $D$ defined in \eqref{eq_definition_decay_constant}. The result is presented in the following proposition.
\begin{proposition}\label{pro_evaluation_decay_constant}
The following inequality holds.
\begin{equation}\label{eq_evaluation_decay_constant}
\begin{split}
&D\le \left(\frac{16}{\beta^2} + \frac{32\pi^2}{3\sqrt{3}\beta} + \frac{16\pi^3}{3\sqrt{3}}\right)\Bigg(\frac{\beta^3}{2}+\frac{(2|t|+4|t'|)e^{\beta \xi}\beta}{1+e^{\beta \xi}}\\
&\qquad + 3\left(\frac{(2|t|+4|t'|)e^{\beta \xi}\beta}{1+e^{\beta \xi}}\right)^2+\left(\frac{(2|t|+4|t'|)e^{\beta \xi}\beta}{1+e^{\beta \xi}}\right)^3\Bigg),
\end{split}
\end{equation}
where $\xi:=4|t|+4|t'|+|\mu|$.
\end{proposition}

The derivation of the inequality \eqref{eq_evaluation_decay_constant} needs the following estimate.
\begin{lemma}\label{lem_discrete_integral}
The following inequality holds.
$$\sum_{\bx\in\G}\frac{1}{1+\sum_{l=1}^2|e^{i2\pi x_l/L}-1|^3L^3 /(8\pi^3\beta^3)}\le 4+\frac{8\pi^2\beta}{3\sqrt{3}}+\frac{4\pi^3\beta^2}{3\sqrt{3}},$$
where $\bx=(x_1,x_2)\in\G$.
\end{lemma}
\begin{proof}
For any $y\in\R$ let $\lfloor y \rfloor$ denote the largest integer which does not exceed $y$. By using the inequality that $|e^{i\theta}-1|\ge 2|\theta|/\pi$ for any $\theta\in [-\pi,\pi]$, we see that
\begin{equation*}
\begin{split}
&\sum_{\bx\in\G}\frac{1}{1+\sum_{l=1}^2|e^{i2\pi x_l/L}-1|^3L^3 /(8\pi^3\beta^3)}\\
&\quad\le 4\sum_{x_1=0}^{\lfloor L/2 \rfloor}\sum_{x_2=0}^{\lfloor L/2 \rfloor}\frac{1}{1+\sum_{l=1}^2|e^{i2\pi x_l/L}-1|^3L^3 /(8\pi^3\beta^3)}\\
&\quad\le 4\sum_{x_1=0}^{\infty}\sum_{x_2=0}^{\infty}\frac{1}{1+8 x_1^3/(\pi^3\beta^3)+8 x_2^3/(\pi^3\beta^3)}\\
&\quad= 4 + 8 \sum_{x_1=1}^{\infty}\frac{1}{1 + 8 x_1^3/(\pi^3\beta^3)}+4 \sum_{x_1=1}^{\infty}\sum_{x_2=1}^{\infty}\frac{1}{1+8 x_1^3/(\pi^3\beta^3)+8 x_2^3/(\pi^3\beta^3)}
\end{split}
\end{equation*}
\begin{equation*}
\begin{split}
&\quad\le 4 + 8 \int_0^{\infty}dx_1\frac{1}{1 + 8 x_1^3/(\pi^3\beta^3)}\\
&\quad\quad+4 \int_0^{\infty}dx_1\int_0^{\infty}dx_2\frac{1}{1+8 x_1^3/(\pi^3\beta^3)+8 x_2^3/(\pi^3\beta^3)}\\
&\quad= 4 + 4\pi\beta \int_0^{\infty}dx\frac{1}{1 + x^3}+\pi^2\beta^2\int_0^{\infty}dx\frac{1}{(1+x^3)^{2/3}}\int_0^{\infty}dx\frac{1}{1+x^3}\\
&\quad\le 4 +  (4\pi\beta + 2\pi^2\beta^2)\int_0^{\infty}dx\frac{1}{1 + x^3},
\end{split}
\end{equation*}
where we used the inequality
$$\int_0^{\infty}dx\frac{1}{(1+x^3)^{2/3}}\le 1 + \int_1^{\infty}\frac{1}{x^2}=2.$$
Then, by using the equality  
$$\int_0^{\infty}dx \frac{1}{1+x^3}=\frac{2\pi}{3\sqrt{3}}$$
(see \cite{GR}), we obtain the desired inequality.
\end{proof}

\begin{proof}[Proof of Proposition \ref{pro_evaluation_decay_constant}]
Let us define a linear operator $d_{l,L}:C^{\infty}(\R^2)\to C^{\infty}(\R^2)$ $(l=1,2)$ by
$$ (d_{l,L}f)(\bk):=\frac{f(\bk+2\pi \be_l/L)-f(\bk)}{2\pi/L}$$
for any $f\in C^{\infty}(\R^2)$. Then, the mean value theorem shows that for all $\bk=(k_1,k_2)\in \R^2$
\begin{equation}\label{eq_proof_decay_constant_1}
\begin{split}
&|(d_{1,L})^3f(k_1,k_2)|\le \sup_{\hat{k}_1\in [k_1,k_1+6\pi/L]}\left|\frac{\partial^3}{\partial k_1^3}f(\hat{k}_1,k_2)\right|,\\
&|(d_{2,L})^3f(k_1,k_2)|\le \sup_{\hat{k}_2\in [k_2,k_2+6\pi/L]}\left|\frac{\partial^3}{\partial k_2^3}f(k_1,\hat{k}_2)\right|.  
\end{split}
\end{equation}

Define a function $F(\bk,x):\R^2\times\R\to \R$ by $F(\bk,x):=e^{x E_{\bk}}/(1+e^{\beta E_{\bk}})$. We see that for any $\bx=(x_1,x_2)\in\G$, $x\in[-\beta,\beta]$ and $l\in \{1,2\}$
\begin{equation}\label{eq_proof_decay_constant_2}
\begin{split}
&\left(\frac{e^{i 2\pi x_l/L}-1}{2\pi/L}\right)^3C(\bx\ua x,\b0\ua 0)\\
&\quad=\frac{1}{L^d}\sum_{\bk\in\G^*}e^{-i\<\bk,\bx\>}((d_{l,L})^3F(\bk,x)1_{x\ge 0}-(d_{l,L})^3F(\bk,x+\beta)1_{x< 0}).
\end{split}
\end{equation}
By \eqref{eq_proof_decay_constant_1} and \eqref{eq_proof_decay_constant_2} we have that for any $\bx=(x_1,x_2)\in\G$, $x\in[-\beta,\beta]$ and $l\in \{1,2\}$
\begin{equation}\label{eq_proof_decay_constant_3}
\left|\left(\frac{e^{i 2\pi x_l/L}-1}{2\pi/L}\right)^3C(\bx\ua x,\b0\ua 0)\right|\le \sup_{x\in [0,\beta]}\sup_{\bk \in [0,2\pi+6\pi/L]\times[0,2\pi+6\pi/L]}\left|\frac{\partial^3}{\partial k_l^3}F(\bk,x)\right|.
\end{equation}

Note that for $l=1,2$
\begin{equation}\label{eq_proof_decay_constant_4}
\begin{split}
\frac{\partial^3}{\partial k_l^3}F(\bk,x)=&\frac{\partial^3 E_{\bk}}{\partial k_l^3}\frac{\partial }{\partial E_{\bk}}\left(\frac{e^{x E_{\bk}}}{1+e^{\beta E_{\bk}}}\right)+3\frac{\partial E_{\bk}}{\partial k_l}\frac{\partial^2 E_{\bk}}{\partial k_l^2}\frac{\partial^2}{\partial E_{\bk}^2}\left(\frac{e^{x E_{\bk}}}{1+e^{\beta E_{\bk}}}\right)\\
&+\left(\frac{\partial E_{\bk}}{\partial k_l}\right)^3\frac{\partial^3}{\partial E_{\bk}^3}\left(\frac{e^{x E_{\bk}}}{1+e^{\beta E_{\bk}}}\right)
\end{split}
\end{equation}
and
\begin{equation}\label{eq_proof_decay_constant_5}
|E_{\bk}|\le 4|t|+4|t'|+|\mu|,\quad 
\left|\frac{\partial E_{\bk}}{\partial k_l}\right|, \left|\frac{\partial^2 E_{\bk}}{\partial k_l^2}\right|, \left|\frac{\partial^3 E_{\bk}}{\partial k_l^3}\right|\le 2|t| + 4|t'|.
\end{equation}

Moreover, we can prove that for $m\in \{1,2,3\}$ and any $E\in\R$
\begin{equation}\label{eq_proof_decay_constant_6}
\sup_{x\in [0,\beta]}\left|\left(\frac{d}{d E}\right)^m\frac{e^{xE}}{1+e^{\beta E}}\right|\le \left(\frac{\beta e^{\beta |E|}}{1+e^{\beta |E|}}\right)^m.
\end{equation}  
By the equalities that $e^{xE}/(1+e^{\beta E})=e^{(\beta-x)(-E)}/(1+e^{\beta (-E)})$ and $d/dE=-d/d(-E)$ we may assume $E\ge 0$ without losing generality in the following argument.
First let us study the case for $m=1$.
$$
\left|\frac{d}{d E}\frac{e^{xE}}{1+e^{\beta E}}\right|=\left|\frac{e^{x E}}{1+e^{\beta E}}\left(x-\frac{\beta e^{\beta E}}{1+e^{\beta E}}\right)\right|\le \frac{\beta e^{\beta E}}{1+e^{\beta E}},$$
which is \eqref{eq_proof_decay_constant_6} for $m=1$.

Next consider the case that $m=2$. We see that
$$\left(\frac{d}{d E}\right)^2\frac{e^{x E}}{1+e^{\beta E}}=\frac{e^{xE}}{1+e^{\beta E}}g_1(x),$$
with 
$$g_1(x):=\left(x -\frac{\beta e^{\beta  E}}{1+e^{\beta E}}\right)^2-\frac{\beta^2e^{\beta  E}}{(1+e^{\beta E})^2}.$$
Note that $|g_1(0)|$, $|g_1(\beta e^{\beta E}/(1+e^{\beta E}))|$, $|g_1(\beta)|\le \beta^2e^{\beta E}/(1+e^{\beta E})$. Thus, we have that
\begin{equation*}
\begin{split}
\left|\frac{d^2}{d E^2}\frac{e^{x_0E}}{1+e^{\beta E}}\right|&\le \frac{e^{\beta E}}{1+e^{\beta E}}\max\{|g_1(0)|, |g_1(\beta e^{\beta E}/(1+e^{\beta E}))|, |g_1(\beta)|\}\\
&\le\left(\frac{\beta e^{\beta E}}{1+e^{\beta E}}\right)^2.
\end{split}
\end{equation*}

Finally, analyze the case for $m=3$. A calculation shows that
\begin{equation}\label{eq_proof_decay_constant_7}
\left(\frac{d}{d E}\right)^3\frac{e^{xE}}{1+e^{\beta E}}=\frac{e^{xE}}{1+e^{\beta E}}g_2(x),\end{equation}
where
\begin{equation*}
g_2(x):=x^3 -\frac{3\beta e^{\beta E}}{1+e^{\beta E}}x^2 + \frac{3\beta^2(e^{\beta E}-1)e^{\beta E}}{(1+e^{\beta E})^2}x -\frac{\beta^3 e^{\beta E}}{(1+e^{\beta E})^3}(e^{2\beta E}-4e^{\beta E}+1).
\end{equation*}
The roots $x_1,x_2$ of the equation $g_2'(x)=0$ are given by
$$x_1=\frac{\beta e^{\beta E/2}}{1+e^{\beta E}}(e^{\beta E/2}-1),\ x_2=\frac{\beta e^{\beta E/2}}{1+e^{\beta E}}(e^{\beta E/2}+1).$$
Since $0\le x_1<\beta \le x_2$, $\max_{x\in [0,\beta]}|g_2(x)|=\max\{|g_2(0)|,|g_2(x_1)|,|g_2(\beta)|\}$. Moreover, we see that
\begin{equation*}
\begin{split}
&|g_2(0)|=\frac{\beta^3e^{\beta E}}{(1+e^{\beta E})^3}|e^{2\beta E}-4e^{\beta E}+1|\le \frac{\beta^3e^{\beta E}}{(1+e^{\beta E})^3}(e^{2\beta E}+e^{\beta E})=\frac{\beta^3e^{2\beta E}}{(1+e^{\beta E})^2},\\  
&|g_2(\beta)|=\frac{\beta^3}{(1+e^{\beta E})^3}|e^{2\beta E}-4e^{\beta E}+1|\le |g_2(0)|,\\  
&|g_2(x_1)|=\frac{\beta^3e^{\beta E}}{(1+e^{\beta E})^3}|e^{\beta E}+2e^{\beta E/2}-1| \le\frac{\beta^3 e^{\beta E}}{(1+e^{\beta E})^3}(e^{2\beta E}+e^{\beta E}) =\frac{\beta^3e^{2\beta E}}{(1+e^{\beta E})^2},
\end{split}
\end{equation*}
which imply that 
\begin{equation}\label{eq_proof_decay_constant_8}
\max_{x_0\in [0,\beta]}|g_2(x_0)|\le \frac{\beta^3e^{2\beta E}}{(1+e^{\beta E})^2}.
\end{equation}
 Combining \eqref{eq_proof_decay_constant_7} with \eqref{eq_proof_decay_constant_8} deduces \eqref{eq_proof_decay_constant_6} for $m=3$. 

Then by \eqref{eq_proof_decay_constant_3}-\eqref{eq_proof_decay_constant_6} we have for $l=1,2$
\begin{equation}\label{eq_proof_decay_constant_9}
\begin{split}
&\left|\left(\frac{e^{i2\pi x_l/L}-1}{2\pi/L}\right)^3C(\bx\ua x,\b0\ua 0)\right|\\
&\quad \le \frac{(2|t|+4|t'|)\beta e^{\beta \xi}}{1+e^{\beta \xi}} + 3 \left(\frac{(2|t|+4|t'|)\beta e^{\beta \xi}}{1+e^{\beta \xi}}\right)^2+\left(\frac{(2|t|+4|t'|)\beta e^{\beta \xi}}{1+e^{\beta \xi}}\right)^3,
\end{split}
\end{equation}
with $\xi:=4|t|+4|t'|+|\mu|$.
By using the inequality that $|C(\bx\ua x,\b0\ua 0)|\le 1$, \eqref{eq_proof_decay_constant_9} and Lemma \ref{lem_discrete_integral} we can derive the inequality \eqref{eq_evaluation_decay_constant}.
\end{proof} 

\subsection{The 2nd order perturbation}
As a preparation for the numerical implementation of the 2nd order perturbation $a_0+a_1U+a_2U^2$, we rewrite the formula \eqref{eq_perturbation_series_P_1_1} for $a_n$ in a more suitable form for practical computation.

By decomposing the determinant of the covariance matrix into the sums over permutations, $G_n$ defined in \eqref{eq_perturbation_series_P_2} can be written as follows.
\begin{equation}\label{eq_computation_G_n}
G_n=\frac{1}{n!}\sum_{\pi,\tau\in S_n}\sgn(\pi)\sgn(\tau)g_n(\pi,\tau),
\end{equation}
with 
\begin{equation*}
\begin{split}
&g_n(\pi,\tau):=\\
&\qquad\opPi^n_{j=1}\Bigg(-\sum_{\bx_1^j,\bx_2^j,\by_1^j,\by_2^j\in \G}\int_0^{\beta}dx_j(\delta_{\bx_1^j,\bx_2^j}\delta_{\by_1^j,\by_2^j}\delta_{\bx_1^j,\by_1^j}+\la_{\bx_1^j,\bx_2^j,\by_1^j,\by_2^j}+\la_{\by_1^j,\by_2^j,\bx_1^j,\bx_2^j})\\
&\qquad\qquad\quad\cdot C(\bx_1^j\ua x_j,\by_1^{\pi(j)}\ua x_{\pi(j)})C(\bx_2^j\da x_j,\by_2^{\tau(j)}\da x_{\tau(j)})\Bigg).
\end{split}
\end{equation*}
The formula \eqref{eq_perturbation_series_P_1_1} for $a_0,a_1,a_2$ is 
\begin{equation}\label{eq_computation_a_n}
\begin{split}
&a_0=-\frac{1}{\beta}\frac{\partial}{\partial \lambda_{\tilde{\X}_1}}G_1\Big|_{\lambda_{\X}=0\atop\forall \X\in\G^4},\ a_1=-\frac{1}{\beta}\frac{\partial}{\partial \lambda_{\tilde{\X}_1}}\left(G_2-\frac{1}{2}G_1^2\right)\Big|_{\lambda_{\X}=0\atop\forall \X\in\G^4},\\&a_2=-\frac{1}{\beta}\frac{\partial}{\partial \lambda_{\tilde{\X}_1}}\left(G_3-G_1G_2+\frac{1}{3}G_1^3\right)\Big|_{\lambda_{\X}=0\atop\forall \X\in\G^4}.
\end{split}
\end{equation}
By substituting \eqref{eq_computation_G_n} into \eqref{eq_computation_a_n} and canceling we can simplify the expressions for $a_1$, $a_2$ as follows.
\begin{equation*}
\begin{split}
&a_1=-\frac{1}{\beta}\frac{\partial}{\partial \lambda_{\tilde{\X}_1}}\frac{1}{2}\Bigg(g_2\left(\left(\begin{array}{cc} 1 & 2 \\ 2 & 1 \end{array}\right), \left(\begin{array}{cc} 1 & 2 \\ 2 & 1 \end{array}\right)\right)\\
&\qquad\qquad\qquad\qquad- g_2\left(\text{Id}_2,\left(\begin{array}{cc} 1 & 2 \\ 2 & 1 \end{array}\right)\right)- g_2\left(\left(\begin{array}{cc} 1 & 2 \\ 2 & 1 \end{array}\right), \text{Id}_2\right)\Bigg)\Bigg|_{\la_{\X}=0\atop\forall \X\in\G^4},\\
&a_2= -\frac{1}{\beta}\frac{\partial}{\partial \la_{\tilde{\X}_1}}\frac{1}{3}\Bigg(g_3\left(\text{Id}_3,\left(\begin{array}{ccc} 1 & 2 & 3 \\ 2 & 3 & 1 \end{array}\right)\right)+g_3\left(\left(\begin{array}{ccc} 1 & 2 & 3 \\ 2 & 3 & 1 \end{array}\right),\text{Id}_3\right)\\
&\quad\quad+g_3\left(\left(\begin{array}{ccc} 1 & 2 & 3 \\ 2 & 3 & 1 \end{array}\right),\left(\begin{array}{ccc} 1 & 2 & 3 \\ 2 & 3 & 1 \end{array}\right)\right)+3g_3\left(\left(\begin{array}{ccc} 1 & 2 & 3 \\ 1 & 3 & 2 \end{array}\right),\left(\begin{array}{ccc} 1 & 2 & 3 \\ 2 & 1 & 3 \end{array}\right)\right)\\
&\quad\quad-3g_3\left(\left(\begin{array}{ccc} 1 & 2 & 3 \\ 1 & 3 & 2 \end{array}\right),\left(\begin{array}{ccc} 1 & 2 & 3 \\ 2 & 3 & 1 \end{array}\right)\right)-3g_3\left(\left(\begin{array}{ccc} 1 & 2 & 3 \\ 2 & 3 & 1 \end{array}\right),\left(\begin{array}{ccc} 1 & 2 & 3 \\ 1 & 3 & 2 \end{array}\right)\right)\\
&\quad\quad + g_3\left(\left(\begin{array}{ccc} 1 & 2 & 3 \\ 2 & 3 & 1 \end{array}\right),\left(\begin{array}{ccc} 1 & 2 & 3 \\ 3 & 1 & 2 \end{array}\right)\right)\Bigg)\Bigg|_{\la_{\X}=0\atop\forall \X\in\G^4},
\end{split}
\end{equation*}
 where
\begin{equation*}
\begin{split}
&\text{Id}_2= \left(\begin{array}{cc} 1 & 2 \\ 1 & 2 \end{array}\right),\left(\begin{array}{cc} 1 & 2 \\ 2 & 1 \end{array}\right)\in S_2,\\
&\text{Id}_3= \left(\begin{array}{ccc} 1 & 2 &3 \\ 1 & 2 &3 \end{array}\right), \left(\begin{array}{ccc} 1 & 2 & 3\\ 1 & 3 & 2 \end{array}\right),  \left(\begin{array}{ccc} 1 & 2 & 3\\ 2 & 3 & 1 \end{array}\right), \left(\begin{array}{ccc} 1 & 2 & 3\\ 3 & 1 & 2 \end{array}\right)\in S_3.    
\end{split}
\end{equation*}

In practice we need to implement the permutation-dependent terms\\ 
$\partial /\partial \la_{\tilde{\X}_1}g_n(\pi,\tau)|_{\la_{\X}=0,\forall \X\in\G^4}$. Let us describe its implementation below.

Note that the multiple integral $\opPi_{j=1}^n\int_0^{\beta}dx_jf(x_1,\cdots,x_n)$ of any integrable function $f$ can be decomposed into a sum of $n!$ integrals as follows.
\begin{equation}\label{eq_computation_decomposition_integral}
\begin{split}
\opPi_{j=1}^n\int_0^{\beta}dx_jf(x_1,\cdots,x_n)&=\opPi_{j=1}^n\int_0^{\beta}dx_j\sum_{\eta\in S_n}1_{x_{\eta(1)}>x_{\eta(2)}>\cdots >x_{\eta(n)}}f(x_1,\cdots,x_n)\\
&=\sum_{\eta\in S_n}\opPi_{j=1}^n\int_0^{x_{\eta(j-1)}}dx_{\eta(j)}f(x_1,\cdots,x_n),
\end{split}
\end{equation}
where $x_{\eta(0)}:=\beta$. 
 
By substituting the expression \eqref{eq_covariance_matrix} of the covariance matrix and decomposing the integral $\opPi_{j=1}^n\int_0^{\beta}dx_j$ as in \eqref{eq_computation_decomposition_integral}, we can expand $\partial /\partial \la_{\tilde{\X}_1}g_n(\pi,\tau)|_{\la_{\X}=0,\forall \X\in\G^4}$ as sums over the momentum space $\G^*$ as follows. For $n\in\N$
\begin{equation}\label{eq_implementation_formula}
\begin{split}
&\frac{\partial}{\partial \la_{\tilde{\X}_1}}g_n(\pi,\tau)\Big|_{\la_{\X}=0\atop\forall \X\in\G^4}=\frac{(-1)^n}{L^{(n+1)d}}\sum_{\bk_1,\cdots,\bk_n,\bp_1,\cdots,\bp_n\in \G^*}\\
&\cdot\sum_{j=1}^n(\cos(\<\bk_j,\bx_1\>+\<\bp_j,\bx_2\>-\<\bk_{\pi^{-1}(j)},\by_1\>-\<\bp_{\tau^{-1}(j)},\by_2\>)\\
&\quad +\cos(\<\bk_j,\by_1\>+\<\bp_j,\by_2\>-\<\bk_{\pi^{-1}(j)},\bx_1\>-\<\bp_{\tau^{-1}(j)},\bx_2\>))\\
&\cdot\opPi_{l=1\atop l\neq j}^n\delta(\bk_l+\bp_l-\bk_{\pi^{-1}(l)}-\bp_{\tau^{-1}(l)})\\
&\cdot\sum_{\eta\in S_n}\opPi_{j=1}^n\int_0^{x_{\eta(j-1)}}dx_{\eta(j)}e^{x_{\eta(j)}(E_{\bk_{\eta(j)}}+E_{\bp_{\eta(j)}} - E_{\bk_{\pi^{-1}(\eta(j))}}-E_{\bp_{\tau^{-1}(\eta(j))}})}\\
&\cdot\opPi_{j=1}^n\left(\frac{1_{x_{\pi(j)}-x_j\le 0}}{1+e^{\beta E_{\bk_j}}}-\frac{1_{x_{\pi(j)}-x_j>0}}{1+e^{-\beta E_{\bk_j}}}\right)
\left(\frac{1_{x_{\tau(j)}-x_j\le 0}}{1+e^{\beta E_{\bp_j}}}-\frac{1_{x_{\tau(j)}-x_j>0}}{1+e^{-\beta E_{\bp_j}}}\right),
\end{split}
\end{equation}
where $x_{\eta(0)}=\beta$. 

Let us sketch how to implement \eqref{eq_implementation_formula}. We prepare a function of real variables $\tilde{E}_1,\cdots,\tilde{E}_n$ returning the exact value of $\opPi_{j=1}^n\int_0^{x_{j-1}}dx_{j}e^{x_{j}\tilde{E}_j}$ with $x_0=\beta$ beforehand. Then we iterate the system with respect to the variables $\bk_1,\cdots,\bk_n$, $\bp_1,\cdots,\bp_n\in\G^*$. For fixed $\bk_1,\cdots,\bk_n,\bp_1,\cdots,\bp_n\in\G^*$ we iterate with respect to the permutation $\eta\in S_n$. For each $\eta\in S_n$ we substitute the variables $\tilde{E}_j=E_{\bk_{\eta(j)}}+E_{\bp_{\eta(j)}} - E_{\bk_{\pi^{-1}(\eta(j))}}-E_{\bp_{\tau^{-1}(\eta(j))}}$ $(j=1,\cdots,n)$ into the function $\opPi_{j=1}^n\int_0^{x_{j-1}}dx_{j}e^{x_{j}\tilde{E}_j}$ and its returning value is then multiplied by the constant
$$
1_{x_{\eta(1)}>\cdots>x_{\eta(n)}}\opPi_{j=1}^n\left(\frac{1_{x_{\pi(j)}-x_j\le 0}}{1+e^{\beta E_{\bk_j}}}-\frac{1_{x_{\pi(j)}-x_j>0}}{1+e^{-\beta E_{\bk_j}}}\right)
\left(\frac{1_{x_{\tau(j)}-x_j\le 0}}{1+e^{\beta E_{\bp_j}}}-\frac{1_{x_{\tau(j)}-x_j>0}}{1+e^{-\beta E_{\bp_j}}}\right).
$$

\subsection{Numerical values}
Here we display our numerical results. In our computation we fix the physical parameters $t,t',\mu,\beta$ to satisfy $t=t'=\mu=0.01$, $\beta=1$. In this configuration the upper bound on $D$ obtained in Proposition \ref{pro_evaluation_decay_constant} is $92.04$. Thus, by Theorem \ref{thm_estimate_correlation_function} the radius of convergence $1/(27|D|)$ of our perturbation series $\sum_{n=0}^{\infty}a_nU^n$ is estimated to be larger than equal to $4.024\times 10^{-4}$.

The errors between the correlation function and the 2nd order perturbation for various $|U|$ less than $4.024\times 10^{-4}$ are exhibited in Table \ref{table_error_estimate}, where $\text{Error}$ is defined by the right hand side of the inequality \eqref{eq_estimate_correlation_function_3} for $m = 2$ and $D=92.04$ and satisfies that
$$|\<\psi_{\bx_1\ua}^*\psi_{\bx_2\da}^*\psi_{\by_2\da}\psi_{\by_1\ua} +\psi_{\by_1\ua}^*\psi_{\by_2\da}^*\psi_{\bx_2\da}\psi_{\bx_1\ua} \>-a_0-a_1U-a_2U^2| \le \text{Error}.$$
\begin{table}
\begin{center}
\begin{tabular}{|c|c|c|c|c|}
\hline
$|U|$ & $1.0\times 10^{-6}$ & $5.0\times10^{-6}$  & $1.0\times 10^{-5}$ & $5.0\times 10^{-5}$\\
 \hline
$\text{Error}$ & $1.408\times 10^{-7}$ & $1.773\times 10^{-5}$  & $1.433\times 10^{-4}$ & $1.942\times 10^{-2}$ \\\hline \hline
$|U|$ & $1.0\times 10^{-4}$ & $2.0\times 10^{-4}$ & $3.0\times 10^{-4}$ & $4.0\times 10^{-4}$ \\
 \hline
$\text{Error}$ & $1.739\times 10^{-1}$ & $1.842$ & $9.454$ & $7.307\times 10$ \\
\hline 
\end{tabular}
\caption{Errors between the correlation function and the 2nd order perturbation}\label{table_error_estimate}
\end{center}
\end{table}

Let us fix $U=1.0\times 10^{-5}$. According to Table \ref{table_error_estimate}, the error between the correlation function and the 2nd order perturbation is estimated as $1.433\times 10^{-4}$. Table \ref{table_same_sites} shows values of $a_0,a_1,a_2$ and $a_0+a_1U+a_2U^2$ in the case that $\bx_1=\bx_2=\by_1=\by_2=(l,l)$ $(l\in\{0,1,\cdots,5\})$ for various lattice size $L$ from $10$ up to $18$. We observe that each of $a_0,a_1,a_2$ respectively takes the same value for any $L\in\{10,\cdots,18\}$ and $l\in\{0,\cdots,5\}$.    
\begin{table}
\begin{center}
\begin{tabular}{|c|c|}
\hline
$L$ & $10,11,\cdots, 18$ \\
 \hline
$a_0$ & $5.050\times 10^{-1}$   \\
\hline
$a_1$ & $-3.774\times 10^{-1}$ \\
\hline
$a_2$ & $9.339\times 10^{-2}$  \\
\hline
$a_0 + a_1 U + a_2 U^2$ & $5.050\times 10^{-1}$ \\
\hline 
\end{tabular}
\caption{2nd order perturbation in the case that $\bx_1=\bx_2=\by_1=\by_2=(l,l)$ $(l\in\{0,1,\cdots,5\})$}\label{table_same_sites}
\end{center}
\end{table}

Table \ref{table_x1_equal_y1_x2_equal_y2} shows the values of $a_0,a_1,a_2$ and $a_0+a_1U+a_2U^2$ in the case that $\bx_1=\by_1=(0,0), \bx_2=\by_2=(l,l)$ $(l\in\{1,\cdots,5\})$ for various lattice size $L$ from $10$ to $18$. Again we see that each of $a_0,a_1,a_2$ respectively takes the same value for any $L\in\{10,\cdots,18\}$ and $l\in\{1,\cdots,5\}$. 
\begin{table}
\begin{center}
\begin{tabular}{|c|c|}
\hline
$L$ & $10,11,\cdots, 18$ \\
 \hline
$a_0$ & $5.050\times 10^{-1}$   \\
\hline
$a_1$ & $-2.524\times 10^{-1}$ \\
\hline
$a_2$ & $9.402\times 10^{-2}$  \\
\hline
$a_0 + a_1 U + a_2 U^2$ & $5.050\times 10^{-1}$ \\
\hline 
\end{tabular}
\caption{2nd order perturbation in the case that $\bx_1=\by_1=(0,0),\bx_2=\by_2=(l,l)$ $(l\in\{1,\cdots,5\})$}\label{table_x1_equal_y1_x2_equal_y2}
\end{center}
\end{table}
Since we have fixed a small $U$ so that Error becomes sufficiently small, the 1st and 2nd order terms do not contribute to the sum $a_0+a_1U+a_2U^2$ much in these numerical simulations. 

We also computed $a_0,a_1,a_2$ in the case that $\bx_1=\bx_2=(0,0), \by_1=\by_2=(l,l)$ for $l\in\{1,\cdots,5\}$ for $L=10,11,\cdots,18$. The result shows that $|a_0|,|a_1|,|a_2|\le 1.5\times10^{-5}$ for any $l\in\{1,\cdots,5\}$ and $L\in\{10,11,\cdots,18\}$. In this case the values of $|a_0|,|a_1|,|a_2|$ are much smaller than those presented in Tables \ref{table_same_sites}-\ref{table_x1_equal_y1_x2_equal_y2}. This result indicates that the 4 point correlation function takes small values if $\bx_1=\bx_2$, $\by_1=\by_2$ and $|\bx_1-\by_1|$ is large and agrees with the decaying property of the 4 point correlation function for the 2 dimensional Hubbard model proved in \cite{KT}.     

\appendix

\section*{Appendices}
In this section we review the definitions of the Fermionic Fock space and the annihilation, creation operators, prove Proposition \ref{pro_perturbation_series} and show that the covariance matrix $C_h$ has a non-zero determinant independent of the parameter $h$. 

We write a matrix $M$ indexed by finite sets $S$, $S'$ with $\sharp S=\sharp S'=n$ as $M=(M(s,s'))_{s\in S,s'\in S'}$. In this notation let us think that each element of $S$ and $S'$ has already been given a number from $1$ to $n$ and $M$ is defined by $M=(M(s_j,s'_k))_{1\le j,k\le n}$ even if the numbering of $S$ and $S'$ is not specified in the context. The main results Proposition \ref{pro_perturbation_series} and Proposition \ref{pro_determinant_covariance} concluded after some argument involving such matrices in this Appendices are independent of how to number the index sets. For any finite set $B$ let $L^2(B;\C)$ denote the complex linear space consisting of complex-valued functions on $B$, even when we do not introduce an inner product in $L^2(B;\C)$.

\section{The Fermionic Fock space}\label{appendix_Fock_space}
The first part of Appendices reviews the definitions of the Fermionic Fock space on the lattice $\G\times \spin$ and the annihilation, creation operators $\psi_{\bx\s},\psi_{\bx\s}^*$.

For any $n\in \N$ we consider the linear space $L^2((\G\times \spin)^n;\C)$ as a Hilbert space equipped with the inner product $\<\cdot,\cdot\>_{L^2((\G\times \spin)^n;\C)}$ defined by
\begin{equation*}
\begin{split}
&\<\phi_1,\phi_2\>_{L^2((\G\times \spin)^n;\C)}\\
&\quad:= \sum_{\bx_1,\cdots,\bx_n\in \G}\sum_{\s_1,\cdots,\s_n\in\spin}\overline{\phi_1(\bx_1\s_1,\cdots,\bx_n\s_n)}\phi_2(\bx_1\s_1,\cdots,\bx_n\s_n).
\end{split}
\end{equation*}
By convention we set $L^2((\G\times \spin)^0;\C):=\C$. 

For $n\in \N$ the anti-symmetrization operator $A_n:L^2((\G\times \spin)^n;\C)\to$\\
$ L^2((\G\times \spin)^n;\C)$ is defined by
$$(A_n\phi)(\bx_1\s_1,\cdots,\bx_n\s_n):= \frac{1}{n!}\sum_{\pi\in S_n}\sgn(\pi)\phi(\bx_{\pi(1)}\s_{\pi(1)},\cdots,\bx_{\pi(n)}\s_{\pi(n)}).$$
The operator $A_0$ is defined as the identity map on $\C$, i.e, $A_0z:=z$ for all $z\in\C$.

The subspace $A_n(L^2((\G\times \spin)^n;\C))$ of $L^2((\G\times \spin)^n;\C)$ is called as the Fermionic $n-$particle space and is a Hilbert space equipped with the inner product of $L^2((\G\times \spin)^n;\C)$. Note that by anti-symmetry for any $n> 2L^d$, \\
$A_n(L^2((\G\times \spin)^n;\C))=\{0\}$.

The Fermionic Fock space $F_f(L^2(\G\times \spin;\C))$ is defined as the direct sum of $A_n(L^2((\G\times \spin)^n;\C))$ $(n=0,\cdots,2L^d)$ as follows.
$$F_f(L^2(\G\times \spin;\C)) := \bigoplus_{n=0}^{2L^d}A_n(L^2((\G\times \spin)^n;\C)).$$
The space $F_f(L^2(\G\times \spin;\C))$ is a Hilbert space with inner product $\<\cdot,\cdot\>_{F_f}$ defined by
$$\<\phi_1,\phi_2\>_{F_f}:= \sum_{n=0}^{2L^d}\<\phi_{1,n},\phi_{2,n}\>_{L^2((\G\times \spin)^n;\C)},$$
for any vectors $\phi_1=(\phi_{1,0},\phi_{1,1},\cdots,\phi_{1,2L^d})$, $\phi_2=(\phi_{2,0},\phi_{2,1},\cdots,\phi_{2,2L^d})$\\
$\in F_f(L^2(\G\times \spin;\C))$.

Define a set of functions
 $\{\phi_{\bk\s}\}_{(\bk,\s)\in \G^*\times \spin}$ $\subset L^2(\G\times\spin;\C)$ by \\
$\phi_{\bk\s}(\bx\tau):=\delta_{\s,\tau}e^{-i\<\bk,\bx\>}/L^{d/2}$. We then define a function
 $\phi_{\bk_1\s_1}\cdots\phi_{\bk_n\s_n}$\\
$\in L^2((\G\times\spin)^n;\C)$
 by
$$
\phi_{\bk_1\s_1}\cdots\phi_{\bk_n\s_n}(\bx_1\tau_1,\cdots,\bx_n\tau_n):=\phi_{\bk_1\s_1}(\bx_1\tau_1)\cdots\phi_{\bk_n\s_n}(\bx_n\tau_n).
$$
An orthonormal basis of $F_f(L^2(\G\times\spin;\C))$ is given by
 $\bigcup_{n=0}^{2L^d}B_n$, where
\begin{equation}\label{eq_orthonormal_basis}
\begin{split}
&B_0:=\{1\}(\subset \C),\\ 
&B_n:=\left\{\sqrt{n!}A_n\left(\opPi_{(\bk,\s)\in
 \G^*\times\spin}\phi_{\bk\s}^{n_{\bk\s}}\right)\ \Big|\ n_{\bk\s}\in \{0,1\},\
 \sum_{\bk\in \G^*}\sum_{\s\in \spin}n_{\bk\s}=n\right\}
\end{split}
\end{equation}
for $n\in\{1,2,\cdots,2L^d\}$.
Thus, we see that $\dim F_f(L^2(\G\times\spin;\C))=\sum_{n=0}^{2L^d}\sharp B_n=2^{2L^d}$.

The annihilation operator $\psi_{\bx\s}:F_f(L^2(\G\times \spin;\C))\to F_f(L^2(\G\times \spin;\C))$ $(\bx\in \G,\s\in \spin)$ is defined in the following steps. For any $n\in\N\cup \{0\}$ and any $\phi\in A_{n+1}(L^2((\G\times \spin)^{n+1};\C))$, $\psi_{\bx\s}\phi\in A_{n}(L^2((\G\times \spin)^{n};\C))$ is defined by 
$$(\psi_{\bx\s}\phi)(\bx_1\s_1,\cdots,\bx_n\s_n):=\sqrt{n+1}\phi(\bx\s,\bx_1\s_1,\cdots,\bx_n\s_n).$$ 
For any $z\in A_{0}(L^2((\G\times \spin)^{0};\C))$, $\psi_{\bx\s} z:=0$. The domain of the operator $\psi_{\bx\s}$ is then extended to the whole space $F_f(L^2(\G\times \spin;\C))$ by linearity.

The creation operator $\psi_{\bx\s}^*:F_f(L^2(\G\times \spin;\C))\to F_f(L^2(\G\times \spin;\C))$ is the adjoint operator of $\psi_{\bx\s}$ and characterized as follows. For any $n\in\N\cup \{0\}$ and any $\phi\in A_{n}(L^2((\G\times \spin)^{n};\C))$, $\psi_{\bx\s}^*\phi\in A_{n+1}(L^2((\G\times \spin)^{n+1};\C))$ and \begin{equation*}
\begin{split}
&(\psi_{\bx\s}^*\phi)(\bx_1\s_1,\cdots,\bx_{n+1}\s_{n+1})\\
&\quad=\frac{1}{\sqrt{n+1}}\sum_{l=1}^{n+1}(-1)^{l-1}\delta_{\bx,\bx_l}\delta_{\s,\s_l}\phi(\bx_1\s_1,\cdots,\widehat{\bx_l\s_l},\cdots,\bx_{n+1}\s_{n+1}),
\end{split}
\end{equation*}
where the notation `$\widehat{\bx_l\s_l}$' stands for the omission of the variable $\bx_l\s_l$. 

For any operators $A$, $B$ let $\{A,B\}$ denote $AB+BA$. The operators $\psi_{\bx\s}$, $\psi_{\bx\s}^*$ $(\bx\in \G,\s\in \spin)$ satisfy the following canonical anti-commutation relations. For all $\bx,\by\in \G$, $\s,\tau\in \spin$,
\begin{equation}\label{eq_CAR}
\{\psi_{\bx\s},\psi_{\by\tau}\} = \{\psi_{\bx\s}^*,\psi_{\by\tau}^*\}=0,\ \{\psi_{\bx\s},\psi_{\by\tau}^*\}=\delta_{\bx,\by}\delta_{\s,\tau}.
\end{equation}
See, e.g, \cite{BR} for more detailed definitions of the Fermionic Fock space and the operators on it.

\section{The temperature-ordered perturbation series}\label{appendix_perturbation_series}
In this section we present the derivation of the perturbation series \eqref{eq_perturbation_series}. Propositions claimed here are standard tools in many-body theory (see, e.g, \\ \cite[\mbox{Chapter 2, Chapter 3}]{M}). This part of Appendices is devoted to show them in a mathematical context.

Let us fix notations used in the analysis below. Let $H_0$, $V_{\lambda}$ be the operators defined in \eqref{eq_H_0}-\eqref{eq_matrix_F} and \eqref{eq_V_lambda}, respectively. In our argument in this section, however, we do not use the relation \eqref{eq_relation_U_lambda} or the condition \eqref{eq_constraint_U} imposed on the parameter $\{U_{\X}\}_{\X\in \G^4}$. One can consider more general $V_{\la}$ of the form \eqref{eq_V_lambda} parameterized by any complex multi-variable $\{U_{\X}\}_{\X\in \G^4}$ in this section.

Define the operators $V_{\lambda}(s)$, $\psi_{\bx\s}(s)$, $\psi_{\bx\s}^*(s)$ $(s\in \R,\bx\in\G,\s\in\spin)$ by
$$V_{\lambda}(s):=e^{sH_0}V_{\lambda}e^{-sH_0},\ \psi_{\bx\s}(s):=e^{sH_0}\psi_{\bx\s}e^{-sH_0},\ \psi_{\bx\s}^*(s):=e^{sH_0}\psi_{\bx\s}^*e^{-sH_0}.$$
For $a\in \{0,1\}$ the operator $\psi_{\bx\s a}(s)$ is defined by
$$\psi_{\bx\s a}(s):=\left\{\begin{array}{ll} \psi_{\bx\s}^*(s)& \text{ if }a=1,\\ \psi_{\bx\s}(s) & \text{ if }a=0.\end{array}\right.$$

Next we define the ordering operators $T_1$, $T_2$.
\begin{definition}\label{def_T_1}
Consider linear operators $M(s_1),\cdots, M(s_n)$\\ $:F_f(L^2(\G\times \spin;\C))\to F_f(L^2(\G\times \spin;\C))$ parameterized by $s_1,\cdots,s_n\in \R$. Assume that $s_j\neq s_k$ for any $j,k\in \{1,\cdots,n\}$ with $j\neq k$. The operator $T_1(M(s_1)\cdots M(s_n))$ is defined by
$$T_1(M(s_1)\cdots M(s_n)):=M(s_{\pi(1)})\cdots M(s_{\pi(n)}),$$
where $\pi\in S_n$ is uniquely determined by the condition that
$$s_{\pi(1)}>s_{\pi(2)}>\cdots > s_{\pi(n)}.$$
\end{definition}

Let us define a relation `$\sim$' in the set $\{\psi_{\bx\s a}(s)\}_{(\bx,\s,a,s)\in \G\times\spin\times\{0,1\}\times \R}$ as follows.
$$\psi_{\bx\s a_1}(s_1)\sim \psi_{\by\tau a_2}(s_2)\text{ if }a_1=a_2\text{ and }s_1=s_2.$$
We see that `$\sim$' is an equivalence relation in $\{\psi_{\bx\s a}(s)\}_{(\bx,\s,a,s)\in \G\times\spin\times\{0,1\}\times \R}$. Let $[\psi_{\bx\s a}(s)]$ denote the equivalent class represented by an element $\psi_{\bx\s a}(s)$. We define relations `$\succ$' and `$\succeq$' in the quotient set $\{\psi_{\bx\s a}(s)\}_{(\bx,\s,a,s)\in \G\times\spin\times\{0,1\}\times \R}/\sim$ as follows.
\begin{equation*}
\begin{split}
&[\psi_{\bx\s a_1}(s_1)]\succ [\psi_{\by\tau a_2}(s_2)]\text{ if }s_1>s_2,\text{ or }s_1=s_2\text{ and }a_1>a_2,\\
&[\psi_{\bx\s a_1}(s_1)]\succeq [\psi_{\by\tau a_2}(s_2)]\text{ if }[\psi_{\bx\s a_1}(s_1)]\succ [\psi_{\by\tau a_2}(s_2)]\text{ or }[\psi_{\bx\s a_1}(s_1)]= [\psi_{\by\tau a_2}(s_2)].
\end{split}
\end{equation*}
The set $\{\psi_{\bx\s a}(s)\}_{(\bx,\s,a,s)\in \G\times\spin\times\{0,1\}\times \R}/\sim$ is totally ordered under the relation `$\succeq$' and the relation `$\succ$' is a strict order in this quotient set. 

\begin{definition}\label{def_T_2} 
For any $\psi_{\bx_j\s_j a_j}(s_j)$ $(j=1,\cdots,n)$ the operator \\
$T_2(\psi_{\bx_1\s_1 a_1}(s_1)\cdots\psi_{\bx_n\s_n a_n}(s_n))$ is defined by
\begin{equation*}
\begin{split}
&T_2(\psi_{\bx_1\s_1 a_1}(s_1)\cdots\psi_{\bx_n\s_n a_n}(s_n))\\
&\quad:=\sgn(\pi)\psi_{\bx_{\pi(1)}\s_{\pi(1)} a_{\pi(1)}}(s_{\pi(1)})\cdots\psi_{\bx_{\pi(n)}\s_{\pi(n)} a_{\pi(n)}}(s_{\pi(n)}),
\end{split}
\end{equation*}
where $\pi\in S_n$ is uniquely determined by the conditions that
$$[\psi_{\bx_{\pi(1)}\s_{\pi(1)} a_{\pi(1)}}(s_{\pi(1)})]\succeq [\psi_{\bx_{\pi(2)}\s_{\pi(2)} a_{\pi(2)}}(s_{\pi(2)})]\succeq \cdots \succeq [\psi_{\bx_{\pi(n)}\s_{\pi(n)} a_{\pi(n)}}(s_{\pi(n)})],$$
and if there exist $l_1,l_2\in \{1,\cdots,n\}$ with $l_1<l_2$ such that 
$$[\psi_{\bx_{l_1}\s_{l_1} a_{l_1}}(s_{l_1})]=[\psi_{\bx_{l_2}\s_{l_2} a_{l_2}}(s_{l_2})],\neq [\psi_{\bx_{j}\s_{j} a_{j}}(s_{j})]\ (\forall j\in \{l_1+1,l_2-1\})$$
and $\pi(m)=l_1$ with $m\in\{1,\cdots,n\}$, then $\pi(m+1)=l_2$.
\end{definition}
Using the ordering operator $T_1$ we have the following expansion. 
\begin{lemma}\label{lem_expansion_T_1}
For any $t_1,t_2\in\R$ with $t_1<t_2$,
\begin{equation*}
\begin{split}
&e^{-(t_2-t_1)(H_0+V_{\lambda})}\\
&\quad=e^{-(t_2-t_1)H_0}+e^{-t_2 H_0}\sum_{n=1}^{\infty}\frac{(-1)^n}{n!}\int_{[t_1,t_2]^n}ds_1\cdots ds_nT_1(V_{\lambda}(s_1)\cdots V_{\lambda}(s_n)) e^{t_1 H_0}.
\end{split}
\end{equation*}
\end{lemma}
\begin{remark}
Though the operator $T_1(V_{\lambda}(s_1)\cdots V_{\lambda}(s_n))$ is defined only for \\
$(s_1,\cdots,s_n)$ with $s_j\neq s_k$ $(j\neq k)$, we can consider $T_1(V_{\lambda}(s_1)\cdots V_{\lambda}(s_n))$ as a Bochner integrable function over $[t_1,t_2]^n$ since the Lebesgue measure of the set 
$\{(s_1,\cdots,s_n)\in [t_1,t_2]^n\ |\ \exists j,k\in \{1,\cdots,n\}\text{ s.t. }j\neq k\text{ and }s_j=s_k\}$
is zero.
\end{remark}
\begin{proof}[Proof of Lemma \ref{lem_expansion_T_1}]
Since the operator-valued function $\xi\mapsto e^{-(t_2-t_1)(H_0+\xi V_{\lambda})}$ is analytic, we have
\begin{equation}\label{eq_expansion_T_1_1}
e^{-(t_2-t_1)(H_0+V_{\lambda})}=\sum_{n=0}^{\infty}\frac{1}{n!}\left(\frac{d}{d\xi}\right)^ne^{-(t_2-t_1)(H_0+\xi V_{\lambda})}\Big|_{\xi =0}.
\end{equation}
It is sufficient to show that for all $n\in\N$ and all $\xi\in \R$
\begin{equation}\label{eq_expansion_T_1_2}
\begin{split}
&\left(\frac{d}{d\xi}\right)^ne^{-(t_2-t_1)(H_0+\xi V_{\lambda})}\\
&\quad=(-1)^n e^{-t_2 (H_0+\xi V_{\lambda})}\int_{[t_1,t_2]^n}ds_1\cdots ds_nT_1(V_{\lambda,\xi}(s_1)\cdots V_{\lambda,\xi}(s_n)) e^{t_1 (H_0+\xi V_{\lambda})},
\end{split}
\end{equation}
where $V_{\lambda,\xi}(s):=e^{s(H_0+\xi V_{\lambda})}V_{\lambda}e^{-s(H_0+\xi V_{\lambda})}$. In fact, substituting \eqref{eq_expansion_T_1_2} for $\xi =0$ into \eqref{eq_expansion_T_1_1} gives the result. We show \eqref{eq_expansion_T_1_2} by induction on $n$.

By Lemma \ref{lem_derivative_exponential} we have
\begin{equation*}
\begin{split}
\frac{d}{d\xi}e^{-(t_2-t_1)(H_0+\xi V_{\lambda})}&=\int_0^1ds e^{-(1-s)(t_2-t_1)(H_0+\xi V_{\lambda})}(t_1-t_2)V_{\lambda}e^{-s(t_2-t_1)(H_0+\xi V_{\lambda})}\\&=-e^{-t_2(H_0+\xi V_{\lambda})}\int_{t_1}^{t_2}ds V_{\lambda,\xi}(s)e^{t_1(H_0+\xi V_{\lambda})},
\end{split}
\end{equation*}
which is \eqref{eq_expansion_T_1_2} for $n=1$.

Let us assume that \eqref{eq_expansion_T_1_2} is true for $n-1$ $(n\ge 2)$. 
\begin{equation*}
\begin{split}
&\left(\frac{d}{d\xi}\right)^{n-1}e^{-(t_2-t_1)(H_0+\xi V_{\lambda})}\\
&\quad=(-1)^{n-1}(n-1)! e^{-t_2 (H_0+\xi V_{\lambda})}\int_{[t_1,t_2]^{n-1}}ds_1ds_2\cdots ds_{n-1}\\
&\qquad\cdot 1_{s_1>s_2>\cdots > s_{n-1}}V_{\lambda,\xi}(s_1)V_{\lambda,\xi}(s_2)\cdots V_{\lambda,\xi}(s_{n-1})e^{t_1(H_0+\xi V_{\lambda})}\\
&\quad=(-1)^{n-1}(n-1)!\int_{t_1}^{t_2}ds_1\int_{t_1}^{s_1}ds_2\cdots\int_{t_1}^{s_{n-2}}ds_{n-1} e^{-(t_2-s_1)(H_0+\xi V_{\lambda})}V_{\lambda}\\
&\qquad\cdot e^{-(s_1-s_2)(H_0+\xi V_{\lambda})}V_{\lambda}\cdots e^{-(s_{n-2}-s_{n-1})(H_0+\xi V_{\lambda})}V_{\lambda}e^{-(s_{n-1}-t_1)(H_0+\xi V_{\lambda})}.\end{split}
\end{equation*}
By writing $t_2=s_0$, $t_1=s_n$ and using Lemma \ref{lem_derivative_exponential} again we observe that
\begin{equation*}
\begin{split}
&\left(\frac{d}{d\xi}\right)^ne^{-(t_2-t_1)(H_0+\xi V_{\lambda})}\\
&\quad=(-1)^{n-1}(n-1)!\int_{s_n}^{s_0}ds_1\cdots\int_{s_n}^{s_{n-2}}ds_{n-1}\sum_{j=0}^{n-1}\\
&\qquad\cdot e^{-(s_0-s_1)(H_0+\xi V_{\lambda})}V_{\lambda}\cdots\frac{d}{d\xi}\left(e^{-(s_j-s_{j+1})(H_0+\xi V_{\lambda})}\right)V_{\lambda}\\&\qquad\cdots e^{-(s_{n-2}-s_{n-1})(H_0+\xi V_{\lambda})}V_{\lambda}e^{-(s_{n-1}-s_n)(H_0+\xi V_{\lambda})}
\end{split}
\end{equation*}
\begin{equation*}
\begin{split}
&\quad = (-1)^{n}(n-1)!\int_{s_n}^{s_0}ds_1\cdots\int_{s_n}^{s_{n-2}}ds_{n-1}\sum_{j=0}^{n-1}\\
&\qquad\cdot e^{-(s_0-s_1)(H_0+\xi V_{\lambda})}V_{\lambda} \cdots \int_{s_{j+1}}^{s_j}ds e^{-(s_j-s)(H_0+\xi V_{\lambda})}V_{\lambda}e^{-(s-s_{j+1})(H_0+\xi V_{\lambda})}V_{\lambda}\\
&\qquad \cdots e^{-(s_{n-2}-s_{n-1})(H_0+\xi V_{\lambda})}V_{\lambda}e^{-(s_{n-1}-s_n)(H_0+\xi V_{\lambda})}\\
&\quad = (-1)^{n}(n-1)!\sum_{j=0}^{n-1}\int_{s_n}^{s_0}ds_1\cdots\int_{s_n}^{s_{n-2}}ds_{n-1}\int_{s_{j+1}}^{s_j}ds1_{s_0>s_1>\cdots>s_j>s>s_{j+1}>\cdots>s_n}\\&\qquad\cdot e^{-(s_0-s_1)(H_0+\xi V_{\lambda})}V_{\lambda}\cdots e^{-(s_j-s)(H_0+\xi V_{\lambda})}V_{\lambda}e^{-(s-s_{j+1})(H_0+\xi V_{\lambda})}V_{\lambda}\\&\qquad \cdots e^{-(s_{n-2}-s_{n-1})(H_0+\xi V_{\lambda})}V_{\lambda}e^{-(s_{n-1}-s_n)(H_0+\xi V_{\lambda})}.
\end{split}
\end{equation*}
Then by changing the index of the variables $\{s_j,s\ |\ j=0,\cdots n\}$ we obtain
\begin{equation*}
\begin{split}
&\left(\frac{d}{d\xi}\right)^ne^{-(t_2-t_1)(H_0+\xi V_{\lambda})}\\
&\ =(-1)^{n}n!e^{-t_2(H_0+\xi V_{\lambda})}\int_{[t_1,t_2]^n}ds_1\cdots ds_n 1_{s_1>\cdots >s_n}V_{\lambda,\xi}(s_1)\cdots V_{\lambda,\xi}(s_n)e^{t_1(H_0+\xi V_{\lambda})}\\
&\ =(-1)^{n}e^{-t_2(H_0+\xi V_{\lambda})}\int_{[t_1,t_2]^n}ds_1\cdots ds_n T_1(V_{\lambda,\xi}(s_1)\cdots V_{\lambda,\xi}(s_n))e^{t_1(H_0+\xi V_{\lambda})},
\end{split}
\end{equation*}
which completes the proof.
\end{proof}

Next we prepare some properties of the operators
$\psi_{\bx\s}^*(s)$ and $\psi_{\bx\s}(s)$. Using the matrix $\{F(\bx\s,\by\tau)\}_{(\bx,\s),(\by,\tau)\in\G\times\spin}$ defined in \eqref{eq_matrix_F}, we define the matrices $F(a)$ $(a=0,1)$ by
$$F(a):=\left\{\begin{array}{ll}F &\text{ if }a=0,\\ -F^t &\text{ if }a=1.\end{array}\right.$$
Lemma \ref{lem_time_evolution_operator} and Lemma \ref{lem_formula_time_evolution_operator} below follow \cite[\mbox{Lemma 3.2.1, Lemma 3.2.2}]{L}. However, we give the proof to make this section self-contained.
\begin{lemma}\label{lem_time_evolution_operator}
The following equalities hold.
\begin{enumerate}[(i)]
\item\label{lem_time_evolution_property_1} For any $(\bx,\s,a,s)\in \G\times\spin\times\{0,1\}\times\R$
\begin{equation*}
\psi_{\bx\s a}(s)=\sum_{\by\in \G}\sum_{\tau\in \spin}e^{-sF(a)}(\bx\s,\by\tau)\psi_{\by\tau a}.
\end{equation*}
\item\label{lem_time_evolution_property_2} For any $(\bx,\s,s),(\by,\tau,t)\in \G\times\spin\times\R$
\begin{equation*}
\{\psi_{\bx\s}(s),\psi_{\by\tau}(t)\}=\{\psi_{\bx\s}^*(s),\psi_{\by\tau}^*(t)\}=0,\ \{\psi_{\bx\s}^*(s),\psi_{\by\tau}(t)\}=e^{(s-t)F}(\by\tau,\bx\s).
\end{equation*}
\end{enumerate}
\end{lemma}
\begin{proof}
We see that for $a\in \{0,1\}$
\begin{equation}\label{eq_time_operator_1}
\frac{d}{ds}\psi_{\bx\s a}(s)=e^{sH_0}(H_0\psi_{\bx\s a}-\psi_{\bx\s a}H_0)e^{-sH_0}.
\end{equation}
By using \eqref{eq_CAR} we can show that for $a\in \{0,1\}$
\begin{equation}\label{eq_time_operator_2}
H_0\psi_{\bx\s a}=-\sum_{\by\in
 \G}\sum_{\tau\in\spin}F(a)(\bx\s,\by\tau)\psi_{\by\tau a}+\psi_{\bx\s a}H_0.
\end{equation}
By combining \eqref{eq_time_operator_1} with \eqref{eq_time_operator_2}, we
 obtain a differential equation
\begin{equation*}
\frac{d}{ds}\psi_{\bx\s a}(s)=-\sum_{\by\in \G}\sum_{\tau\in
 \spin}F(a)(\bx\s,\by\tau)\psi_{\by\tau a}(s),
\end{equation*}
for $a\in \{0,1\}$, which gives \eqref{lem_time_evolution_property_1}.

By using \eqref{lem_time_evolution_property_1} and \eqref{eq_CAR}, the first equalities of \eqref{lem_time_evolution_property_2} can be proved. Moreover, we see that
\begin{equation*}
\begin{split}
\{\psi_{\bx\s}^*(s),\psi_{\by\tau}(t)\}&=\sum_{\bx_1,\bx_2\in
 \G}\sum_{\s_1,\s_2\in\spin}e^{sF}(\bx_1\s_1,\bx\s)e^{-tF}(\by\tau,\bx_2\s_2)\{\psi_{\bx_1\s_1}^*,\psi_{\bx_2\s_2}\}\\
&=\sum_{\bx_1\in\G}\sum_{\s_1\in\spin}e^{-tF}(\by\tau,\bx_1\s_1)e^{sF}(\bx_1\s_1,\bx\s)=e^{(s-t)F}(\by\tau,\bx\s).
\end{split}
\end{equation*}
\end{proof}

For any linear operator $A:F_f(L^2(\G\times\spin;\C))\to
F_f(L^2(\G\times\spin;\C))$, let $\<A\>_0$ denote $\Tr(e^{-\beta H_0}A)/\Tr
e^{-\beta H_0}$. For a set of the operators $\{\psi_{\bx_j\s_ja_j}(s_j)\}_{j=1}^n$,\\
let $\psi_{\bx_1\s_1a_1}(s_1)\cdots\widehat{\psi_{\bx_j\s_ja_j}(s_j)}\cdots\psi_{\bx_n\s_na_n}(s_n)$ denote the product obtained by eliminating $\psi_{\bx_j\s_ja_j}(s_j)$ from the product $\psi_{\bx_1\s_1a_1}(s_1)\cdots\psi_{\bx_n\s_na_n}(s_n)$. 
\begin{lemma}\label{lem_formula_time_evolution_operator}
If $n\in \N$ is odd, for any $(\bx_j,\s_j,a_j,s_j)\in \G\times\spin\times\{0,1\}\times\R$ $(j=1,\cdots,n)$
$$\<\psi_{\bx_1\s_1a_1}(s_1)\cdots\psi_{\bx_n\s_na_n}(s_n)\>_0=0.$$
If $n\in \N$ is even, for any $(\bx_j,\s_j,a_j,s_j)\in \G\times\spin\times\{0,1\}\times\R$ $(j=1,\cdots,n)$
\begin{equation}\label{eq_formula_time_evolution_operator_1}
\begin{split}
&\<\psi_{\bx_1\s_1a_1}(s_1)\cdots\psi_{\bx_n\s_na_n}(s_n)\>_0\\
&=\sum_{j=2}^n(-1)^j\<\psi_{\bx_1\s_1a_1}(s_1)\psi_{\bx_j\s_ja_j}(s_j)\>_0\<\psi_{\bx_2\s_2a_2}(s_2)\cdots\widehat{\psi_{\bx_{j}\s_{j}a_{j}}(s_{j})}
\cdots\psi_{\bx_n\s_na_n}(s_n)\>_0.
\end{split}
\end{equation}
Moreover,
\begin{equation}\label{eq_formula_time_evolution_operator_2}
\begin{split}
&\<\psi_{\bx_1\s_1a_1}(s_1)\psi_{\bx_2\s_2a_2}(s_2)\>_0\\
&\quad=\sum_{\by\in
 \G}\sum_{\tau\in \spin}\left(I+e^{-\beta
			 F(a_1)}\right)^{-1}(\bx_1\s_1,\by\tau)\{\psi_{\by\tau a_1}(s_1),\psi_{\bx_2\s_2 a_2}(s_2)\}.
\end{split}
\end{equation}
\end{lemma}
\begin{proof} 
By using the orthonormal basis $\bigcup_{m=0}^{2L^d}B_m$ defined in \eqref{eq_orthonormal_basis}, we can write
\begin{equation}\label{eq_formula_time_evolution_operator_3}
\begin{split}
&\Tr(e^{-\beta
 H_0}\psi_{\bx_1\s_1a_1}(s_1)\cdots\psi_{\bx_n\s_na_n}(s_n))\\
&\quad=\sum_{m=0}^{2L^d}\sum_{\phi\in
 B_m}\<\phi,e^{-\beta
 H_0}\psi_{\bx_1\s_1a_1}(s_1)\cdots\psi_{\bx_n\s_na_n}(s_n)\phi\>_{F_f}.
\end{split}
\end{equation}
Since for all $s\in \R$ and $m\in \{0,1,\cdots,2L^d\}$
$$
e^{sH_0}\left(A_m(L^2((\G\times\spin)^m;\C))\right)\subset A_m(L^2((\G\times\spin)^m;\C)),
$$
we see that if $n$ is odd, for any $m\in \{0,1,\cdots,2L^d\}$ and $\phi \in B_m$$$e^{-\beta H_0}\psi_{\bx_1\s_1a_1}(s_1)\cdots\psi_{\bx_n\s_na_n}(s_n)\phi=0,$$
or
$$
e^{-\beta
 H_0}\psi_{\bx_1\s_1a_1}(s_1)\cdots\psi_{\bx_n\s_na_n}(s_n)\phi\in A_l(L^2((\G\times\spin)^l;\C))
$$
with $l\neq m$, which implies that
\begin{equation}\label{eq_formula_time_evolution_operator_4}
\<\phi,e^{-\beta
 H_0}\psi_{\bx_1\s_1a_1}(s_1)\cdots\psi_{\bx_n\s_na_n}(s_n)\phi\>_{F_f}=0.
\end{equation}
The first statement follows from \eqref{eq_formula_time_evolution_operator_3} and \eqref{eq_formula_time_evolution_operator_4}.

Let us assume that $n$ is even. We see that
\begin{equation*}
\begin{split}
&\psi_{\bx_1\s_1a_1}(s_1)\psi_{\bx_2\s_2a_2}(s_2)\cdots\psi_{\bx_n\s_na_n}(s_n)\\
&\quad=\{\psi_{\bx_1\s_1a_1}(s_1),\psi_{\bx_2\s_2a_2}(s_2)\}\psi_{\bx_3\s_3a_3}(s_3)\cdots\psi_{\bx_n\s_na_n}(s_n)\\
&\qquad -\psi_{\bx_2\s_2a_2}(s_2)\psi_{\bx_1\s_1a_1}(s_1)\psi_{\bx_3\s_3a_3}(s_3)\cdots\psi_{\bx_n\s_na_n}(s_n)\\
&\quad =\{\psi_{\bx_1\s_1a_1}(s_1),\psi_{\bx_2\s_2a_2}(s_2)\}\psi_{\bx_3\s_3a_3}(s_3)\cdots\psi_{\bx_n\s_na_n}(s_n)\\
&\qquad -\psi_{\bx_2\s_2a_2}(s_2)\{\psi_{\bx_1\s_1a_1}(s_1),\psi_{\bx_3\s_3a_3}(s_3)\}\psi_{\bx_4\s_4a_4}(s_4)\cdots\psi_{\bx_n\s_na_n}(s_n)\\
&\qquad+\psi_{\bx_2\s_2a_2}(s_2)\psi_{\bx_3\s_3a_3}(s_3)\psi_{\bx_1\s_1a_1}(s_1)\psi_{\bx_4\s_4a_4}(s_4)\cdots\psi_{\bx_n\s_na_n}(s_n)\\
&\quad=\sum_{j=2}^n\{\psi_{\bx_1\s_1a_1}(s_1),\psi_{\bx_j\s_ja_j}(s_j)\}(-1)^j\psi_{\bx_2\s_2a_2}(s_2)\cdots\widehat{\psi_{\bx_{j}\s_{j}a_{j}}(s_{j})}\cdots\psi_{\bx_n\s_na_n}(s_n)\\
&\qquad -(-1)^n\psi_{\bx_2\s_2a_2}(s_2)\cdots\psi_{\bx_n\s_na_n}(s_n)\psi_{\bx_1\s_1a_1}(s_1),
\end{split}
\end{equation*}
which yields
\begin{equation}\label{eq_formula_time_evolution_operator_5}
\begin{split}
&\<\psi_{\bx_1\s_1a_1}(s_1)\psi_{\bx_2\s_2a_2}(s_2)\cdots\psi_{\bx_n\s_na_n}(s_n)\>_0\\
&\qquad\qquad\qquad\qquad+\<\psi_{\bx_2\s_2a_2}(s_2)\cdots\psi_{\bx_n\s_na_n}(s_n)\psi_{\bx_1\s_1a_1}(s_1)\>_0\\
&\quad=\sum_{j=2}^n\{\psi_{\bx_1\s_1a_1}(s_1),\psi_{\bx_j\s_ja_j}(s_j)\}\\
&\qquad\cdot (-1)^j\<\psi_{\bx_2\s_2a_2}(s_2)\cdots\widehat{\psi_{\bx_{j}\s_{j}a_{j}}(s_{j})}\cdots\psi_{\bx_n\s_na_n}(s_n)\>_0.
\end{split}
\end{equation}

On the other hand, by Lemma \ref{lem_time_evolution_operator} \eqref{lem_time_evolution_property_1},
\begin{equation}\label{eq_formula_time_evolution_operator_6}
\psi_{\bx\s a}(s+\beta)=\sum_{\by\in \G}\sum_{\tau\in \spin}e^{-\beta
 F(a)}(\bx\s,\by\tau)\psi_{\by\tau a}(s).
\end{equation}
By using \eqref{eq_formula_time_evolution_operator_6} and the equality that $\Tr(AB)=\Tr(BA)$ for any operators $A$, $B$, we observe that
\begin{equation}\label{eq_formula_time_evolution_operator_7}
\begin{split}
&\<\psi_{\bx_2\s_2a_2}(s_2)\cdots\psi_{\bx_n\s_na_n}(s_n)\psi_{\bx_1\s_1a_1}(s_1)\>_0 \\
&\quad=\<\psi_{\bx_1\s_1a_1}(s_1+\beta)\psi_{\bx_2\s_2a_2}(s_2)\cdots\psi_{\bx_n\s_na_n}(s_n)\>_0\\
&\quad =\sum_{\by\in \G}\sum_{\tau\in \spin}e^{-\beta
 F(a_1)}(\bx_1\s_1,\by\tau)\<\psi_{\by\tau
 a_1}(s_1)\psi_{\bx_2\s_2a_2}(s_2)\cdots\psi_{\bx_n\s_na_n}(s_n)\>_0.
\end{split}
\end{equation}

By substituting \eqref{eq_formula_time_evolution_operator_7} into \eqref{eq_formula_time_evolution_operator_5} we
obtain
\begin{equation}\label{eq_formula_time_evolution_operator_8}
\begin{split}
&\sum_{\by\in
 \G}\sum_{\tau\in\spin}(\delta_{\bx_1,\by}\delta_{\s_1,\tau}+e^{-\beta F(a_1)}(\bx_1\s_1,\by\tau))\\
&\qquad\cdot\<\psi_{\by\tau
 a_1}(s_1)\psi_{\bx_2\s_2a_2}(s_2)\cdots\psi_{\bx_n\s_na_n}(s_n)\>_0\\
& =\sum_{j=2}^n\{\psi_{\bx_1\s_1a_1(s_1)},\psi_{\bx_j\s_ja_j(s_j)}\}(-1)^j\<\psi_{\bx_2\s_2a_2}(s_2)\cdots\widehat{\psi_{\bx_{j}\s_{j}a_{j}}(s_{j})}\cdots\psi_{\bx_n\s_na_n}(s_n)\>_0.
\end{split}
\end{equation}

Let us define a unitary matrix $M=(M(\bk\tau,\bx\s))_{(\bk,\tau)\in
\G^*\times\spin, (\bx,\s)\in \G\times\spin}$ by
$$
M(\bk\tau,\bx\s):=\frac{\delta_{\s,\tau}}{L^{d/2}}e^{-i\<\bk,\bx\>}.
$$
Then we have for all $(\bk,\tau),(\hat{\bk},\hat{\tau})\in\G^*\times\spin$
\begin{equation}\label{eq_MFM}
MFM^*(\bk\tau,\hat{\bk}\hat{\tau})=\overline{M}F^tM^t(\bk\tau,\hat{\bk}\hat{\tau})=\delta_{\bk,\hat{\bk}}\delta_{\tau,\hat{\tau}}E_{\hat{\bk}},
\end{equation}
where $E_{\bk}$ is defined in \eqref{eq_dispersion_relation}.
The equality \eqref{eq_MFM} implies that
\begin{equation*}
\det(I+e^{-\beta F})=\det(I+e^{-\beta MFM^*})\neq 0,\ \det(I+e^{\beta F^t})=\det(I+e^{\beta \overline{M}F^tM^t})\neq 0.
\end{equation*}
Thus, for $a = 0,1$ the matrix $I+e^{-\beta F(a)}$ is invertible. The equality
\eqref{eq_formula_time_evolution_operator_8} leads to 
\begin{equation}\label{eq_formula_time_evolution_operator_9}
\begin{split}
&\<\psi_{\bx_1\s_1
 a_1}(s_1)\psi_{\bx_2\s_2a_2}(s_2)\cdots\psi_{\bx_n\s_na_n}(s_n)\>_0\\
&\quad = \sum_{j=2}^n\sum_{\by\in \G}\sum_{\tau\in\spin}\left(I+e^{-\beta
 F(a_1)}\right)^{-1}(\bx_1\s_1,\by\tau)\{\psi_{\by\tau
 a_1}(s_1),\psi_{\bx_j\s_ja_j}(s_j)\}(-1)^j\\
&\qquad\cdot\<\psi_{\bx_2\s_2a_2}(s_2)\cdots\widehat{\psi_{\bx_{j}\s_{j}a_{j}}(s_{j})}\cdots\psi_{\bx_n\s_na_n}(s_n)\>_0.
\end{split}
\end{equation}
The equality \eqref{eq_formula_time_evolution_operator_2} is \eqref{eq_formula_time_evolution_operator_9} for $n=2$. Then, by substituting \eqref{eq_formula_time_evolution_operator_2} into
 \eqref{eq_formula_time_evolution_operator_9} we obtain \eqref{eq_formula_time_evolution_operator_1}.
\end{proof}

From now we show some lemmas involving the ordering operator $T_2$. To simplify notations, let $\psi_j$ denote $\psi_{\bx_j \s_{j} a_{j}}(s_{j})$ for fixed variables $(\bx_j,\s_j,a_j,s_j)\in\G\times\spin\times\{0,1\}\times\R$ $(j=1,\cdots,n)$.
\begin{lemma}\label{lem_permutation_T_2}
For any $\pi\in S_n$,
\begin{equation}\label{eq_permutation_T_2_1}
T_2(\psi_1\psi_2\cdots\psi_n)=\sgn(\pi)T_2(\psi_{\pi(1)}\psi_{\pi(2)}\cdots\psi_{\pi(n)}).
\end{equation}
\end{lemma}
\begin{proof}
It is sufficient to show \eqref{eq_permutation_T_2_1} for any transposition $\pi$ as any permutation is a product of transpositions. Let us assume that $\pi=(j,k)$, $1\le j<k \le n$. Let $\tau,\eta\in S_n$ be the unique permutations associated with the definitions of $T_2(\psi_1\cdots\psi_n)$ and $T_2(\psi_{\pi(1)}\cdots\psi_{\pi(n)})$, respectively.
\begin{align}
&T_2(\psi_1\cdots\psi_n)=\sgn(\tau)\psi_{\tau(1)}\cdots\psi_{\tau(n)},\label{eq_permutation_T_2_2}\\
&T_2(\psi_{\pi(1)}\cdots\psi_{\pi(n)})=\sgn(\eta)\psi_{\pi(\eta(1))}\cdots\psi_{\pi(\eta(n))}.\label{eq_permutation_T_2_3}
\end{align}

First consider the case that $[\psi_j]\neq[\psi_k]$. Let $A,B\subset \{j+1,\cdots,k-1\}$ satisfy that $[\psi_j]=[\psi_{\alpha}]$ for any $\alpha \in A$, $[\psi_k]=[\psi_{\gamma}]$ for any $\gamma \in B$ and $[\psi_j], [\psi_k]\neq [\psi_p]$ for any $p\in \{j+1,\cdots,k-1\}\backslash A\cup B$. 

If $A,B\neq \emptyset$, we can write $A=\{\alpha_1,\cdots,\alpha_l\}$, $B=\{\gamma_1,\cdots,\gamma_m\}$ with $j+1\le \alpha_1 <\cdots <\alpha_l \le k-1$, $j+1\le \gamma_1 <\cdots <\gamma_m \le k-1$. By the definition of $T_2$ the product $\psi_{\pi(\eta(1))}\cdots\psi_{\pi(\eta(n))}$ is obtained by replacing $\psi_j\psi_{\alpha_1}\cdots\psi_{\alpha_l}$ and $\psi_{\gamma_1}\cdots\psi_{\gamma_m}\psi_k$ by $\psi_{\alpha_1}\cdots\psi_{\alpha_l}\psi_j$ and $\psi_k\psi_{\gamma_1}\cdots\psi_{\gamma_m}$ respectively in the product $\psi_{\tau(1)}\cdots\psi_{\tau(n)}$. Thus, if we define cycles $\zeta_1,\zeta_2\in S_n$ by 
\begin{equation*}
\zeta_1=\left(\begin{array}{ccccc}j & \alpha_1 &\cdots & \alpha_{l-1} & \alpha_l\\ \alpha_1 & \alpha_2 &\cdots & \alpha_{l} & j \end{array}\right),\ \zeta_2=\left(\begin{array}{ccccc}\gamma_1 & \gamma_2 &\cdots & \gamma_{m} & k \\ k & \gamma_1 &\cdots & \gamma_{m-1} & \gamma_m \end{array}\right),
\end{equation*}       
the permutation $\eta$ is written as 
\begin{equation}\label{eq_permutation_T_2_4}
\eta = \pi^{-1}\zeta_1\zeta_2\tau.
\end{equation}

On the other hand, Lemma \ref{lem_time_evolution_operator} (ii) ensures that
\begin{equation}\label{eq_permutation_T_2_5}
\psi_{\alpha_1}\cdots\psi_{\alpha_l}\psi_j=(-1)^l\psi_j\psi_{\alpha_1}\cdots\psi_{\alpha_l},\ \psi_k\psi_{\gamma_1}\cdots\psi_{\gamma_m}=(-1)^m\psi_{\gamma_1}\cdots\psi_{\gamma_m}\psi_k.
\end{equation}
By \eqref{eq_permutation_T_2_2}-\eqref{eq_permutation_T_2_5} we see that
\begin{equation}\label{eq_permutation_T_2_6}
\begin{split}
T_2(\psi_{\pi(1)}\cdots\psi_{\pi(n)})&= \sgn(\pi^{-1}\zeta_1\zeta_2\tau)\psi_{\zeta_1(\zeta_2(\tau(1)))}\cdots\psi_{\zeta_1(\zeta_2(\tau(n)))}\\
&= (-1)^{1+l+m}\sgn(\tau)(-1)^{l+m}\psi_{\tau(1)}\cdots\psi_{\tau(n)}\\
&= -T_2(\psi_1\cdots\psi_n).
\end{split}
\end{equation}

If $A=\emptyset$ or $B=\emptyset$, by setting $\zeta_1 = \text{Id}$ and $l=0$ or $\zeta_2 = \text{Id}$ and $m=0$, respectively, we see that the equalities \eqref{eq_permutation_T_2_4} and \eqref{eq_permutation_T_2_6} hold true. 

Next consider the case that $[\psi_j]=[\psi_k]$. Let $\tilde{A}\subset\{j+1,\cdots,k-1\}$ be such that $[\psi_j]=[\psi_q]$ for any $q\in \tilde{A}$ and $[\psi_j]\neq [\psi_q]$ for any $q\in \{j+1,\cdots,k-1\}\backslash\tilde{A}$.

If $\tilde{A}\neq\emptyset$, we write $\tilde{A}$ as $\tilde{A}=\{q_1,\cdots,q_r\}$ with $j+1\le q_1 < \cdots < q_r\le k-1$. By the definition of $T_2$ the product $\psi_{\pi(\eta(1))}\cdots\psi_{\pi(\eta(n))}$ is obtained by replacing $\psi_{j}\psi_{q_1}\cdots\psi_{q_r}\psi_k$ by $\psi_{k}\psi_{q_1}\cdots\psi_{q_r}\psi_j$ in the product $\psi_{\tau(1)}\cdots\psi_{\tau(n)}$. Thus, the permutation $\eta$ satisfies the equality 
\begin{equation}\label{eq_permutation_T_2_7}
\eta = \tau.
\end{equation}
By Lemma \ref{lem_time_evolution_operator} (ii) the following equality holds.
\begin{equation}\label{eq_permutation_T_2_8}
\psi_{k}\psi_{q_1}\cdots\psi_{q_r}\psi_j = - \psi_{j}\psi_{q_1}\cdots\psi_{q_r}\psi_k.
\end{equation}
By combining \eqref{eq_permutation_T_2_2}-\eqref{eq_permutation_T_2_3} with \eqref{eq_permutation_T_2_7}-\eqref{eq_permutation_T_2_8} we have 
\begin{equation}\label{eq_permutation_T_2_9}
\begin{split}
T_2(\psi_{\pi(1)}\cdots\psi_{\pi(n)})&= \sgn(\tau)\psi_{\pi(\tau(1))}\cdots\psi_{\pi(\tau(n))}\\
&= - \sgn(\tau)\psi_{\tau(1)}\cdots\psi_{\tau(n)}\\
&= -T_2(\psi_1\cdots\psi_n).
\end{split}
\end{equation} 

By repeating the same argument as above without the term $\psi_{q_1}\cdots\psi_{q_r}$ we can prove the equalities \eqref{eq_permutation_T_2_9} for the case that $\tilde{A}=\emptyset$, which completes the proof. 
\end{proof}
 
\begin{lemma}\label{lem_formula_T_2}
Assume that $n\in \N$ is even and $[\psi_1]\succeq [\psi_j]$ $(\forall j\in\{2,3,\cdots,n\})$. The following equality holds.
\begin{equation}\label{eq_formula_T_2_1}
\<T_2(\psi_1\cdots\psi_n)\>_0=\sum_{j=2}^n(-1)^j\<T_2(\psi_1\psi_j)\>_0\<T_2(\psi_2\cdots\widehat{\psi_j}\cdots\psi_n)\>_0.
\end{equation}
\end{lemma}
\begin{proof}
For $n=2$ the equality \eqref{eq_formula_T_2_1} is trivial. Assume that $n\ge 4$. Let $\tau\in S_n$ be the unique permutation associated with the definition of $T_2(\psi_1\cdots\psi_n)$.
$$T_2(\psi_1\cdots\psi_n)=\sgn(\tau)\psi_{\tau(1)}\cdots\psi_{\tau(n)}.$$
By assumption $\tau(1)=1$. Moreover, Lemma \ref{lem_formula_time_evolution_operator} ensures that
\begin{equation}\label{eq_formula_T_2_2}
\<T_2(\psi_1\cdots\psi_n)\>_0=\sgn(\tau)\sum_{j=2}^n(-1)^j\<\psi_1\psi_{\tau(j)}\>_0\<\psi_{\tau(2)}\cdots\widehat{\psi_{\tau(j)}}\cdots\psi_{\tau(n)}\>_0.
\end{equation}

Let us fix $j\in \{2,3,\cdots,n\}$. Let $\pi\in S_n$ be such that
\begin{equation}\label{eq_formula_T_2_3}
(\pi(1),\pi(2),\pi(3),\cdots,\pi(n))=(1,\tau(j),\tau(2),\cdots,\widehat{\tau(j)},\cdots,\tau(n)),
\end{equation}
where `$\widehat{\tau(j)}$' stands for the omission of the number $\tau(j)$ from the row \\
$(\tau(2),\tau(3),\cdots,\tau(n))$. Then, we have
\begin{equation}\label{eq_formula_T_2_4}
\sgn(\pi)=(-1)^{j-2}\sgn(\tau)=(-1)^j\sgn(\tau).
\end{equation}

On the other hand, we can write $\{1,\cdots,n\}\backslash\{1,\tau(j)\}=\{l_1,\cdots,l_{n-2}\}$ with $2\le l_1<l_2<\cdots<l_{n-2}\le n$. There exists $\eta\in S_{n-2}$ such that
\begin{equation}\label{eq_formula_T_2_5}
(l_{\eta(1)},l_{\eta(2)},\cdots,l_{\eta(n-2)})=(\tau(2),\tau(3),\cdots,\widehat{\tau(j)},\cdots,\tau(n)).
\end{equation}
By \eqref{eq_formula_T_2_3} and \eqref{eq_formula_T_2_5} we obtain
$$(\pi(1),\pi(2),\pi(3),\cdots,\pi(n))=(1,\tau(j),l_{\eta(1)},l_{\eta(2)},\cdots,l_{\eta(n-2)}),$$
which implies that
\begin{equation}\label{eq_formula_T_2_6}
\sgn(\pi)=(-1)^{\tau(j)-2}\sgn(\eta)=(-1)^{\tau(j)}\sgn(\eta).
\end{equation}
By \eqref{eq_formula_T_2_4} and \eqref{eq_formula_T_2_6} we have
\begin{equation}\label{eq_formula_T_2_7}
(-1)^j\sgn(\tau)=(-1)^{\tau(j)}\sgn(\eta).
\end{equation}
Note the equalities that
\begin{equation}\label{eq_formula_T_2_8}
\begin{split}
&\<T_2(\psi_1\psi_{\tau(j)})\>_0=\<\psi_1\psi_{\tau(j)}\>_0,\\
&\<T_2(\psi_2\cdots\widehat{\psi_{\tau(j)}}\cdots\psi_n)\>_0=\sgn(\eta)\<\psi_{\tau(2)}\cdots\widehat{\psi_{\tau(j)}}\cdots\psi_{\tau(n)}\>_0.
\end{split}
\end{equation}
By substituting \eqref{eq_formula_T_2_7} and  \eqref{eq_formula_T_2_8} into \eqref{eq_formula_T_2_2} we see that
\begin{equation*}
\begin{split}
\<T_2(\psi_1\cdots\psi_n)\>_0&=\sum_{j=2}^n(-1)^{\tau(j)}\<T_2(\psi_1\psi_{\tau(j)})\>_0\<T_2(\psi_2\cdots\widehat{\psi_{\tau(j)}}\cdots\psi_n)\>_0\\
&=\sum_{j=2}^n(-1)^{j}\<T_2(\psi_1\psi_{j})\>_0\<T_2(\psi_2\cdots\widehat{\psi_{j}}\cdots\psi_n)\>_0,
\end{split}
\end{equation*}
which is \eqref{eq_formula_T_2_1}.
\end{proof}

\begin{lemma}\label{lem_determinant_T_2}
For all $\bx_j,\by_j\in \G$, $\s_j,\tau_j\in \spin$, $s_j,t_j\in \R$
 $(j=1,2,\cdots,n)$,
\begin{equation}\label{eq_determinant_T_2_1}
\begin{split}
&\<T_2(\psi_{\bx_1\s_1}^*(s_1)\psi_{\by_1\tau_1}(t_1)\cdots\psi_{\bx_n\s_n}^*(s_n)\psi_{\by_n\tau_n}(t_n))\>_0\\
&\qquad=\det(\<T_2(\psi_{\bx_j\s_j}^*(s_j)\psi_{\by_k\tau_k}(t_k))\>_0)_{1\le
 j,k\le n}.
\end{split}
\end{equation}
\end{lemma}

\begin{proof}
We show \eqref{eq_determinant_T_2_1} by induction on $n$.
The equality \eqref{eq_determinant_T_2_1} is obviously true when
 $n=1$. Let us assume that \eqref{eq_determinant_T_2_1} is true for
 $n-1$ $(n\ge 2)$.

Lemma \ref{lem_permutation_T_2} implies that for all $\pi\in S_n$
\begin{equation}\label{eq_determinant_T_2_2}
\begin{split}
&\<T_2(\psi_{\bx_1\s_1}^*(s_1)\psi_{\by_1\tau_1}(t_1)\cdots\psi_{\bx_n\s_n}^*(s_n)\psi_{\by_n\tau_n}(t_n))\>_0\\
&\quad =\<T_2(\psi_{\bx_{\pi(1)}\s_{\pi(1)}}^*(s_{\pi(1)})\psi_{\by_{\pi(1)}\tau_{\pi(1)}}(t_{\pi(1)})\cdots\psi_{\bx_{\pi(n)}\s_{\pi(n)}}^*(s_{\pi(n)})\psi_{\by_{\pi(n)}\tau_{\pi(n)}}(t_{\pi(n)}))\>_0\\
&\quad =(-1)^n\<T_2(\psi_{\by_{\pi(1)}\tau_{\pi(1)}}(t_{\pi(1)})\psi_{\bx_{\pi(1)}\s_{\pi(1)}}^*(s_{\pi(1)})\\
&\qquad\qquad\qquad\quad\cdots\psi_{\by_{\pi(n)}\tau_{\pi(n)}}(t_{\pi(n)})\psi_{\bx_{\pi(n)}\s_{\pi(n)}}^*(s_{\pi(n)}))\>_0,
\end{split}
\end{equation}
and 
\begin{equation}\label{eq_determinant_T_2_3}
\begin{split}
&\det(\<T_2(\psi_{\bx_j\s_j}^*(s_j)\psi_{\by_k\tau_k}(t_k))\>_0)_{1\le
 j,k\le n}\\
&\quad=\det(\<T_2(\psi_{\bx_{\pi(j)}\s_{\pi(j)}}^*(s_{\pi(j)})\psi_{\by_{\pi(k)}\tau_{\pi(k)}}(t_{\pi(k)}))\>_0)_{1\le
 j,k\le n}\\
&\quad=(-1)^n\det(\<T_2(\psi_{\by_{\pi(k)}\tau_{\pi(k)}}(t_{\pi(k)})\psi_{\bx_{\pi(j)}\s_{\pi(j)}}^*(s_{\pi(j)}))\>_0)_{1\le
 j,k\le n}.
\end{split}
\end{equation}
The equalities \eqref{eq_determinant_T_2_2} and
 \eqref{eq_determinant_T_2_3} enable us to assume that\\
 $[\psi_{\bx_1\s_1}^*(s_1)]\succeq [\psi_{\bx_j\s_j}^*(s_j)],[\psi_{\by_j\tau_j}(t_j)]$ $(\forall j\in\{1,\cdots,n\})$ without losing generality in the following argument.

 By using Lemma \ref{lem_permutation_T_2}, Lemma
 \ref{lem_formula_T_2}, the hypothesis of induction, and the fact
 that $\<\psi_{\bx\s}^*(t)\psi_{\bx'\s'}^*(t')\>_0=0$, we have
\begin{equation*}
\begin{split}
&\<T_2(\psi_{\bx_1\s_1}^*(s_1)\psi_{\by_1\tau_1}(t_1)\cdots\psi_{\bx_n\s_n}^*(s_n)\psi_{\by_n\tau_n}(t_n))\>_0\\
& =\sum_{j=1}^n\<T_2(\psi_{\bx_1\s_1}^*(s_1)\psi_{\by_j\tau_j}(t_j))\>_0\\
&\ \cdot\<T_2(\psi_{\by_1\tau_1}(t_1)\psi_{\bx_2\s_2}^*(s_2)\psi_{\by_2\tau_2}(t_2)\cdots\psi_{\bx_j\s_j}^*(s_j)\widehat{\psi_{\by_{j}\tau_{j}}(t_{j})}\cdots\psi_{\bx_n\s_n}^*(s_n)\psi_{\by_n\tau_n}(t_n))\>_0\\
&=\sum_{j=1}^n(-1)^{j-1}\<T_2(\psi_{\bx_1\s_1}^*(s_1)\psi_{\by_j\tau_j}(t_j))\>_0
\det(\<T_2(\psi_{\bx_l\s_l}^*(s_l)\psi_{\by_k\tau_k}(t_k))\>_0)_{1\le
 l,k\le n\atop l\neq 1, k\neq j}\\
&=\det(\<T_2(\psi_{\bx_j\s_j}^*(s_j)\psi_{\by_k\tau_k}(t_k))\>_0)_{1\le
 j,k\le n},
\end{split}
\end{equation*}
which concludes the proof.
\end{proof}

\begin{lemma}\label{lem_derivation_covariance}
For all $\bx,\by\in\G$, $\s,\tau\in\spin$, $x,y\in \R$
\begin{equation}\label{eq_derivation_covariance_1}
\<T_2(\psi_{\bx\s}^*(x)\psi_{\by\tau}(y))\>_0=C(\bx\s x,\by\tau
 y),
\end{equation}
where $C(\bx\s x,\by\tau y)$ is defined in \eqref{eq_covariance_matrix}.
\end{lemma}

\begin{proof}
By the definition of $T_2$, we have
\begin{equation}\label{eq_derivation_covariance_2}
\<T_2(\psi_{\bx\s}^*(x)\psi_{\by\tau}(y))\>_0=\<\psi_{\bx\s}^*(x)\psi_{\by\tau}(y)\>_01_{x-y\ge
 0}-\<\psi_{\by\tau}(y)\psi_{\bx\s}^*(x)\>_01_{x-y< 0}.
\end{equation}
By Lemma \ref{lem_time_evolution_operator} \eqref{lem_time_evolution_property_2}, \eqref{eq_formula_time_evolution_operator_2}
and \eqref{eq_MFM}, we see that
\begin{equation}\label{eq_derivation_covariance_3}
\begin{split}
&\<\psi_{\bx\s}^*(x)\psi_{\by\tau}(y)\>_0\\
&\quad=\sum_{\bx'\in
 \G}\sum_{\s'\in \spin}\left(I+e^{\beta
 F^t}\right)^{-1}(\bx\s,\bx'\s')\{\psi_{\bx'\s'}^*(x),\psi_{\by\tau}(y)\}\\
&\quad=\sum_{\bx'\in
 \G}\sum_{\s'\in \spin}\left(I+e^{\beta
 F^t}\right)^{-1}(\bx\s,\bx'\s')e^{(x-y)F}(\by\tau,\bx'\s')\\
&\quad=\left(\left(I+e^{\beta
 F^t}\right)^{-1}e^{(x-y)F^t}\right)(\bx\s,\by\tau)\\
&\quad=\left(M^t\left(I+e^{\beta \overline{M}
 F^tM^t}\right)^{-1}e^{(x-y)\overline{M}F^tM^t}\overline{M}\right)(\bx\s,\by\tau)\\
&\quad=\frac{\delta_{\s,\tau}}{L^d}\sum_{\bk,\hat{\bk}\in\G^*}\delta_{\bk,\hat{\bk}}e^{-i\<\bx,\bk\>}e^{i\<\by,\hat{\bk}\>}\frac{e^{(x-y)E_{\hat{\bk}}}}{1+e^{\beta
 E_{\hat{\bk}}}}=\frac{\delta_{\s,\tau}}{L^d}\sum_{\bk\in\G^*}e^{i\<\bk,\by-\bx\>}\frac{e^{-(y-x)E_{\bk}}}{1+e^{\beta
 E_{\bk}}},
\end{split}
\end{equation}
\begin{equation}\label{eq_derivation_covariance_4}
\begin{split}
&\<\psi_{\by\tau}(y)\psi_{\bx\s}^*(x)\>_0\\
&\quad=\sum_{\by'\in
 \G}\sum_{\tau'\in \spin}\left(I+e^{-\beta
 F}\right)^{-1}(\by\tau,\by'\tau')\{\psi_{\by'\tau'}(y),\psi_{\bx\s}^*(x)\}\\
&\quad=\sum_{\by'\in
 \G}\sum_{\tau'\in \spin}\left(I+e^{-\beta
 F}\right)^{-1}(\by\tau,\by'\tau')e^{(x-y)F}(\by'\tau',\bx\s)\\
&\quad=\left(\left(I+e^{-\beta F}\right)^{-1}e^{(x-y)F}\right)(\by\tau,\bx\s)\\
&\quad=\left(M^*\left(I+e^{-\beta M
 FM^*}\right)^{-1}e^{(x-y)MFM^*}M\right)(\by\tau,\bx\s)\\
&\quad=\frac{\delta_{\s,\tau}}{L^d}\sum_{\bk,\hat{\bk}\in\G^*}\delta_{\bk,\hat{\bk}}e^{i\<\by,\bk\>}e^{-i\<\bx,\hat{\bk}\>}\frac{e^{(x-y)E_{\hat{\bk}}}}{1+e^{-\beta
 E_{\hat{\bk}}}}=\frac{\delta_{\s,\tau}}{L^d}\sum_{\bk\in\G^*}e^{i\<\bk,\by-\bx\>}\frac{e^{-(y-x)E_{\bk}}}{1+e^{-\beta E_{\bk}}}.
\end{split}
\end{equation}
By combining \eqref{eq_derivation_covariance_3} and \eqref{eq_derivation_covariance_4} with
 \eqref{eq_derivation_covariance_2}, we obtain \eqref{eq_derivation_covariance_1}.
\end{proof}

We have prepared all the lemmas necessary to prove Proposition \ref{pro_perturbation_series}.
\begin{proof}[Proof of Proposition \ref{pro_perturbation_series}]
By applying Lemma \ref{lem_expansion_T_1} for $t_1=0$, $t_2=\beta$ we have
\begin{equation}\label{eq_proof_perturbation_series_1}
\begin{split}
e^{-\beta H_{\lambda}}&=e^{-\beta H_0}+e^{-\beta H_0}\sum_{n=1}^{\infty}\frac{(-1)^n}{n!}\int_{[0,\beta]^n}ds_1\cdots ds_nT_1(V_{\lambda}(s_1)\cdots V_{\lambda}(s_n))\\
&=e^{-\beta H_0}+e^{-\beta H_0}\sum_{n=1}^{\infty}(-1)^n\int_{[0,\beta]^n}ds_1\cdots ds_n1_{s_1>\cdots>s_n}V_{\lambda}(s_1)\cdots V_{\lambda}(s_n).
\end{split}
\end{equation}
By \eqref{eq_proof_perturbation_series_1}, Lemma \ref{lem_time_evolution_operator} \eqref{lem_time_evolution_property_2}, the definition of $T_2$ and Lemma \ref{lem_permutation_T_2}, we see that
\begin{equation*}
\begin{split}
&\frac{\Tr e^{-\beta H_{\lambda}}}{\Tr e^{-\beta H_0}}\\
&=1+\sum_{n=1}^{\infty}\opPi_{j=1}^n\left(-\sum_{\bx_j,\by_j,\bz_j,\bw_j\in\G}\int_0^{\beta}ds_jU_{\bx_j,\by_j,\bz_j,\bw_j}\right)1_{s_1>\cdots>s_n}\\
&\qquad\cdot \<\psi_{\bx_1\ua}^*(s_1)\psi_{\by_1\da}^*(s_1)\psi_{\bw_1\da}(s_1)\psi_{\bz_1\ua}(s_1)\cdots \psi_{\bx_n\ua}^*(s_n)\psi_{\by_n\da}^*(s_n)\psi_{\bw_n\da}(s_n)\psi_{\bz_n\ua}(s_n)\>_0\\
&=1+\sum_{n=1}^{\infty}\opPi_{j=1}^n\left(-\sum_{\bx_{2j-1},\bx_{2j},\by_{2j-1},\by_{2j}\in\G}\int_0^{\beta}dx_{2j-1}U_{\bx_{2j-1},\bx_{2j},\by_{2j-1},\by_{2j}}\right)\\
&\qquad\cdot 1_{x_{1}>x_{3}>\cdots>x_{2n-1}}(-1)^n \<\psi_{\bx_1\ua}^*(x_{1})\psi_{\bx_2\da}^*(x_{1})\psi_{\by_1\ua}(x_{1})\psi_{\by_2\da}(x_{1})\cdots\\
&\qquad\qquad\qquad\cdot \psi_{\bx_{2n-1}\ua}^*(x_{2n-1})\psi_{\bx_{2n}\da}^*(x_{2n-1})\psi_{\by_{2n-1}\ua}(x_{2n-1})\psi_{\by_{2n}\da}(x_{2n-1})\>_0
\end{split}
\end{equation*}
\begin{equation*}
\begin{split}
&=1+\sum_{n=1}^{\infty}\opPi_{j=1}^n\left(-\sum_{\bx_{2j-1},\bx_{2j},\by_{2j-1},\by_{2j}\in\G}\int_0^{\beta}dx_{2j-1}U_{\bx_{2j-1},\bx_{2j},\by_{2j-1},,\by_{2j}}\right)\\
&\qquad\cdot 1_{x_{1}>x_{3}>\cdots>x_{2n-1}}\<T_2(\psi_{\bx_1\ua}^*(x_{1})\psi_{\by_1\ua}(x_{1})\psi_{\bx_2\da}^*(x_{1})\psi_{\by_2\da}(x_{1})\cdots\\
&\qquad\qquad\cdot\psi_{\bx_{2n-1}\ua}^*(x_{2n-1})\psi_{\by_{2n-1}\ua}(x_{2n-1})\psi_{\bx_{2n}\da}^*(x_{2n-1})\psi_{\by_{2n}\da}(x_{2n-1}))\>_0\\
&=1+\sum_{n=1}^{\infty}\frac{1}{n!}\opPi_{j=1}^n\\
&\left(-\sum_{\bx_{2j-1},\bx_{2j},\by_{2j-1},\by_{2j}\in\G}\sum_{\s_{2j-1},\s_{2j}\in\spin}\int_0^{\beta}dx_{2j-1}\delta_{\s_{2j-1},\ua}\delta_{\s_{2j},\da}U_{\bx_{2j-1},\bx_{2j},\by_{2j-1},,\by_{2j}}\right)\\
&\cdot\<T_2(\psi_{\bx_1\s_1}^*(x_{1})\psi_{\by_1\s_1}(x_{1})\psi_{\bx_2\s_2}^*(x_{2})\psi_{\by_2\s_2}(x_{2})\cdots\\
&\quad\cdot\psi_{\bx_{2n-1}\s_{2n-1}}^*(x_{2n-1})\psi_{\by_{2n-1}\s_{2n-1}}(x_{2n-1})\psi_{\bx_{2n}\s_{2n}}^*(x_{2n})\psi_{\by_{2n}\s_{2n}}(x_{2n}))\>_0\Big|_{x_{2j}=x_{2j-1}\atop \forall j\in\{1,\cdots,n\}}.
\end{split}
\end{equation*}
Then by using Lemma \ref{lem_determinant_T_2} and Lemma \ref{lem_derivation_covariance} we obtain the series \eqref{eq_perturbation_series}.
\end{proof}

\section{Diagonalization of the covariance matrix}\label{appendix_covariance_matrix}
In this part of Appendices we diagonalize the covariance matrix \\
$(C_h(\bx\s x,\by\tau y))_{(\bx,\s, x),(\by, \tau, y)\in
\G\times\spin \times [0,\beta)_h}$ and calculate its determinant. The fact that the determinant of the covariance matrix is non-zero, which is to be proved in Proposition \ref{pro_determinant_covariance}, verifies the well-posedness of the Grassmann Gaussian integral defined in Definition \ref{defn_grassmann_gaussian_integral}.

For convenience of calculation we assume that $h\in 2\N/\beta$. Define the sets $W_h$ and $M_h$ by
\begin{equation*}
W_h := \left\{\o \in \frac{\pi}{\beta}\Z\ \Big|\ -\pi h\le \o < \pi h
 \right\},\ M_h := \left\{\o \in \frac{\pi}{\beta}(2\Z+1)\ \Big|\ -\pi h< \o < \pi h
 \right\}.
\end{equation*}
Note that $\sharp W_h=2\beta h$ and $\sharp M_h=\beta h$.
The assumption that $h\in 2\N/\beta$ ensures the equality 
\begin{equation}\label{eq_matsubara}
M_h = W_h\backslash \frac{2\pi \Z}{\beta}.
\end{equation}
The set $M_h$ is seen as a set of the Matsubara frequencies with cut-off.

For $f\in L^2([-\beta,\beta)_h;\C)$ we define $\hat{f}\in L^2(W_h;\C)$
by
$$\hat{f}(\o):=\frac{1}{h}\sum_{t \in [-\beta,\beta)_h}e^{-i\o t}f(t).$$

\begin{lemma}\label{lem_fourier_inversion}
For any $f\in L^2([-\beta,\beta)_h;\C)$
$$f(t)=\frac{1}{2\beta}\sum_{\o\in W_h}e^{i\o t}\hat{f}(\o),\ \forall t\in [-\beta,\beta)_h.$$
\end{lemma}
\begin{proof}
If $t = -\beta +s/h$ with $s\in \{0,\cdots,2\beta h-1\}$,
\begin{equation*}
\begin{split}
\frac{1}{2\beta}\sum_{\o\in W_h}e^{i\o t}\hat{f}(\o)&=\frac{1}{2\beta
 h}\sum_{\o\in W_h}\sum_{u \in [-\beta,\beta)_h}e^{i\o t}e^{-i\o
 u}f(u)\\
&= \frac{1}{2\beta h}\sum_{m= 0}^{2\beta h-1}\sum_{l=0}^{2\beta h
 -1}e^{i(-\pi h+\pi m/\beta)(s/h-l/h)}f\left(-\beta +\frac{l}{h}\right)\\
&= \frac{1}{2\beta h}\sum_{l=0}^{2\beta h
 -1}e^{-i\pi(s-l)}\sum_{m=0}^{2\beta h-1}
e^{i\pi m(s-l)/(\beta h)}f\left(-\beta +\frac{l}{h}\right)\\
&= \sum_{l=0}^{2\beta h -1}e^{-i\pi(s-l)}\delta_{s,l}f\left(-\beta +\frac{l}{h}\right)=f\left(-\beta +\frac{s}{h} \right)=f(t).
\end{split}
\end{equation*}
\end{proof}

\begin{lemma}\label{lem_fourier_inversion_matsubara}
If $f\in L^2([-\beta,\beta)_h;\C)$ satisfies $f(t)=-f(t+\beta)$ for all
 $t\in [-\beta,\beta)_h$ with $t< 0$,
\begin{equation}\label{eq_fourier_inversion_matsubara_1}
f(t)=\frac{1}{2\beta}\sum_{\o\in M_h}e^{i\o t}\hat{f}(\o),\ \forall t\in [-\beta,\beta)_h.
\end{equation}
\end{lemma}  

\begin{proof}
Take any $\o\in W_h\cap 2\pi\Z/\beta$. By assumption we see that 
\begin{equation}\label{eq_fourier_inversion_matsubara_2}
\begin{split}
\hat{f}(\o)&=\frac{1}{h}\sum_{t\in [-\beta,\beta)_h\backslash
 [0,\beta)_h}e^{-i\o t}f(t)+\frac{1}{h}\sum_{t\in [0,\beta)_h}e^{-i\o
 t}f(t)\\
&=-\frac{1}{h}\sum_{t\in [-\beta,\beta)_h\backslash
 [0,\beta)_h}e^{-i\o t}f(t+\beta)+\frac{1}{h}\sum_{t\in [0,\beta)_h}e^{-i\o
 t}f(t)\\
&=-\frac{1}{h}\sum_{t\in [0,\beta)_h}e^{-i\o (t-\beta)}f(t)+\frac{1}{h}\sum_{t\in [0,\beta)_h}e^{-i\o
 t}f(t)=0.
\end{split}
\end{equation}
Then, by \eqref{eq_matsubara}, \eqref{eq_fourier_inversion_matsubara_2} and Lemma
 \ref{lem_fourier_inversion} we obtain \eqref{eq_fourier_inversion_matsubara_1}.\end{proof}

Let us define $g_{\bk}\in L^2([-\beta,\beta)_h;\C)$ $(\bk\in\G^*)$ by 
$$g_{\bk}(t):=e^{t E_{\bk}}\left\{\frac{1_{t\ge 0}}{1+e^{\beta
 E_{\bk}}}-\frac{1_{t<0}}{1+e^{-\beta
 E_{\bk}}}\right\}.$$
Note that the function $g_{\bk}$ satisfies the anti-periodic property $g_{\bk}(t)=-g_{\bk}(t+\beta)$ for all $t\in [-\beta,\beta)_h$ with $t <0$.

\begin{lemma}\label{lem_matsubara_application}
For all $t\in [-\beta,\beta)_h$
\begin{equation}\label{eq_matsubara_application_1}
g_{\bk}(t)=\frac{1}{\beta}\sum_{\o\in M_h}\frac{e^{i\o t}}{h(1- e^{-i\o/h+E_{\bk}/h})}.
\end{equation}
\end{lemma}

\begin{proof}
By Lemma \ref{lem_fourier_inversion_matsubara},
\begin{equation}\label{eq_matsubara_application_2}
g_{\bk}(t)=\frac{1}{2\beta}\sum_{\o\in M_h}e^{i\o t}\hat{g}_{\bk}(\o).
\end{equation}
Moreover, we observe that for $\o\in M_h$
\begin{equation}\label{eq_matsubara_application_3}
\begin{split}
\hat{g}_{\bk}(\o)&=-\frac{1}{h}\sum_{t\in [-\beta,\beta)_h\backslash
 [0,\beta)_h}e^{-i\o t}\frac{e^{tE_{\bk}}}{1+e^{-\beta E_{\bk}}}
+\frac{1}{h}\sum_{t\in [0,\beta)_h}e^{-i\o
 t}\frac{e^{t E_{\bk}}}{1+e^{\beta E_{\bk}}}\\
&=-\frac{1}{h}\sum_{t\in [0,\beta)_h}e^{-i\o (t-\beta)}\frac{e^{t E_{\bk}}}{1+e^{\beta E_{\bk}}}
+\frac{1}{h}\sum_{t\in [0,\beta)_h}e^{-i\o
 t}\frac{e^{tE_{\bk}}}{1+e^{\beta E_{\bk}}}\\
&=\frac{2}{h}\sum_{t\in [0,\beta)_h}e^{-i\o
 t}\frac{e^{tE_{\bk}}}{1+e^{\beta E_{\bk}}}=\frac{2}{h(1+e^{\beta E_{\bk}})}\sum_{t\in
 [0,\beta)_h}e^{t(-i\o+E_{\bk})}\\
&=\frac{2}{h(1-e^{-i\o/h+E_{\bk}/h})}.
\end{split}
\end{equation}
The equality \eqref{eq_matsubara_application_1} follows from \eqref{eq_matsubara_application_2} and \eqref{eq_matsubara_application_3}.
\end{proof}

By substituting the characterization of $g_{\bk}$ given in Lemma
 \ref{lem_matsubara_application} into 
$$C_h(\bx\s x,\by\tau y)=\frac{\delta_{\s,\tau}}{L^d}\sum_{\bk\in\G^*}e^{i\<\bk,\by-\bx\>}g_{\bk}(x-y)
$$
we obtain
\begin{lemma}\label{lem_covariance_characterization}
For any $(\bx,\s, x), (\by,\tau, y) \in \G\times \spin \times [0,\beta)_h$,
\begin{equation}\label{eq_diagonalized_covariance}
C_h(\bx\s x, \by\tau y) = \frac{\delta_{\s,\tau}}{\beta
 L^d}\sum_{\bk\in \G^*}\sum_{\o\in
 M_h}\frac{e^{i\<\bk,\by-\bx\>}e^{-i\o(y-x)}}{h(1-e^{-i\o/h+E_{\bk}/h})}.  
\end{equation}
\end{lemma}

In order to diagonalize $C_h$, we define a matrix\\ 
$Y=(Y(\bk\tau \o,\bx\s x))_{(\bk,\tau,\o)\in
 \G^*\times \spin \times M_h, (\bx,\s,x)\in
 \G\times \spin \times [0,\beta)_h}$ by
\begin{equation*}
Y(\bk\tau \o,\bx\s x):=\frac{\delta_{\tau,\s}}{\sqrt{\beta h
 L^d}}e^{i\<\bk,\bx\>}e^{-i\o x}.
\end{equation*}

\begin{lemma}\label{lem_unitary_Y}
The matrix $Y$ is unitary.
\end{lemma}
\begin{proof}
Assume that $\o=-\pi h+\pi/\beta +2\pi m/\beta$, $\hat{\o}=-\pi
 h+\pi/\beta +2\pi \hat{m}/\beta$ with $m,\hat{m}\in \{0,1,\cdots,\beta
 h-1\}$. Then we observe that
\begin{equation*}
\begin{split}
YY^*(\bk\tau\o,\hat{\bk}\hat{\tau}\hat{\o})&=\frac{\delta_{\tau,\hat{\tau}}}{\beta
 h L^d}\sum_{\bx\in \G}\sum_{x\in
 [0,\beta)_h}e^{i\<\bx,\bk-\hat{\bk}\>}e^{-ix(\o-\hat{\o})}\\
&=\frac{\delta_{\tau,\hat{\tau}}\delta_{\bk,\hat{\bk}}}{\beta
 h }\sum_{l = 0}^{\beta h-1}e^{-i2\pi l(m-\hat{m})/(\beta h)}\\
&=\delta_{\tau,\hat{\tau}}\delta_{\bk,\hat{\bk}}\delta_{m,\hat{m}}=\delta_{\tau,\hat{\tau}}\delta_{\bk,\hat{\bk}}\delta_{\o,\hat{\o}}.
\end{split}
\end{equation*}
Let $x=s/h$, $\hat{x}=\hat{s}/h$ with $s,\hat{s}\in \{0,1,\cdots,\beta h-1\}$.
\begin{equation*}
\begin{split}
Y^*Y(\bx\s x,\hat{\bx}\hat{\s}\hat{x})&=\frac{\delta_{\s,\hat{\s}}}{\beta
 h L^d}\sum_{\bk\in \G^*}\sum_{\o\in
 M_h}e^{-i\<\bk,\bx-\hat{\bx}\>}e^{i\o(x-\hat{x})}\\
&=\frac{\delta_{\s,\hat{\s}}\delta_{\bx,\hat{\bx}}}{\beta
 h }\sum_{m = 0}^{\beta h-1}e^{i(-\pi h+\pi/\beta+2\pi
 m/\beta)(s/h-\hat{s}/h)}\\
&=\frac{\delta_{\s,\hat{\s}}\delta_{\bx,\hat{\bx}}}{\beta
 h }e^{i(-\pi h+\pi/\beta)(s/h-\hat{s}/h)}
\sum_{m = 0}^{\beta h-1}e^{i2\pi
 m(s-\hat{s})/(\beta h)}\\
&=\delta_{\s,\hat{\s}}\delta_{\bx,\hat{\bx}}\delta_{s,\hat{s}}
e^{i(-\pi h+\pi/\beta)(s/h-\hat{s}/h)}=\delta_{\s,\hat{\s}}\delta_{\bx,\hat{\bx}}\delta_{x,\hat{x}}.
\end{split}
\end{equation*}
\end{proof}

By using the matrix $Y$ and \eqref{eq_diagonalized_covariance} we can diagonalize $C_h$ as follows.
\begin{lemma}\label{lem_diagonal_covariance}
For all $(\bk,\tau,\o),(\hat{\bk},\hat{\tau},\hat{\o})\in \G^*\times
 \spin \times M_h$,
$$
(YC_hY^*)(\bk\tau\o,\hat{\bk}\hat{\tau}\hat{\o})=\delta_{\tau,\hat{\tau}}\delta_{\bk,\hat{\bk}}\delta_{\o,\hat{\o}}\frac{1}{1-e^{-i\o/h+E_{\bk}/h}}.  
$$
 \end{lemma}

Finally we calculate the determinant of the covariance matrix $C_h$.
\begin{proposition}\label{pro_determinant_covariance}
For any $h\in 2\N/\beta$
$$\det C_h=\frac{1}{\opPi_{\bk\in \G^*}(1+e^{\beta E_{\bk}})^2}.$$
\end{proposition}
\begin{proof}
Since $\{e^{-i\o/h+E_{\bk}/h}\ |\ \o \in M_h\}$ is the set of all the
 $\beta h$th roots of $-e^{\beta E_{\bk}}$, 
$$z^{\beta h}+e^{\beta E_{\bk}}=\opPi_{\o\in M_h}(z-e^{-i\o/h+E_{\bk}/h})$$
for all $z\in \C$. Especially,
\begin{equation}\label{eq_determinant_covariance_1}
\opPi_{\o\in M_h}(1-e^{-i\o/h+E_{\bk}/h})=1+e^{\beta E_{\bk}}.
\end{equation}
By Lemma \ref{lem_unitary_Y}, Lemma \ref{lem_diagonal_covariance} and
 \eqref{eq_determinant_covariance_1}, we see that
$$
\det C_h=\det (YC_hY^*)=\opPi_{\bk\in \G^*}\opPi_{\s\in \spin}\opPi_{\o\in
 M_h}\frac{1}{1-e^{-i\o/h+E_{\bk}/h}}=\frac{1}{\opPi_{\bk\in \G^*}(1+e^{\beta E_{\bk}})^2}.
$$
\end{proof}

\section*{Acknowledgments}
I would like to thank M. Salmhofer for discussions and an important hint for the proof of Lemma \ref{lem_bound_a_h_n}, as well as for support during the completion of this work. I also wish to thank the referees for their careful reading of the manuscript.

\end{document}